\documentclass[fleqn,11pt,twoside]{article}

\usepackage{amsthm,amsthm,amssymb, color, xcolor,epsfig, graphics, subfigure}

\usepackage{amsmath, graphicx, latexsym, lscape}

\usepackage{float}
\usepackage[hidelinks]{hyperref}
\usepackage{indentfirst}
\newcommand{\zu}{\hspace{0.1mm}}
\newcommand{\zt}{\hspace{0.3mm}}
\newcommand{\zc}{\hspace{0.5mm}}
\usepackage{mathrsfs}
\usepackage{bbold}
\newcommand{\nlb}{\nolinebreak}
\usepackage{tikz}
\usepackage{caption}
\usepackage{enumitem}
\newcommand{\overbar}[1]{\mkern 1.5mu\overline{\mkern-1.5mu#1}}
\usepackage{placeins}

\def\Tr{\mathop{\rm Tr}}
\usepackage{orcidlink}

\makeatletter
\newcommand{\copyrightnote}[2]{{\renewcommand{\thefootnote}{}
 \footnotetext{\small\it
\begin{flushleft}
 \copyright \ #1   #2  
\end{flushleft}}}}

\newcommand{\Name}[1]{\begin{flushleft}
                       \LARGE \bf #1
                       \end{flushleft}\vspace{-3mm}}

\newcommand{\Author}[1]{\begin{flushleft}
                       \it #1 \end{flushleft}}

\newcommand{\Address}[1]{\begin{flushleft}
                       \it #1 \end{flushleft}}

\newcommand{\Date}[1]{\begin{flushleft}
                      \small  \it #1 \end{flushleft}}

\newcommand{\evenhead}{Author \ name}
\newcommand{\oddhead}{Article \ name}

\renewcommand{\@evenhead}{
\hspace*{-3pt}\raisebox{-15pt}[\headheight][0pt]{\vbox{\hbox to \textwidth
{\thepage \hfil \evenhead}\vskip4pt \hrule}}}
\renewcommand{\@oddhead}{
\hspace*{-3pt}\raisebox{-15pt}[\headheight][0pt]{\vbox{\hbox to \textwidth
{\oddhead \hfil \thepage}\vskip4pt\hrule}}}
\renewcommand{\@evenfoot}{}
\renewcommand{\@oddfoot}{}

\long\def\@makecaption#1#2{%
  \vskip\abovecaptionskip
  \sbox\@tempboxa{\small \textbf{#1.}\ \ #2}%
  \ifdim \wd\@tempboxa >\hsize
    {\small \textbf{#1.}\ \ #2}\par
  \else
    \global \@minipagefalse
    \hb@xt@\hsize{\hfil\box\@tempboxa\hfil}%
  \fi
  \vskip\belowcaptionskip}

\newcommand{\JNMPnumberwithin}[3][\arabic]{%
  \@ifundefined{c@#2}{\@nocounterr{#2}}{%
    \@ifundefined{c@#3}{\@nocnterr{#3}}{%
      \@addtoreset{#2}{#3}%
      \@xp\xdef\csname the#2\endcsname{%
        \@xp\@nx\csname the#3\endcsname .\@nx#1{#2}}}}%
}

\newcommand{\resetfootnoterule} {
  \renewcommand\footnoterule{%
  \kern-3\p@
  \hrule\@width.4\columnwidth
  \kern2.6\p@}
}

\renewcommand{\footnoterule}{}

\makeatother

\theoremstyle{definition}

\begin{document}

\renewcommand{\evenhead}{ {\LARGE\textcolor{blue!10!black!40!green}{{\sf \ \ \ ]ocnmp[}}}\strut\hfill 
C.A. da Silva and L.A. Ferreira
}
\renewcommand{\oddhead}{ {\LARGE\textcolor{blue!10!black!40!green}{{\sf ]ocnmp[}}}\ \ \ \ \  
Gauge-invariant dynamical charges and the 1-instanton solution
}

\thispagestyle{empty}
\newcommand{\FistPageHead}[3]{
\begin{flushleft}
\raisebox{8mm}[0pt][0pt]
{\footnotesize \sf
\parbox{150mm}{{\textcolor{blue!10!black!40!green}{{\bf Open Communications in Nonlinear Mathematical Physics}}}
\ \ {Special Issue: Hietarinta}, 2026\\[0.1cm]
\strut\hfill 
ocnmp:18516
pp #2\hfill {\sc #3}}}\vspace{-13mm}
\end{flushleft}}

\FistPageHead{1}{\pageref{firstpage}--\pageref{lastpage}}{ \ \ }

\strut\hfill

\strut\hfill

\copyrightnote{The authors. Distributed under a Creative Commons Attribution 4.0 International License}

\begin{center}

{\bf {\large A Special OCNMP Issue in Honour of Jarmo Hietarinta}}\\[0.2cm]
{\bf {\large on the Occasion of his 80th Birthday}}
\end{center}

\smallskip

\Name{On the gauge-invariant dynamical charges and densities of the 1-instanton solution}

\Author{Caio Angelo da Silva$^{\hspace{0.5mm}\ast\hspace{0.25mm}a}$ and Luiz Agostinho Ferreira$^{\hspace{0.5mm}\dagger\hspace{0.15mm}a}$}

\Address{{}$^{\ast\hspace{0.3mm}}$\orcidlink{0009-0008-2507-9508} \hspace{-2mm} caioangelodasilva@usp.br , {}$^{\dagger\hspace{0.25mm}}$\orcidlink{0000-0003-3679-4722} \hspace{-2mm} laf@ifsc.usp.br\\
{}$^{a}$Instituto de Física de São Carlos, Universidade de São Paulo, IFSC - USP, 13566-590, São Carlos, SP, Brasil.}

\Date{Received June 15, 2026; Accepted June 23, 2026}

\setcounter{equation}{0}

\smallskip

\noindent
{\bf Citation format for this Article:}\newline
 C.A. da Silva and L.A. Ferreira,
On the gauge-invariant dynamical charges and densities of the 1-instanton solution,
{\it Open Commun. Nonlinear Math. Phys.}, Special Issue:\,Hietarinta, ocnmp:18516, \pageref{firstpage}--\pageref{lastpage}, 2026.

\strut\hfill

\noindent
{\bf The permanent Digital Object Identifier (DOI) for this Article:}\newline
{\it 10.46298/ocnmp.18516}
\strut\hfill

\begin{abstract}

\noindent 
We study the gauge-invariant dynamically conserved charges, and their corresponding densities, for instanton solutions of Yang-Mills theories in four dimensional Euclidean space, for the gauge group $SU(2)$. Those charges were constructed in \cite{ym1,ym2} through the integral equations of Yang-Mills theory, using techniques on generalized loop spaces. We use the integral non-Abelian Gauss law to evaluate the gauge-invariant  flux of the magnetic and electric non-Abelian fields through spherical surfaces centered at the origin of the instanton solution. From such a flux, we define  gauge-invariant charge densities by considering the charge within an infinitesimal spherical shell of  radius $r\equiv\sqrt{x_i \zt x^i}/\lambda$, with $\lambda$ being the parameter of the instanton solution, defining its size, and $x_i \zt x^i = (x^1)^2 + (x^{2\zu})^2 + (x^{3\zu})^2$. We discuss the issue of the reparameterization invariance of the charges and densities, and show that the magnetic and electric fluxes for the instanton and anti-instanton, at $r=1$ and $x^4 = 0$, $x^4$ being the Euclidean time, are non-zero and observable. Our results give an interesting picture of the internal structure of the instanton, and may be important for the properties of the Yang-Mills $\theta$-vacuum.

\end{abstract}

\label{firstpage}


\section{Introduction}
\label{sec:introduction}
\setcounter{equation}{0}

The integral Yang-Mills equations, constructed in \cite{ym1,ym2}, are the generalization, to non-Abelian gauge theories, of the well known integral laws (Gauss, Faraday, etc) of Maxwell theory. They provide powerful tools to investigate the global aspects of gauge theories, and solve a long standing problem in Yang-Mills theory, namely, the construction of gauge invariant conserved electric and magnetic non-Abelian charges. Those charges correspond to the eigenvalues of the holonomy of a flat one-form connection ${\cal A}$ on the generalized loop space of mappings from the two-sphere to space-time $M$, $S^2\rightarrow M$. The techniques involved closely resemble those of integrable field theories \cite{lax,zakharov,faddeevbook}, as they show that the classical dynamics of Yang-Mills theory, coupled to spin $0$ and $1/2$ matter fields, can be cast in the form of a zero curvature condition for the connection ${\cal A}$ on that generalized loop space. In particular, such results apply equally well to the Standard Model of the fundamental interactions, namely Quantum Chromodynamics (QCD) and the Weinberg-Salam model. In fact, it has been shown in \cite{fourym} that such theories possess an exact integrable structure, on the internal space of the non-Abelian electric and magnetic charges, where the flat connection ${\cal A}$ satisfies a Fundamental Poisson Relation (FPR), and the charge operator (its holonomy) a Sklyanin relation \cite{faddeevleshouches,retore}. The FPR gives the Poisson brackets among the entries of the matrix ${\cal A}$ as a Lie bracket of a classical Yang-Baxter matrix and ${\cal A}$. In its turn, the Sklyanin relation gives the Poisson bracket of the matrix entries of that charge operator as a Lie bracket of itself with the Yang-Baxter matrix. The Sklyanin relation leads to the involution of the conserved charges, which are the eigenvalues of the charge operator. The symmetries associated to such integrable structures  are canonical transformations generated by the conserved charges through the Poisson bracket \cite{slavnovbook}, and may relate to the so-called generalized global symmetries \cite{ggs1,ggs2,ggs3,ggs4}. Clearly, Yang-Mills theories are not integrable in the usual sense, as its scattering $S$-matrix does not factorize into two particle $S$-matrices, i.e. it does present particle production. However, Yang-Mills theories do posses a new kind of integrability on the internal space of non-Abelian electric and magnetic charges. 

\phantom{Paragraph}

In this work, we study those gauge invariant conserved charges in the context of the instanton solutions of the pure Yang-Mills theory in four dimensional Euclidean space-time, for the gauge group $SU(2)$.  We will begin with the presentation of the non-Abelian Stokes theorem for a 2-form $B_{\mu\nu} \hspace{0.6mm} dx^\mu \hspace{-0.2mm} \wedge \hspace{-0.2mm} dx^\nu$ defined on spacetime. Then we will show, by substituting the Yang-Mills differential equations and the Bianchi identity, and by making use of two arbitrary parameters $\alpha$ and $\beta$ that can be introduced in the integral formulation, that one is led to an infinite number of integral equations for Yang-Mills theories. We will see that this non-Abelian Stokes theorem also employs a 1-form $A_\mu \zc dx^\mu$, defined on spacetime, through a conjugation by the Wilson line $W$, associated to such  $A_\mu$\zc{\nlb}. It will be shown that, because of this conjugation, all the infinite integral equations  transforms covariantly by gauge transformations. Thus, gauge-invariant quantities can be easily defined by taking traces of such operators or by considering their eigenvalues. The surfaces and volumes to be employed in the integral equations must be scanned with loops (closed 1-dimensional paths on spacetime), based on a fixed reference point $x_R$\zc . The integral formulation to be employed here is in $(3+1)$-dimensions (more precisely, in a 4-dimensional Euclidean spacetime, since we are dealing with the instanton solution), and it will be explained that the quantities calculated are related with the generalized loop space $\mathcal{L}^{(2)}$ of maps from the $2$-sphere to space-time, with a base point.

\phantom{Paragraph}

The formulation on loop space leads in fact to an infinite number of integral equations, due to the appearance of two arbitrary parameters $\alpha$ and $\beta$. For the present work, we will focus on the `first-order' integral equations, which are the non-Abelian generalizations of the integral laws of Maxwell's  Electrodynamics. We will consider a purely spatial volume $\Omega$ on Euclidean spacetime, in particular a sphere of radius $r$, centered at the origin, and whose parameterization by loops was constructed in \cite{directtest}. With a completely spatial integration volume $\Omega$, these first-order integral equations will correspond to the non-Abelian versions of Gauss law for the magnetic and electric fields. We will make use of the instanton and anti-instanton solutions of an $SU(2)$ pure Yang-Mills theory in 4-dimensional Euclidean spacetime, and since these solutions are self-dual, the two non-Abelian Gauss laws will coincide. Thus, the results we shall obtain will correspond to both magnetic and electric \nlb ones.

\phantom{Paragraph}

For the non-Abelian Gauss law for the magnetic (or electric) fields, we have that this equation transforms covariantly by gauge transformations, in particular by a conjugation of a fixed group element, evaluated at the reference point $x_R$\zc . Therefore, we will define a gauge-invariant magnetic (or electric) flux, calculated on $\partial\Omega$, by considering the trace of the surface-integral term of the non-Abelian Gauss law. This gauge-invariant flux will be evaluated and plotted for the instanton and anti-instanton solutions, for different radii of $\partial\Omega$ and different Euclidean times $x^4$. For these results to be considered observable, one needs to check if the quantities obtained are invariant by a reparameterization of the surfaces employed. Indeed, for such results to be considered physical, they evidently must not depend on the arbitrary choice of scanning of the surfaces by loops. All these concepts will be explained in detail, and the paper is self-contained. For the gauge-invariant fluxes plotted for the instanton and anti-instanton, it will be shown that the result for $r=1$ and $x^4=0$ is observable, since it is gauge-invariant and also reparameterization-invariant. This observable magnetic (or electric) flux, which corresponds to an enclosed magnetic (or electric) charge for the instanton and anti-instanton, indicates that these solutions have an internal charge configuration, coming from the non-Abelian Gauss law obtained from our integral equations. For other radii and Euclidean times, it will be shown that the gauge-invariant fluxes obtained are not reparameterization-invariant, and thus not observable. For these cases, we will see that such results do not live in the physical spacetime, but rather on the aforementioned generalized loop space $\mathcal{L}^{(2)}$.

\phantom{Paragraph}

Making use of the gauge-invariant flux above, obtained from the non-Abelian version of Gauss law, we will define a gauge-invariant charge density by considering the charge within an infinitesimal spherical shell, of radius $r$. Then, we will calculate this quantity for the instanton and anti-instanton solutions, for different radii and Euclidean times. We will see that the results obtained are not reparameterization-invariant, apart from some critical radii where the charge density is zero. These critical points are observable, but of less physical relevance since \nlb the charge density vanishes. For the remaining radii and Euclidean times, the gauge-invariant \nlb magnetic (or electric) charge density is not observable, but it may still be of interest since it indicates an overall internal charge configuration for the instanton and anti-instanton. From the observable flux mentioned above, such a charge configuration must indeed be present for these solutions.

\phantom{Paragraph}

With this brief summary of what we shall discuss, let us start with the presentation of the integral formulation of Yang-Mills theories, and with the construction of the integral equations.

\section{The integral formulation of Yang-Mills theories}
\label{sec:integral formulation}
\setcounter{equation}{0}

\subsection{Yang-Mills integral equations}
\label{subsec:integral equations}

The first step to obtain an integral formulation for Yang-Mills theories is to derive the non-Abelian version of Stokes theorem for a 2-form \zu $B_{\mu\nu} \hspace{0.8mm} dx^\mu \hspace{-0.5mm} \wedge \hspace{-0.2mm} dx^\nu$\zt , defined on spacetime. This theorem was first obtained in \cite{afg1}, and it starts with a `Wilson surface' $V$, defined by the equation:

\begin{equation}\label{eq: wilson surface definition}
\frac{dV}{d\tau} - ie V T = 0 \hspace{1cm} \text{with} \hspace{1cm} T \equiv \int_{\sigma_i}^{\sigma_f} d\sigma \zc W^{-1} B_{\mu\nu} W \zc \frac{dx^\mu}{d\sigma} \frac{dx^\nu}{d\tau}
\end{equation}

\vspace{.2cm}

One very important thing to note about this definition is that it involves a conjugation of $B_{\mu\nu}$ by the Wilson line $W$, which is given by its usual definition:

\begin{equation}\label{eq: wilson line definition}
\frac{dW}{d\sigma} + ie A_\mu \frac{dx^\mu}{d\sigma}W = 0 \hspace{0.75cm} \Rightarrow \hspace{.75cm} W \equiv P_1 \exp \Bigg( \hspace{-0.7mm} -ie \hspace{-0.5mm} \int_{\gamma} d\sigma A_\mu \frac{dx^\mu}{d\sigma} \hspace{0.3mm} \Bigg) W_R
\end{equation}

\vspace{0.225cm}

\noindent{}and where the path-ordered exponential $P_1 \exp$ corresponds to the following expression for $W$:

\vspace{-0.275cm}

\begin{equation}\label{eq: wilson line path ordered expression}
\begin{split}
W = & \zc \Bigg( \mathbb{1} - ie \int_{\sigma_i}^{\sigma_f} d\sigma A_{\mu_1} (\sigma) \frac{dx^{\mu_1}}{d\sigma} + (ie)^2 \int_{\sigma_i}^{\sigma_f} d\sigma A_{\mu_1} (\sigma) \frac{dx^{\mu_1}}{d\sigma} \int_{\sigma_i}^{\sigma} d\sigma^\prime A_{\mu_2} (\sigma^\prime) \frac{dx^{\mu_2}}{d\sigma^\prime} \zt - \\[0.3cm]
& - (ie)^3 \int_{\sigma_i}^{\sigma_f} d\sigma A_{\mu_1} (\sigma) \frac{dx^{\mu_1}}{d\sigma} \int_{\sigma_i}^{\sigma} d\sigma^\prime A_{\mu_2} (\sigma^\prime) \frac{dx^{\mu_2}}{d\sigma^\prime} \int_{\sigma_i}^{\sigma^\prime} d\sigma^{\prime\prime} A_{\mu_3} (\sigma^{\prime\prime}) \frac{dx^{\mu_3}}{d\sigma^{\prime\prime}} +  \cdots \hspace{-0.5mm}  \Bigg) \zc W_R
\end{split}
\end{equation}

\vspace{0.225cm}

In the above equations, $\gamma$ is a 1-dimensional path on spacetime, parameterized by $\sigma$, that goes from a reference point $x_R$ to an arbitrary point $x$. Such paramater runs from $\sigma_i$\zc , corresponding to the initial point $x_R$\zc , to $\sigma_f$\zc , corresponding to the final point $x$. The symbol $P_1$ denotes the usual path ordering of the solution, on the parameter $\sigma$. $W_R$ is the integration constant, which is the initial value of $W$ at the reference point $x_R$\zc , and $e$ is the gauge coupling constant.

From (\ref{eq: wilson line definition}), we see that our starting equation (\ref{eq: wilson surface definition}) involves also a 1-form $A_\mu \zc dx^\mu$, defined on spacetime and introduced by the conjugation of $W$. This conjugation will be present throughout the paper, and thus, we shall make use from now on of the following notation:

\begin{equation}\label{eq: W conjugation notation}
X^{W} \equiv W^{-1} X W
\end{equation}

\vspace{0.2cm}

The presence of $W$ on definition (\ref{eq: wilson surface definition}) means that we need to have well defined paths on spacetime for the Wilson lines to be calculated. In fact, the surfaces on which our Wilson surface $V$ will be defined must be constructed in the following way: we shall begin with closed paths, or \textit{loops}, beginning and ending on a \textit{fixed} reference point $x_R$ on spacetime. For a given loop, we have its points parameterized by $\sigma$, with $\sigma_i$ and $\sigma_f$ corresponding to $x_R$\zc . Next, we construct our 2-dimensional surface $\Sigma$ on spacetime by considering a \textit{family} of such loops. Each loop in this family is parameterized by the parameter $\tau$, whith $\tau_i$ being the infinitesimal loop around the fixed point $x_R$\zc , and $\tau_f$ being the final loop which is the border $\partial\Sigma$ of $\Sigma$.

Therefore, we see that the quantity $T$ on (\ref{eq: wilson surface definition}) is defined on a particular loop of parameter $\tau$, since we have an integration on $\sigma$, from $\sigma_i$ to $\sigma_f$\zc . So, we have the Wilson line $W$ being calculated on the points of each loop, and the Wilson surface $V$ is then obtained by integrating (\ref{eq: wilson surface definition}) on the parameter $\tau$, from $\tau_i$ to $\tau_f$\zc . We have $V$ thus being calculated on a surface (built with loops), which motivate its name. The solution of (\ref{eq: wilson surface definition}) is very similar to (\ref{eq: wilson line path ordered expression}):

\vspace{-0.25cm}

\begin{equation}\label{eq: V solution on surface}
\begin{gathered}
\hspace{-4.5cm} V =  V_R \hspace{0.9mm} \Bigg( \hspace{.4mm} \mathbb{1} + ie \hspace{-0.3mm} \int_{\tau_i}^{\tau_f} d\tau \hspace{0.5mm} T (\tau) + (ie)^2 \hspace{-0.5mm} \int_{\tau_i}^{\tau_f} d\tau \hspace{-0.5mm} \int_{\tau_i}^{\tau} d\tau^\prime \hspace{0.5mm} T \big(\tau^\prime\big) \hspace{0.3mm} T(\tau) \hspace{0.5mm} + \\[3.0mm]
\hspace{1.5cm} + \hspace{0.3mm} (ie)^3 \hspace{-0.3mm} \int_{\tau_i}^{\tau_f} d\tau \hspace{-0.5mm} \int_{\tau_i}^{\tau} d\tau^\prime \hspace{-0.5mm} \int_{\tau_i}^{\tau^\prime} \hspace{-0.5mm} d\tau^{\prime\prime} \hspace{0.5mm} T\big(\tau^{\prime\prime}\big) \hspace{0.3mm} T\big(\tau^\prime\big) \hspace{0.3mm} T(\tau) + \cdots \hspace{-0.2mm} \Bigg) \hspace{0.2mm} \equiv \hspace{0.2mm} V_R \hspace{0.4mm} P_2 \exp \Bigg( ie \int_{\Sigma} d\tau \hspace{0.5mm} T \Bigg) 
\end{gathered}
\end{equation}

\vspace{0.25cm}

\noindent{}but here, we see that the ordering is the opposite of (\ref{eq: wilson line path ordered expression}), and because of that the integration constant $V_R$ (which is the initial value of $V$ at the infinitesimal surface around $x_R$) appears on the left. Also, we have here the symbol $P_2$ denoting the surface ordering of the solution, in this case on the parameter $\tau$ as indicated on the explicit expression above.

\phantom{Paragraph}

Having defined our quantity $V$, the derivation of the non-Abelian Stokes theorem for $B_{\mu\nu}$ is actually quite straightforward. The ideia now is to consider a \textit{closed} surface on spacetime, built with loops based on a fixed reference point $x_R$ as explained above. From equation (\ref{eq: wilson surface definition}), we can calculate $V$ on this closed surface, which is given by expression (\ref{eq: V solution on surface}).

The procedure then, for obtaing the non-Abelian Stokes theorem, is to consider a \textit{deformation} of this closed surface, in such a way that it sweeps a 3-dimensional volume $\Omega$ in spacetime. We start with an infinitesimal closed surface around $x_R$\zc , and inflate it until it reaches a final surface, which corresponds to the border $\partial\Omega$ of the volume $\Omega$ that was scanned. Our non-Abelian Stokes theorem will then equate the Wilson surface $V$ calculated directly on $\partial\Omega$, using equation (\ref{eq: wilson surface definition}), with the (same) quantity $V$ calculated over the volume $\Omega$ that has $\partial\Omega$ as its border.

\phantom{Paragraph}

What we can do then is consider a \textit{variation} of equation (\ref{eq: wilson surface definition}), related with the deformation $x^\mu \rightarrow x^\mu + \delta x^\mu$ of the points of an arbitrary closed surface that scans the 3-dimensional volume. In fact, what we want is to obtain an equation for the corresponding variation $\delta V$, which we are then going to relate with the volume-scanning procedure described above.

The easiest way to obtain $\delta V$ is not by varying its expression in (\ref{eq: V solution on surface}), but instead by varying its defining equation (\ref{eq: wilson surface definition}). Doing so, we obtain:

\begin{equation}\label{eq: V variation manipulation 1}
\delta \Bigg( \frac{dV}{d\tau} - ie V T \Bigg) = 0 \hspace{0.75cm} \Rightarrow \hspace{.7cm} \frac{d\delta V}{d\tau} V^{-1} - ie (\delta V) T V^{-1} -ie V (\delta T) V^{-1} = 0
\end{equation}

\vspace{.2cm}

\noindent{}where we have multiplied the expression by $V^{-1}$ from the right. Considering then the differential equation for $V^{-1}$ given below, and multiplying it by $\delta V$ from the left:

\begin{equation}\label{eq: V variation manipulation 2}
\frac{dV^{-1}}{d\tau} + ie T V^{-1} = 0 \hspace{0.8cm} \Rightarrow \hspace{.75cm} \delta V \frac{dV^{-1}}{d\tau} + ie (\delta V ) T V^{-1} = 0
\end{equation}

\vspace{.2cm}

\noindent{}we see that the sum of (\ref{eq: V variation manipulation 1}) with (\ref{eq: V variation manipulation 2}) gives, after a cancelation of two of the terms:

\vspace{-0.15cm}

\begin{equation}\label{eq: V variation manipulation 3}
\frac{d}{d\tau} \big( \delta V . V^{-1} \big) = ie V \big( \delta T \big) V^{-1} \hspace{0.8cm} \Rightarrow \hspace{.75cm} \big( \delta V . V^{-1} \big) (\tau_f) = ie \int_{\tau_i}^{\tau_f} d\tau \hspace{0.4mm} V \big( \delta T \big) V^{-1}
\end{equation}

\vspace{.2cm}

The second expression above was obtained from the first by integrating in $\tau$ and using that $\delta V (\tau_i) = 0$. This relation is true because we consider $x_R$ to be fixed, apart from the deformation of the rest of the surface. So, at $\tau_i$\zc , there is yet no change in the value of $V$, and thus $\delta V (\tau_i) = 0$.

\phantom{Paragraph}

The next step is to evaluate the variation $\delta T$ present in expression (\ref{eq: V variation manipulation 3}) above, using the definition of $T$ in (\ref{eq: wilson surface definition}), and to simplify the expression obtained (performing, for instance, the integrations in both $\tau$ and $\sigma$, which will eliminate some terms of total derivative in these variables). This is a long, but straightforward calculation, which have now been detailed in a handful of papers by our group, and which we shall not repeat here. The only detail that we must mention is that, for the variation of the Wilson line (and its inverse), we again consider the variation not of its path-ordered solution, but instead of its defining equation given in (\ref{eq: wilson line definition}). After varying this differential equation, we must then integrate it until some arbitrary $\sigma$ (and not $\sigma_f$\zc !\zu), since we are going to need $\delta W$ (and $\delta W^{-1}$) inside the $\sigma$ integral on $T$, defined in equation (\ref{eq: wilson surface definition}). For more details about these calculations, one can consult the references \cite{afg1,afg2,ym1,ym2}.

The result that is obtained, after all appropriate simplifications, and after multiplying the second expression of (\ref{eq: V variation manipulation 3}) by $V(\tau_f)$ from the right, is the expression:

\begin{equation}\label{eq: V variation manipulation 4}
\delta V = ie \hspace{0.3mm} \scalebox{1.5}{$\kappa$}_\delta \hspace{0.1mm} V
\end{equation}

\vspace{0.2cm}

\noindent{}where we have dropped the $(\tau_f)$ notation, since all the terms are evaluated at it, and where:

\vspace{-0.3cm}

\begin{equation}\label{eq: kappa delta definition}
\begin{gathered}
\hspace{-1.6cm}\scalebox{1.5}{$\kappa$}_\delta \equiv \int_{\tau_i}^{\tau_f} \hspace{-0.3mm} d\tau \hspace{0.5mm} V \hspace{0.3mm} \Bigg\{ \hspace{-0.9mm} -ie \bigg[ \int_{\sigma_i}^{\sigma_f} \hspace{-0.3mm} d\sigma^\prime B_{\kappa\rho}^{\hspace{0.2mm} W} (\sigma^\prime) \frac{dx^\kappa}{d\sigma^\prime} \frac{dx^\rho}{d\tau} (\sigma^\prime) \hspace{1mm} , \int_{\sigma_i}^{\sigma_f} \hspace{-0.3mm} d\sigma B_{\mu\nu}^{\hspace{0.2mm} W} (\sigma) \frac{dx^\mu}{d\sigma} \delta x^\nu (\sigma) \hspace{0.3mm} \bigg] + \\[3.5mm]
\hspace{-0.1cm} + \int_{\sigma_i}^{\sigma_f} d\sigma \bigg\{ \big( D_{\mu} B_{\nu\lambda} + D_{\lambda} B_{\mu\nu} + D_\nu B_{\lambda\mu}  \big)^W \frac{dx^\mu}{d\sigma} \frac{dx^\nu}{d\tau} \delta x^\lambda + \\[3.5mm]
\hspace{3.7cm} + \hspace{0.3mm} ie \bigg[ B_{\mu\nu}^{\hspace{0.4mm}W} \hspace{0.3mm} (\sigma) \hspace{0.3mm} , \int_{\sigma_i}^{\sigma} d\sigma^\prime F_{\kappa\rho}^{\hspace{0.2mm} W} \hspace{0.3mm} (\sigma^\prime) \frac{dx^\kappa}{d\sigma^\prime} \delta x^{\rho} (\sigma^\prime)  \bigg] \frac{dx^\mu}{d\sigma} \frac{dx^\nu}{d\tau} (\sigma) - \\[3.5mm] 
\hspace{4.7cm} - \hspace{0.3mm} ie \bigg[ B_{\mu\nu}^{\hspace{0.4mm}W} \hspace{0.3mm} (\sigma) \hspace{0.3mm} , \int_{\sigma_i}^{\sigma} d\sigma^\prime F_{\kappa\rho}^{\hspace{0.2mm} W} \hspace{0.3mm} (\sigma^\prime) \frac{dx^\kappa}{d\sigma^\prime} \frac{dx^\rho}{d\tau} (\sigma^\prime)  \bigg] \frac{dx^\mu}{d\sigma} \delta x^\nu (\sigma) \bigg\} \Bigg\} \hspace{0.5mm} V^{-1}
\end{gathered}
\end{equation}

\vspace{0.25cm}

This big quantity deserves some important comments. First, note that all terms in this equation appear conjugated by the Wilson line, following our $X^W$ notation introduced in (\ref{eq: W conjugation notation}). Second, note the natural appearance of the quantities $F_{\mu\nu} \equiv \partial_\mu A_\nu - \partial_\nu A_\mu + ie \zu\zu [\zu A_\mu \zt , A_\nu \zu\zu]$ and $D_\lambda B_{\mu\nu} \equiv \partial_\lambda B_{\mu\nu} + ie \hspace{0.3mm} [ \hspace{0.3mm} A_\lambda \hspace{0.3mm} , B_{\mu\nu} \hspace{0.3mm}  ]$\zc , defined as $F_{\mu\nu}$ and $D_{\lambda} B_{\mu\nu}$ for obvious reasons. Most importantly, note the natural appearence of the structure $D_{\mu} B_{\nu\lambda} + D_{\lambda} B_{\mu\nu} + D_\nu B_{\lambda\mu}$\zc , which presents itself ready for the substitution of the Bianchi identity and Yang-Mills equations, if we substitute $B_{\mu\nu}$ by $F_{\mu\nu}$ and its Hodge dual $\widetilde{F}_{\mu\nu}$ (which is exactly what we are going to do). Also, note in the expression above the presence of commutators, as well as of coupled integrals (in the parameter $\sigma$) for the last two terms. For the first term of (\ref{eq: kappa delta definition}), the two $\sigma$-integrals appear \nlb decoupled.

\phantom{Paragraph}

Having obtained equation (\ref{eq: V variation manipulation 4}) for the variation of $V$, as a result of the deformation of our surface, we now turn back to the volume-scanning procedure described before and consider this deformation to be the inflation of the surface, scanning some volume in spacetime. For this scanning we consider a parameter $\zeta$, in such a way that $\zeta_i$ corresponds to the initial infinitesimal surface around $x_R$\zc , and $\zeta_f$ corresponds to the final surface which is the border $\partial\Omega$ of the volume $\Omega$ that is being scanned. In other words, we are scanning a 3-dimensional volume in spacetime with a family of 2-dimensional closed surfaces, parameterized by $\zeta$. Each of these closed surfaces are scanned by 1-dimensional closed paths (loops), based on $x_R$ and parameterized by $\tau$. Finally, each of these loops have its points parameterized by $\sigma$. This loop-based scheme is fundamental for the integral formulation that is being described here, and has very important consequences for Yang-Mills theories as noted in some recent works by our group \cite{fourym,threeym,twoym}.

\phantom{Paragraph}

So, considering our variation to be in the $\zeta$-direction, we have that equation (\ref{eq: V variation manipulation 4}) turns into:

\begin{equation}\label{eq: wilson surface zeta equation}
\frac{dV}{d\zeta} - ie \hspace{0.3mm} \scalebox{1.5}{$\kappa$} \hspace{0.1mm} V = 0
\end{equation}

\vspace{0.25cm}

\noindent{}where the quantity $\scalebox{1.5}{$\kappa$}$ above is obtained from (\ref{eq: kappa delta definition}) by changing the $\delta x^\mu$\zc 's to $dx^\mu/d\zeta$\zc\zt 's\zt{}.

Now, we have two differential equations for calculating $V$: we can employ (\ref{eq: wilson surface definition}) and calculate it directly on $\partial\Omega$, or we can employ equation (\ref{eq: wilson surface zeta equation}) above and obtain $V$ through an integration over the volume $\Omega$. Equating these two solutions we obtain the important result:

\begin{equation}\label{eq: stokes two form}
V_R \hspace{0.3mm} P_2 \exp \Bigg( ie \int_{\partial \Omega} d\tau \hspace{0.5mm} T \Bigg) = P_3 \exp \Bigg( ie \hspace{-0.5mm} \int_{\Omega} d\zeta \hspace{0.3mm} \scalebox{1.5}{$\kappa$}  \Bigg) V_R 
\end{equation}

\vspace{0.25cm}

This is the non-Abelian Stokes theorem for the 2-form $B_{\mu\nu}$\zc (with a 1-form $A_\mu$ introduced via the Wilson line). Note that the two quantities appear with opposite orderings, and that for the last quantity $P_3$ denotes a volume ordering, in this case on the parameter $\zeta$.

For this result, we may denote its left-hand side by $V(\partial\Omega)$, which corresponds to $V$ calculated directly on $\partial\Omega$, and its right-hand side by $V(\Omega)$, which corresponds to the same quantity $V$ but calculated with the volume integral over $\Omega$. With this notation, (\ref{eq: stokes two form}) may be written as:

\begin{equation}\label{eq: stokes two form compact}
V(\zu\partial\Omega\zu) = V(\Omega)
\end{equation}

\vspace{0.25cm}

With the theorem (\ref{eq: stokes two form}) at hand, it is easy to obtain the integral equations for Yang-Mills theories: we just substitute the 2-form $B_{\mu\nu}$ for $F_{\mu\nu}$ and $\widetilde{F}_{\mu\nu}$\zc , and consider the 1-form $A_\mu$ to be the gauge fields associated with $F_{\mu\nu}$\zc . In particular, we make the following substitution:

\begin{equation}\label{eq: 2-form substitution}
B_{\mu\nu} \equiv \alpha F_{\mu\nu} + \beta \widetilde{F}_{\mu\nu}
\end{equation}

\vspace{0.25cm}

\noindent{}where $\alpha$ and $\beta$ are arbitrary parameters whose practical purpose will be clear in a moment.

The calculation then is to make the substitution (\ref{eq: 2-form substitution}) on both sides of the Stokes theorem (\ref{eq: stokes two form}), and make use of the Bianchi identity and Yang-Mills equations given below:

\begin{equation}\label{eq: bianchi ym}
D_{\lambda} F_{\mu\nu} + D_{\nu} F_{\lambda\mu} + D_{\mu} F_{\nu\lambda} = 0 \hspace{.85cm} , \hspace{.75cm} D_{\lambda} \widetilde{F}_{\mu\nu} + D_{\nu} \widetilde{F}_{\lambda\mu} + D_{\mu} \widetilde{F}_{\nu\lambda} = \widetilde{J}_{\lambda\mu\nu}
\end{equation}

\vspace{0.25cm}

\noindent{}where $\widetilde{J}_{\lambda\mu\nu} \equiv \varepsilon_{\lambda\mu\nu\rho} J^{\rho}$ and where $\widetilde{F}_{\mu\nu} \equiv \varepsilon_{\mu\nu\rho\sigma} \zu\zu F^{\rho\sigma}/2$\zc . In fact, one can easily show that the second expression above is equivalent to $D_{\mu} F^{\mu\nu} = J^\nu$, which are indeed the Yang-Mills equations.

In particular, it is the substitution of (\ref{eq: bianchi ym}) that turn the (mathematical) non-Abelian Stokes theorem into the (dynamical) integral equations of Yang-Mills theories. Considering an infinitesimal volume at $x_R$\zc , one can show \cite{ym1,ym2} that these integral equations reduce to the usual differential equations (\ref{eq: bianchi ym}), at the point $x_R$\zc . Thus, we see that the integral formulation presented here, obtained for the first time by our group in 2012 \cite{ym1,ym2}, is indeed equivalent to the differential formulation proposed by C. N. Yang and R. L. Mills in 1954 \cite{ymoriginal}.

\phantom{Paragraph}

Now, we have the Yang-Mills integral equations given by an equality between two ordered exponentials. Since we introduced $\alpha$ and $\beta$ in equation ($\ref{eq: 2-form substitution}$), what we can do is expand both sides of (\ref{eq: stokes two form}) into two series on these parameters. Doing so, we obtain expressions of the form:

\vspace{-0.25cm}

\begin{equation}\label{eq: V border expansion}
V(\partial\Omega) = V_R + \alpha \zt V_\alpha(\partial\Omega) + \beta \zc V_\beta\zt (\partial\Omega) + \alpha^2 \zt V_{\alpha^2 \zu\zu}(\partial\Omega) + \beta^2 \zt V_{\beta^2 \zu\zu}(\partial\Omega) + \alpha\beta \zt V_{\alpha\beta \zt}(\partial\Omega) + \cdots
\end{equation}

\vspace{-0.6cm}

\begin{equation}\label{eq: V volume expansion}
V(\Omega) = V_R + \alpha \zt V_\alpha(\Omega) + \beta \zc V_\beta\zt (\Omega) + \alpha^2 \zt V_{\alpha^2 \zu\zu}(\Omega) + \beta^2 \zt V_{\beta^2 \zu\zu}(\Omega) + \alpha\beta \zt V_{\alpha\beta \zt}(\Omega) + \cdots
\end{equation}

\vspace{0.2cm}

\noindent{}where the coefficients $V_\alpha$\zc , $V_\beta$\zc , etc. must be determined from the corresponding expansions of the left- and right-hand sides of equation (\ref{eq: stokes two form}), considering (\ref{eq: 2-form substitution}). From equation (\ref{eq: V solution on surface}), note that all such coefficients will have the integration constant $V_R$ present inside their expressions.

With the expansions (\ref{eq: V border expansion}) and (\ref{eq: V volume expansion}) for the two sides of (\ref{eq: stokes two form}), one then makes the formal identification of the coefficients of same order in $\alpha$, $\beta$, etc. Doing so, the integral formulation of Yang-Mills theories finds itself given in terms of an \textit{infinite} number of integral equations.

\phantom{Paragraph}

The first of these equations, which we shall refer to as the equations of ``order $\alpha$'' and of ``order $\beta$'', can be easily obtained. From (\ref{eq: V border expansion}) and (\ref{eq: V volume expansion}), they correspond to the equalities:

\begin{equation}\label{eq: alpha and beta equations definition}
V_\alpha(\partial\Omega) = V_\alpha(\Omega) \hspace{0.8cm} , \hspace{.75cm} V_\beta\zt (\partial\Omega) = V_\beta\zt (\Omega)
\end{equation}

\vspace{0.25cm}

Thus, by consulting equation (\ref{eq: V solution on surface}), checking the expression for $T$ in (\ref{eq: wilson surface definition}), and substituting (\ref{eq: 2-form substitution}) for $B_{\mu\nu}$\zc , one easily obtains the coefficients $V_\alpha(\partial\Omega)$ and $V_\beta\zt (\partial\Omega)$ above being given by:

\vspace{0.1cm}

\begin{equation}\label{eq: left side alpha}
V_\alpha(\partial\Omega) = ie \hspace{0.3mm} V_R \int_{\tau_i}^{\tau_f} \hspace{-0.5mm} d\tau \hspace{-0.3mm} \int_{\sigma_i}^{\sigma_f} \hspace{-0.3mm} d\sigma \hspace{0.2mm} F_{\mu\nu}^{\hspace{0.3mm}W} \hspace{0.2mm} \frac{dx^\mu}{d\sigma} \frac{dx^\nu}{d\tau} \bigg|_{\zeta = \zeta_f}
\end{equation}

\vspace{0.2cm}

\begin{equation}\label{eq: left side beta}
V_\beta\zt (\partial\Omega) = ie \hspace{0.3mm} V_R \int_{\tau_i}^{\tau_f} \hspace{-0.5mm} d\tau \hspace{-0.3mm} \int_{\sigma_i}^{\sigma_f} \hspace{-0.3mm} d\sigma \hspace{0.2mm} \widetilde{F}_{\mu\nu}^{\hspace{0.3mm}W} \hspace{0.2mm} \frac{dx^\mu}{d\sigma} \frac{dx^\nu}{d\tau}\bigg|_{\zeta = \zeta_f}
\end{equation}

\vspace{0.35cm}

\noindent{}where the notation $|_{\zeta = \zt \zeta_f}$ indicates that the expressions are indeed being evaluated at $\partial\Omega$.

For the coefficients $V_\alpha(\Omega)$ and $V_\beta\zt (\Omega)$ of (\ref{eq: alpha and beta equations definition}), one must expand the right-hand side of equation (\ref{eq: stokes two form}), which involves $\scalebox{1.5}{$\kappa$}$. Consulting equation (\ref{eq: kappa delta definition}) for this quantity, we see that we have a Wilson surface $V$ (and its inverse $V^{-1}$) inside of the $\tau$-integral. These quantities must be expanded with expressions similar to (\ref{eq: V solution on surface}), and so we see that both $V$ and $V^{-1}$ contribute with terms of all orders in $\alpha$ and $\beta$. Since we want the integral equations of lowest order in these parameters, one sees that we must then consider only the identity terms of $V$ and $V^{-1}$. More precisely, by equation (\ref{eq: V solution on surface}), we see that we need to consider not $\mathbb{1}$, but actually $V_R$ for $V$ and $V^{-1}_R$ for $V^{-1}$. Since $V^{-1}$ appears on the right of the quantity $\scalebox{1.5}{$\kappa$}$, we see that the $V^{-1}_R$ obtained will then cancel out the integration constant $V_R$ that appears on the right-hand side of (\ref{eq: stokes two form}).

Having said that, now we need only to identify the terms of first order in $\alpha$ and $\beta$ that appear on the quantity inside the $V$-conjugation in (\ref{eq: kappa delta definition}). On doing so, we obtain that the contribution of order $\alpha$ for the right-hand side of (\ref{eq: stokes two form}) (namely, $V_\alpha(\Omega)$) is thus given by\footnote{The commutator given below, refering to the contribution in first order in $\alpha$, is not explicitly written inside the quantity $\scalebox{1.5}{$\kappa$}$ on equation (\ref{eq: kappa delta definition}). To obtain it, one must consider the last two terms of (\ref{eq: kappa delta definition}), combine them and manipulate the integration regions to decouple the integrals in $\sigma$ and $\sigma^\prime$. We will not do this calculation \nlb here.}:

\vspace{-0.25cm}

\begin{equation}\label{eq: right side alpha}
\hspace{-0.3cm}V_\alpha(\Omega) = (ie)^2 \hspace{0.3mm} V_R \int_{\zeta_i}^{\zeta_f} \hspace{-0.5mm} d\zeta \hspace{-0.3mm} \int_{\tau_i}^{\tau_f} \hspace{-0.5mm} d\tau \hspace{0.3mm} \bigg[ \int_{\sigma_i}^{\sigma_f} d\sigma^\prime F_{\kappa\rho}^{\hspace{0.1mm}W} (\sigma^\prime) \frac{dx^\kappa}{d\sigma^\prime} \frac{dx^\rho}{d\tau} (\sigma^\prime) \hspace{0.9mm} , \hspace{-0.2mm} \int_{\sigma_i}^{\sigma_f} d\sigma F_{\mu\nu}^{\hspace{0.1mm}W} (\sigma) \frac{dx^\mu}{d\sigma} \frac{dx^\nu}{d\zeta} (\sigma) \hspace{0.3mm} \bigg]
\end{equation}

\vspace{0.25cm}

\noindent{}and that the contribution of first order in $\beta$ (namely, $V_\beta\zt (\Omega)$) is given by:

\vspace{-0.25cm}

\begin{equation}\label{eq: right side beta}
\begin{gathered}
\hspace{-5.95cm}V_\beta\zt (\Omega) = ie \hspace{0.3mm} V_R \int_{\zeta_i}^{\zeta_f} \hspace{-0.5mm} d\zeta \hspace{-0.3mm} \int_{\tau_i}^{\tau_f} \hspace{-0.5mm} d\tau \hspace{0.3mm}  \int_{\sigma_i}^{\sigma_f} d\sigma \Bigg\{ \widetilde{J}_{\mu\nu\lambda}^{\hspace{0.8mm}W} \frac{dx^\mu}{d\sigma} \frac{dx^\nu}{d\tau} \frac{dx^\lambda}{d\zeta} + \\[3.5mm]
\hspace{4.8cm}+ ie \hspace{0.3mm} \bigg[ \int_{\sigma_i}^{\sigma} d\sigma^\prime F_{\kappa\rho}^{\hspace{0.1mm}W} (\sigma^\prime) \frac{dx^\kappa}{d\sigma^\prime} \frac{dx^\rho}{d\tau} (\sigma^\prime)  \hspace{0.9mm} , \hspace{-0.2mm} \widetilde{F}_{\mu\nu}^{\hspace{0.1mm}W} (\sigma) \frac{dx^\mu}{d\sigma} \frac{dx^\nu}{d\zeta} (\sigma) \hspace{0.3mm} \bigg] + \\[3.5mm]
\hspace{5.05cm}\hspace{-0.325cm}+ ie \hspace{0.3mm} \bigg[ \widetilde{F}_{\mu\nu}^{\hspace{0.1mm}W} (\sigma) \frac{dx^\mu}{d\sigma} \frac{dx^\nu}{d\tau} (\sigma) \hspace{0.9mm} , \hspace{-0.2mm} \int_{\sigma_i}^{\sigma} d\sigma^\prime F_{\kappa\rho}^{\hspace{0.1mm}W} (\sigma^\prime) \frac{dx^\kappa}{d\sigma^\prime} \frac{dx^\rho}{d\zeta} (\sigma^\prime) \hspace{0.3mm} \bigg] \Bigg\}
\end{gathered}
\end{equation}

\vspace{0.2cm}

Note that the integration constant $V_R$ present above (as well as a factor of $ie$) will cancel out the $V_R$ (and the factor of $ie$) present on the left-hand side contributions given in (\ref{eq: left side alpha}) and (\ref{eq: left side beta}), since in all these cases the constant $V_R$ appear on the left of the expressions.

\phantom{Paragraph}

With the coefficients of (\ref{eq: alpha and beta equations definition}), we then have the integral equation of order $\alpha$ given by:

\vspace{-0.25cm}

\begin{equation}\label{eq: alpha equation}
\begin{split}
&\int_{\tau_i}^{\tau_f} \hspace{-0.5mm} d\tau \hspace{-0.3mm} \int_{\sigma_i}^{\sigma_f} \hspace{-0.3mm} d\sigma \hspace{0.2mm} F_{\mu\nu}^{\hspace{0.3mm}W} \hspace{0.2mm} \frac{dx^\mu}{d\sigma} \frac{dx^\nu}{d\tau} \bigg|_{\zeta = \zeta_f} = \\[3.5mm]
&= ie \hspace{-0.3mm} \int_{\zeta_i}^{\zeta_f} \hspace{-0.5mm} d\zeta \hspace{-0.3mm} \int_{\tau_i}^{\tau_f} \hspace{-0.5mm} d\tau \hspace{0.3mm} \bigg[ \int_{\sigma_i}^{\sigma_f} d\sigma^\prime F_{\kappa\rho}^{\hspace{0.1mm}W} (\sigma^\prime) \frac{dx^\kappa}{d\sigma^\prime} \frac{dx^\rho}{d\tau} (\sigma^\prime) \hspace{0.9mm} , \hspace{-0.2mm} \int_{\sigma_i}^{\sigma_f} d\sigma F_{\mu\nu}^{\hspace{0.1mm}W} (\sigma) \frac{dx^\mu}{d\sigma} \frac{dx^\nu}{d\zeta} (\sigma) \hspace{0.3mm} \bigg] \hspace{0.435cm}
\end{split}
\end{equation}

\vspace{0.25cm}

\noindent{}and the integral equation of order $\beta$ given by:

\vspace{-0.25cm}

\begin{equation}\label{eq: beta equation}
\begin{split}
&\hspace{0.2cm}\int_{\tau_i}^{\tau_f} \hspace{-0.5mm} d\tau \hspace{-0.3mm} \int_{\sigma_i}^{\sigma_f} \hspace{-0.3mm} d\sigma \hspace{0.2mm} \widetilde{F}_{\mu\nu}^{\hspace{0.3mm}W} \hspace{0.2mm} \frac{dx^\mu}{d\sigma} \frac{dx^\nu}{d\tau} \bigg|_{\zeta = \zeta_f} =
\\[3.5mm]
&\hspace{0.2cm}= \int_{\zeta_i}^{\zeta_f} \hspace{-0.5mm} d\zeta \hspace{-0.3mm} \int_{\tau_i}^{\tau_f} \hspace{-0.5mm} d\tau \hspace{-0.3mm} \int_{\sigma_i}^{\sigma_f} \hspace{-0.5mm} d\sigma \hspace{0.3mm} \Bigg\{ \widetilde{J}_{\mu\nu\lambda}^{\hspace{0.8mm}W} \frac{dx^\mu}{d\sigma} \frac{dx^\nu}{d\tau} \frac{dx^\lambda}{d\zeta} \hspace{0.5mm}+ \\[3.5mm] 
&\hspace{4.7cm}+ ie \hspace{0.3mm} \bigg[ \int_{\sigma_i}^{\sigma} d\sigma^\prime F_{\kappa\rho}^{\hspace{0.1mm}W} (\sigma^\prime) \frac{dx^\kappa}{d\sigma^\prime} \frac{dx^\rho}{d\tau} (\sigma^\prime)  \hspace{0.9mm} , \widetilde{F}_{\mu\nu}^{\hspace{0.1mm}W} (\sigma) \frac{dx^\mu}{d\sigma} \frac{dx^\nu}{d\zeta} (\sigma) \hspace{0.3mm} \bigg] + \\[3.5mm]
&\hspace{4.7cm}+ ie \hspace{0.3mm} \bigg[ \widetilde{F}_{\mu\nu}^{\hspace{0.1mm}W} (\sigma) \frac{dx^\mu}{d\sigma} \frac{dx^\nu}{d\tau} (\sigma) \hspace{0.9mm} , \hspace{-0.2mm} \int_{\sigma_i}^{\sigma} d\sigma^\prime F_{\kappa\rho}^{\hspace{0.1mm}W} (\sigma^\prime) \frac{dx^\kappa}{d\sigma^\prime} \frac{dx^\rho}{d\zeta} (\sigma^\prime) \hspace{0.3mm} \bigg] \Bigg\}
\end{split}
\end{equation}

\vspace{0.2cm}

Considering an Abelian gauge theory, one sees that the equations obtained above reduce to the usual Maxwell equations of Electrodynamics.\footnote{Actually, to obtain each one of Maxwell's equations, one must also specify the integration volume $\Omega$ on spacetime. With a completely spatial volume $\Omega$, equations (\ref{eq: alpha equation}) and (\ref{eq: beta equation}) give the two Gauss laws for the magnetic and eletric fields, respectively. With an integration volume $\Omega$ that has also a time component, equations (\ref{eq: alpha equation}) and (\ref{eq: beta equation}) give, respectively, the Faraday and Ampère laws of Electrodynamics.}  The next integral equations of Yang-Mills theories (of higher orders in $\alpha$ and $\beta$) can also be obtained by collecting its appropriate terms, as described above. Nonetheless, we will not be interested in those equations here.

\phantom{Paragraph}

One of the most important properties of equations (\ref{eq: alpha equation}) and (\ref{eq: beta equation}) (and, most generally, of all the infinite Yang-Mills integral equations obtained from the formulation presented above) is how they transform by applying a gauge transformation. As we know,  the field strenght $F_{\mu\nu}$\zc, its Hodge dual $\widetilde{F}_{\mu\nu}$ and the Hodge dual of the current, $\widetilde{J}_{\mu\nu\lambda}$\zc , all transform locally by conjugation as $F_{\mu\nu}(x) \rightarrow g(x) F_{\mu\nu}(x) \hspace{0.3mm} g^{-1}(x)$. On the other hand, since all the loops employed on the scanning of $\Omega$ are based on the same reference point $x_R$\zc , we have that the Wilson line $W$ (on any point $x$ of $\Omega$) transforms globally as $W(x) \rightarrow g(x) W(x) \hspace{0.3mm} g^{-1}_R$\zc , where $g_R \equiv g\zu\zu(x_R)$\zc . Therefore, it is easy to see that the conjugated quantity $F_{\mu\nu}^{\hspace{0.2mm}W}(x)$ transforms by gauge transformations as:

\begin{equation}\label{eq: FW transformation}
F_{\mu\nu}^{\hspace{0.2mm}W}(x) \zu \equiv W^{-1} (x) F_{\mu\nu}(x) \zu\zu W(x) \rightarrow  g_R \zu F_{\mu\nu}^{\hspace{0.2mm}W} (x) \zu\zu g^{-1}_R
\end{equation}

\vspace{.25cm}

\noindent{}which is the same way the other quantities $\widetilde{F}_{\mu\nu}^{\hspace{0.3mm}W}$ and $\widetilde{J}_{\mu\nu\lambda}^{\hspace{0.8mm}W}$ will also transform. So, since $g_R$ is a constant (and the same for all loops), we have that it drops out of all the integrations performed. Then, by looking at equations (\ref{eq: alpha equation}) and (\ref{eq: beta equation}), it is easy to see that all of its terms will become conjugated by $g_R$\zc , which then means that these integral equations transform covariantly by gauge transformations, with a conjugation of the constant group element $g \zu\zu (x_R)$. From the same arguments above, this will also be true for all the infinity of integral equations.

This very important property, in the context of Yang-Mills theories, is one of the main reasons why the Wilson line $W$ was introduced in our starting equation (\ref{eq: wilson surface definition}). From this, it also becomes clear the necessity of the scanning of $\Omega$ with loops based on a fixed reference point \nlb $x_R$\zc \nlb . Since our integral equations transform by a conjugation of $g_R$\zc , we see that gauge-invariant quantities can be easily obtained by taking traces or by considering eigenvalues.

\phantom{Paragraph}

On the next subsection, we will define the gauge-invariant magnetic fluxes and charge density that we are interested in this paper. But before that, let us mention an important last detail, and an important simplification that can be made with the integral equations obtained above.

\phantom{Paragraph}

The detail has to do with the BPST instanton solution, which we shall make use in a moment. This solution, obtained in 1975 by Belavin, Polyakov, Schwartz and Tyupkin \cite{bpstinstanton}, is an important example of the so-called self-dual solutions of (pure) Yang-Mills theories, satisfying:

\begin{equation}\label{eq: self-duality equation}
F_{\mu\nu} = \kappa \widetilde{F}_{\mu\nu}
\end{equation}

\vspace{0.2cm}

With this equation, called the self-duality equation, one sees that for a pure Yang-Mills theory (where $J^\nu = 0$) the Yang-Mills equations given in (\ref{eq: bianchi ym}) become automatically solved by the Bianchi identity, since by (\ref{eq: self-duality equation}) we can substitute $F_{\mu\nu}$ by $\widetilde{F}_{\mu\nu}$\zc . Therefore, we are obtaining solutions of second-order differential equations (the Yang-Mills equations) by solving the first-order differential equation (\ref{eq: self-duality equation}). These particular solutions of Yang-Mills theories have very important properties, presented for example in \cite{Manton:2004tk,shnirbook}, which we will not discuss here.

For the self-duality equation (\ref{eq: self-duality equation}), it is easy to show that $\kappa$ must be $\pm i$ for a Minkowski spacetime and $\pm 1$ for an Euclidean spacetime.\footnote{Note that we made no use of the metric when obtaining the integral equations for Yang-Mills theories, which means that they are equally valid for both Minkowski and Euclidean spacetimes!} The imaginary unit $i$, obtained in the Minkowski case, is problematic for the solutions of (\ref{eq: self-duality equation}). The most interesting solutions of (\ref{eq: self-duality equation}) are those obtained in Euclidean spacetime, which is the case of the BPST instanton. From this equation, we can actually associate two separate solutions with the constant $\kappa$: for $\kappa = 1$ we have the instanton, or 1-instanton, and for $\kappa=-1$ we have the anti-instanton.

The consequence of self-duality for the Yang-Mills integral equations is that the $\beta$-equations will turn themselves into the $\alpha$-equations (which is the integral analogue of the Yang-Mills equations in (\ref{eq: bianchi ym}) turning into the Biachi indentity). This can be seen by brute force substitution, remembering that for self-duality we have $J^\nu = 0$. However, the easiest way to see it is by considering the definition $B_{\mu\nu} \equiv \alpha F_{\mu\nu} + \beta \widetilde{F}_{\mu\nu}$ given in (\ref{eq: 2-form substitution}). Using equation (\ref{eq: self-duality equation}), we see that $B_{\mu\nu}$ becomes $B_{\mu\nu} = \big(\alpha + \beta\kappa^{-1}\big) F_{\mu\nu} \equiv \gamma F_{\mu\nu}$\zc . Thus, we are left with an expansion on this new parameter $\gamma$, which will yield the exact same equations as making $\beta=0$ from the beginning and expanding on the $\alpha$-parameter only. In other words, for self-dual solutions such as the instanton, we lose no information at all by considering only the $\alpha$-equations.

\phantom{Paragraph}

So, considering the $\alpha$-equations, the important simplification that was mentioned before is that these integral equations can be written exclusively in terms of the Wilson line $W$ and its derivatives. By considering a deformation of the loop on which $W$ is being calculated, one can make use of its definition in (\ref{eq: wilson line definition}) and perform a variation of this equation to obtain:

\begin{equation}\label{eq: variation wilson final}
\delta W (\sigma_f) = ie \zc W(\sigma_f) \hspace{-0.5mm} \int_{\sigma_i}^{\sigma_f} d\sigma F_{\mu\nu}^{\hspace{0.1mm}W} \frac{dx^\mu}{d\sigma} \delta x^\nu 
\end{equation}

\vspace{0.2cm}

\noindent{}where the $\sigma_f$ dependence indicates that we have our quantities calculated at the end of the loop. For convenience, we shall change this notation to $W_c$\zc , where the $c$ indicates that we have $W$ calculated on a closed path (in other words, on the whole loop). Considering then the variation of the loop to be in the $\tau$- or $\zeta$-direction, the equation above acquires the derivative $d/d\tau$ or $d/d\zeta$, in the place of the $\delta$'s. Then, by isolating the $\sigma$-integrals obtained in these equations and comparing them with the terms of (\ref{eq: alpha equation}), one concludes that the $\alpha$-equation may be written as:

\begin{equation}\label{eq: alpha equation wilson line}
\int_{\tau_i}^{\tau_f}\hspace{-0.3mm} d\tau \hspace{0.3mm} W_c^{-1} \hspace{0.3mm} \frac{dW_c}{d\tau} \bigg|_{\zeta = \zeta_f}  = \int_{\zeta_i}^{\zeta_f}\hspace{-0.5mm} d\zeta \hspace{-0.2mm} \int_{\tau_i}^{\tau_f}\hspace{-0.5mm} d\tau \bigg[ \hspace{0.3mm} W_c^{-1} \hspace{0.3mm} \frac{dW_c}{d\tau} \hspace{0.5mm} , \hspace{0.3mm} W_c^{-1} \hspace{0.3mm} \frac{dW_c}{d\zeta} \hspace{0.5mm} \bigg]
\end{equation}

\vspace{0.25cm}

Thus, for the $\alpha$-equation, we see that the only thing needed is the value of the Wilson line on the loops employed on the scanning of $\Omega$. From the same arguments above, this will also be true for all the other integral equations of higher orders in the parameter $\alpha$.

\subsection{Magnetic fluxes and magnetic charge density}
\label{subsec:fluxes and charge density}

In this paper we will consider $\Omega$ to be a completely spatial sphere (with `spatial' meaning on the coordinates $x^1$, $x^2$ and $x^3$, with $x^4$ being the Euclidean time), centered at the origin and of radius $0 \leq \zeta_f < \infty$. We will make use of a particular parameterization, firstly constructed in \cite{directtest}, where $\Omega$ is scanned by spherical surfaces (centered at the origin) of radius $\zeta$, and where the reference point $x_R$ is positioned at infinity, at $(x^1,x^2,x^3) = (-\infty,0,0)$. This parameterization and its loops will be better explained on the beginning of the next section.

\phantom{Paragraph}

Since we are dealing with a spatial volume $\Omega$, we have our equation of order $\alpha$ corresponding to the non-Abelian version of Gauss law for the magnetic field (since we have $F_{ij} = - \varepsilon_{ijk} B^k$). For the case of the BPST instanton, this integral equation will also correspond to the non-Abelian version of Gauss law for the electric field, because of the self-dual nature of this solution.

So, having a Gauss law, the interpretation of the $\alpha$-equation (\ref{eq: alpha equation}) is direct: we have, on its left-hand side, a \textit{flux} of the non-Abelian magnetic field (conjugated by $W$) through the surface $\partial\Omega$, which is then equal to a \textit{magnetic charge} inside $\Omega$, given by the right-hand side of (\ref{eq: alpha equation}).

With that in mind, and remembering that the left-hand side of (\ref{eq: alpha equation}) transforms by a conjugation of $g_R$ when applying a gauge transformation, we make the following definition:

\begin{equation}\label{eq: alpha fluxes definition}
\Phi_B^{(\alpha \zu , \zu N \zu)} \equiv \frac{1}{N} \Tr \Bigg( \int_{\tau_i}^{\tau_f} \hspace{-0.5mm} d\tau \hspace{-0.3mm} \int_{\sigma_i}^{\sigma_f} \hspace{-0.3mm} d\sigma \hspace{0.2mm} F_{ij}^{\zu W} \hspace{0.2mm} \frac{dx^i}{d\sigma} \frac{dx^j}{d\tau} \bigg|_{\zeta = \zeta_f} \Bigg)^{\hspace{-0.8mm}N}
\end{equation}

\vspace{0.25cm}

\noindent{}which is gauge-invariant, and where $\Phi_B$ thus have the meaning of a magnetic flux through $\partial\Omega$. Here, the index $\alpha$ means that we are considering the magnetic flux that comes from the integral equation of order $\alpha$. Similar fluxes $\Phi_B^{(\alpha^m \hspace{-0.2mm}  , \zu\zu N \zu)}$ can be defined considering the left-hand sides of the integral equations of higher orders in $\alpha$, but until now we have no interpretation for the corresponding `enclosed charges', on the right-hand side of such equations. The index $N$ refers to the $N$ matrix multiplications that can be considered inside the trace, which was introduced since it does not alter the gauge-invariance property of the magnetic flux that is being defined.\footnote{On trying to define gauge-invariant quantities, one could also opt for eigenvalues, instead of (\ref{eq: alpha fluxes definition}). However, as can be shown, the eigenvalues of a matrix $A$ can be given entirely in terms of its traces $\Tr \big(A^N\zu \big)$, depending on the dimension of $A$. Therefore, it is indeed important to consider the index $N$ in definition (\ref{eq: alpha fluxes definition}).}

The magnetic flux through $\partial\Omega$ defined in (\ref{eq: alpha fluxes definition}) can be identified, because of Gauss law, with the magnetic charge contained in $\Omega$. This enclosed magnetic charge is rigorously given by the right-hand side of (\ref{eq: alpha equation}), since this is the one corresponding to the volume integral over $\Omega$. Nevertheless, since we have the equality on (\ref{eq: alpha equation}), we shall use the flux (\ref{eq: alpha fluxes definition}) defined above to refer as well to the enclosed magnetic charge on $\Omega$. We shall do this because the left-hand side of (\ref{eq: alpha equation}) (and thus the flux above) is much more simple to calculate than its right-hand side.

\phantom{Paragraph}

Before continuing, let us define a mathematical structure which is of great importance for the integral formulation that we just built. As we have seen, for our integral equations we have to consider $\Omega$ as being scanned by closed surfaces, which in turn are scanned by loops based on a fixed reference point $x_R$\zc . These closed surfaces are specified by the parameter $\zeta$, where $\zeta_i$ corresponds to the infinitesimal surface at $x_R$ and $\zeta_f$ corresponds to the border $\partial\Omega$.

Mathematically, what is being done is the consideration of a map, from $S^2$ to spacetime $M$, with the north-pole of $S^2$ being mapped on the fixed spacetime event $x_R$\zc{\nlb}. The set of all such mappings corresponds to the generalized loop space $\mathcal{L}^{(2)}$, defined as the space of functions:

\begin{equation}\label{eq: loop space definition}
\mathcal{L}^{(2)} \equiv \big\{ \hspace{0.7mm} f : S^2 \rightarrow M \hspace{1.25mm} | \hspace{1.25mm} \text{north-pole} \rightarrow x_R \hspace{0.5mm} \big\}
\end{equation}

\vspace{0.2cm}

With this definition we thus see that a scanning of $\Omega$ by closed surfaces corresponds to a 1-dimensional path on $\mathcal{L}^{(2)}$, from an initial point, corresponding to the infinitesimal closed surface at $x_R$\zc , to a final point, corresponding to the closed surface $\partial\Omega$. Also, we see that this path on $\mathcal{L}^{(2)}$ is dependent on the choice of scanning for $\Omega$. This is true because a change of scanning corresponds to a change of the functions from $S^2$ to $M$, which thus changes the corresponding path on $\mathcal{L}^{(2)}$, even though the initial and final points remain the same.

Actually, consider for example the final point of the path, corresponding to $\partial\Omega$. This particular point on $\mathcal{L}^{(2)}$ is itself also dependent on the choice of parameterization, since the closed surface $\partial\Omega$ can be scanned by loops in many different ways. Each one of these scannings of $\partial\Omega$ correspond to different functions from $S^2$ to $M$, which thus mean that they will correspond to different points on $\mathcal{L}^{(2)}$, even though the physical surface on $M$ is the same!

With that in mind, it becomes clear that the integral formulation of Yang-Mills theories in $(3+1)$-dimensions that was presented is actually defined on the generalized loop space $\mathcal{L}^{(2)}$, since it is intimately tied to the scanning of $\Omega$ by closed surfaces, with loops based on $x_R$\zc .\footnote{In $(d+1)$-dimensions, one is led to the generalized loop space $\mathcal{L}^{(d-1)}$, as explained in \cite{ym2}.}

So, if one is trying to extract observable results for Yang-Mills theories from its integral formulation, it must be checked (in addition to gauge invariance, which was already discussed) if these results are \textit{reparameterization-invariant}. This must be checked in order to see if such results are indeed tied to the physical surfaces and volumes considered in spacetime $M$, or if they are parameterization-dependent, being in this case defined only on $\mathcal{L}^{(2)}$.

\phantom{Paragraph}

This check of the reparameterization-invariance condition is actually very much expected, even without identifying the structure of the generalized loop space $\mathcal{L}^{(2)}$. Indeed, being the choice of parameterization a mathematical arbitrariness involved in the calculations only, the physical results to be extracted from the integral equations must evidently not depend on it.

Nevertheless, the identification of the generalized loop space helps and, as shown in recent developments by our group \cite{fourym,threeym,twoym}, it is increasingly proving to be a structure of great importance not only for its integral formulation, but for Yang-Mills theory itself.

\phantom{Paragraph}

Considering then the problem of the reparameterization of $\Omega$, one very important result, shown in \cite{ym1,ym2}, is that the quantity $P_3 \exp \big( \zu\zu ie \hspace{-0.5mm} \int_{\Omega} d\zeta \hspace{0.3mm} \scalebox{1.5}{$\kappa$} \zt \big)$ that appears on the right-hand side of (\ref{eq: stokes two form compact}) is in fact \textit{independent} of the path on $\mathcal{L}^{(2)}$, as long as its initial and final points, corresponding to the surfaces at $x_R$ and at $\partial\Omega$, remain fixed.\footnote{In fact, consider two different volumes $\Omega$ and $\Omega^\prime$ on spacetime, that share the same border: $\partial\Omega = \partial\Omega^\prime$. From the integral equation (\ref{eq: stokes two form compact}), one sees that if $\partial\Omega = \partial\Omega^\prime$ we then obtain $V(\Omega) = V(\Omega^\prime)$. Now, since $\Omega$ and $\Omega^\prime$ are different, the corresponding paths on $\mathcal{L}^{(2)}$, from $x_R$ to $\partial\Omega = \partial\Omega^\prime$, will be different. Nevertheless, $V(\Omega) = V(\Omega^\prime)$, and thus we conclude that the quantity $P_3 \exp \big( \zu\zu ie \hspace{-0.5mm} \int_{\Omega} d\zeta \hspace{0.3mm} \scalebox{1.5}{$\kappa$} \zt \big)$ is indeed independent of the path on $\mathcal{L}^{(2)}$. Note that we could consider two different parameterizations of one same volume $\Omega$. These two parameterizations also correspond to different paths on $\mathcal{L}^{(2)}$, but since $x_R$ and $\partial\Omega$ are fixed, we may apply the same argument above.} Since the reparameterization of $\Omega$ corresponds to a change of path on $\mathcal{L}^{(2)}$, this result establishes the invariance of all the integral equations obtained from (\ref{eq: stokes two form compact}) by reparameterization of the integration volume $\Omega$.

From this path independency on $\mathcal{L}^{(2)}$, it can also be shown \cite{ym1,ym2} that the gauge-invariant charges obtained from the integral formulation of Yang-Mills theories are actually \textit{conserved} in time, when one considers $\Omega$ to be purely spatial. This is a very important result, which does not come from a Noether-type theorem nor a topological-type conservation law. It is a dynamical conservation of the charges, associated indeed with the path independency of the charge operator on $\mathcal{L}^{(2)}$.

In the present work we are interested in the internal magnetic fluxes (enclosed magnetic charges) and in the magnetic charge density of the instanton solution. These are not conserved quantities, and thus we shall not present here the derivation of the result mentioned above. 

\phantom{Paragraph}

So, we have established the invariance with respect to the reparameterization of the integration volume $\Omega$. Nevertheless, we need as well to check if our results will be invariant with respect to the reparameterization of the final surface $\partial\Omega$ by the loops that scan it. Indeed, in expression (\ref{eq: alpha fluxes definition}), we have defined a gauge-invariant quantity which is being calculated on $\partial\Omega$. For this surface, we need to choose a particular scanning with loops based on $x_R$\zc . Therefore, if we want $\Phi_B^{(\alpha \zu , \zu N \zu)}$ to be physical, this quantity must clearly not depend on this choice of parameterization.

Denoting by $L_\alpha$ the left-hand side of the $\alpha$-equation (\ref{eq: alpha equation}), which is what is being considering inside the trace operation on (\ref{eq: alpha fluxes definition}), the first thing to determine is how this quantity $L_\alpha$ change by a reparameterization of $\partial\Omega$. The way to do it is by considering a deformation of the loops that scan $\partial\Omega$, which can be related with a variation $\delta L_\alpha$\zc , in a very similar way that was done before to derive the non-Abelian Stokes theorem. Performing this variation, one obtains:

\begin{equation}\label{eq: variation left alpha}
\delta L_\alpha = ie \hspace{-0.3mm} \int_{\tau_i}^{\tau_f} \hspace{-0.5mm} d\tau \hspace{0.3mm} \bigg[ \int_{\sigma_i}^{\sigma_f} d\sigma F_{\mu\nu}^{\hspace{0.1mm}W} (\sigma) \frac{dx^\mu}{d\sigma} \frac{dx^\nu}{d\tau} (\sigma) \hspace{0.9mm} , \hspace{-0.2mm} \int_{\sigma_i}^{\sigma_f} d\sigma^\prime F_{\kappa\rho}^{\hspace{0.1mm}W} (\sigma^\prime) \frac{dx^\kappa}{d\sigma^\prime} \zt \delta x^\rho (\sigma^\prime) \hspace{0.3mm} \bigg]
\end{equation}

\vspace{0.25cm}

This expression should be expected. By considering the deformation of the loops to be in the $\zeta$-direction, one sees that the expression above yields directly the $\alpha$-equation (\ref{eq: alpha equation}), which in this paper was obtained through a much longer derivation, making use of the non-Abelian Stokes theorem for $B_{\mu\nu}$\zc . Nevertheless, for the case we are now considering, the deformations of the loops must be in the $\sigma$- and $\tau$-directions only, since we want deformations representing a reparameterization of the surface $\partial\Omega$, which thus can not be perpendicular to it.

Having said that, and considering that a reparameterization of the surface must involve an arbitrary deformation of the loops that scan it, we consider $\delta x^\mu$ to be given by:

\begin{equation}\label{eq: arbitrary variation}
\delta x^\mu \equiv a \zu\zu (\sigma,\tau) \zt \frac{dx^\mu}{d\sigma} + b \zu\zu (\sigma,\tau) \zt \frac{dx^\mu}{d\tau}
\end{equation}

\vspace{0.2cm}

\noindent{}where $a$ and $b$ are the arbitrary parameters giving the arbitrary reparameterization of $\partial\Omega$. 

Inserting this expression into (\ref{eq: variation left alpha}), we have that its first term will not contribute because of the symmetric contractions with the indices of $F_{\kappa\rho}$\zc . Therefore, we are left with:

\vspace{-0.25cm}

\begin{equation}\label{eq: variation left alpha final}
\delta L_\alpha = ie \hspace{-0.3mm} \int_{\tau_i}^{\tau_f} \hspace{-0.5mm} d\tau \hspace{0.3mm} \bigg[ \int_{\sigma_i}^{\sigma_f} d\sigma F_{\mu\nu}^{\hspace{0.1mm}W} (\sigma) \frac{dx^\mu}{d\sigma} \frac{dx^\nu}{d\tau} (\sigma) \hspace{0.9mm} , \hspace{-0.2mm} \int_{\sigma_i}^{\sigma_f} d\sigma^\prime \zt b(\sigma^\prime) \zu\zu F_{\kappa\rho}^{\hspace{0.1mm}W} (\sigma^\prime) \frac{dx^\kappa}{d\sigma^\prime} \frac{dx^\rho}{d\tau} (\sigma^\prime) \hspace{0.3mm} \bigg]
\end{equation}

\vspace{0.2cm}

\noindent{}where we have dropped the $\tau$-dependece for convenience. With this expression at hand, one can then check if the gauge-invariant quantity defined on (\ref{eq: alpha fluxes definition}) is also invariant by the reparameterization of $\partial\Omega$. Considering a variation of $\Phi_B^{(\alpha \zu , \zu N \zu)}$, we thus obtain the expression:

\begin{equation}\label{eq: variation alpha fluxes}
\delta \zu\zu \Phi_B^{(\alpha \zu , \zu N \zu)} = \Tr \Big( L_\alpha^{N-1} \zc \delta L_\alpha \Big)
\end{equation}

\vspace{0.2cm}

\noindent{}which must then be checked in order to see if the quantity $\Phi_B^{(\alpha \zu , \zu N \zu)}$ changes by reparameterization. Note that we are not \textit{imposing} $\delta \zu\zu \Phi_B^{(\alpha \zu , \zu N \zu)}$ to be equal to zero, since this would lead to conditions which are not satisfied by the instanton solution over all surfaces $\partial\Omega$, as we shall see. Instead, we are considering the gauge-invariant fluxes $\Phi_B^{(\alpha \zu , \zu N \zu)}$ as they are, and only after checking if $\delta \zu\zu \Phi_B^{(\alpha \zu , \zu N \zu)} = 0$, to see if the results obtained are physical, or if they are only defined on $\mathcal{L}^{(2)}$.

\phantom{Paragraph}

Another interesting quantity that we shall consider is a gauge-invariant magnetic charge density, which can be easily defined by making use of (\ref{eq: alpha fluxes definition}). As said before, when $\Omega$ is completely spatial, we have that the $\alpha$-equation (\ref{eq: alpha equation}) corresponds to the the non-Abelian version of Gauss law for the magnetic field. Therefore, the $\alpha$-fluxes through $\partial\Omega$, given in (\ref{eq: alpha fluxes definition}), correspond in this case to the the non-Abelian magnetic charges contained in $\Omega$, which are rigorously given by the right-hand side of (\ref{eq: alpha equation}). So, let us now make this indentification explicit by defining:

\begin{equation}\label{eq: enclosed charge definition}
Q^{\zu (\hspace{-0.1mm} N \zu)}_{enc} \equiv \Phi_B^{(\alpha \zu , \zu N \zu)}
\end{equation}

\vspace{0.2cm}

\noindent{}which corresponds to the enclosed magnetic charge on $\Omega$, associated with the index $N$ as explained before. As already pointed out, we shall define this quantity $Q^{\zu (\hspace{-0.1mm} N \zu)}_{enc}$ in terms of the left-hand side of (\ref{eq: alpha equation}) because it is much more simple to calculate.

So, what we do is consider $\Omega$ to be a sphere, of radius $\zeta_f$\zc . From (\ref{eq: enclosed charge definition}), we thus have the magnetic charge enclosed on this sphere, which we shall call $Q^{\zu (\hspace{-0.1mm} N \zu)}_{enc} (\zeta_f)$. Next, we consider another sphere, infinitesimally larger than the previous one, of radius $\zeta_f^\prime = \zeta_f + h$. With this sphere we have the corresponding enclosed charge being given by $Q^{\zu (\hspace{-0.1mm} N \zu)}_{enc} (\zeta_f + h)$. From this, we write the quantity $Q^{\zu (\hspace{-0.1mm} N \zu)}_{enc} (r + h) - Q^{\zu (\hspace{-0.1mm} N \zu)}_{enc} (r)$ as being the total magnetic charge contained in the infinitesimal volume between the two spheres, where we have renamed $\zeta_f$ by $r$, for convenience.

Dividing this total charge by the volume $4\pi r^2 h$ where it is distributed, and taking the limit as $h \rightarrow 0$, we thus obtain the magnetic charge density being defined as:

\begin{equation}\label{eq: charge density definition}
\rho^{\zu\zu (\hspace{-0.1mm} N \zu)} \equiv \frac{1}{4\pi r^2}\frac{d \zu\zu Q^{\zu (\hspace{-0.1mm} N \zu)}_{enc}}{dr}
\end{equation}

\vspace{0.2cm}

\noindent{}where $Q^{\zu (\hspace{-0.1mm} N \zu)}_{enc}$ must be calculated considering a spatial sphere of radius $r$. Note that this $\rho^{\zu\zu (\hspace{-0.1mm} N \zu)}$ is gauge-invariant, since by (\ref{eq: enclosed charge definition}) it is constructed making use of the gauge-invariant fluxes (\ref{eq: alpha fluxes definition}).

\phantom{Paragraph}

By the same discussion made with the fluxes (\ref{eq: alpha fluxes definition}), if one wants to obtain a physical charge density, it must be checked whether $\rho^{\zu\zu (\hspace{-0.1mm} N \zu)}$ is reparameterization-invariant. Considering the definition (\ref{eq: enclosed charge definition}), and making use of (\ref{eq: variation alpha fluxes}), one obtains for the variation of $\rho^{\zu\zu (\hspace{-0.1mm} N \zu)}$:

\begin{equation}\label{eq: charge density variation}
\delta \zu \rho^{\zu\zu (\hspace{-0.1mm} N \zu)} = \frac{1}{4\pi r^2} \Tr \hspace{-0.3mm} \Bigg( \frac{d \zu L_\alpha^{N-1}}{dr} \zt \delta L_\alpha + L_\alpha^{N-1} \zt \frac{d \zt \delta \hspace{-0.2mm} L_\alpha}{dr} \hspace{-0.2mm}  \Bigg)
\end{equation}

\vspace{0.2cm}

If $\delta \zu \rho^{\zu\zu (\hspace{-0.1mm} N \zu)} = 0$, then the quantity $\rho^{\zu\zu (\hspace{-0.1mm} N \zu)}$ does not change by reparameterization of the surfaces, and thus the corresponding result is physical, since it is already gauge-invariant. In the same way as before, we will not impose such condition, but instead verify if it is satisfied or not.

\phantom{Paragraph}

With this last paragraph, we have explained everything we will need. In the next section, we shall present the parameterization of the spatial spheres that will be employed in the calculations. Following that, we will proceed with the evaluation of the Wilson lines, which shall be used to calculate the left-hand side of the $\alpha$-equation, as given in (\ref{eq: alpha equation wilson line}).

\section{The Wilson lines}
\label{sec:parameterization and wilson lines}
\setcounter{equation}{0}

\subsection{The parameterization of the spatial spheres}
\label{subsec:parameterization}

In this paper, we will make use of the parameterization constructed in \cite{directtest} for $\Omega$ as a spatial sphere of radius $\zeta_f$\zc , centered at the origin of $x^1$, $x^2$ and $x^3$. The reference point $x_R$ is positioned at infinity, at $(x^1,x^2,x^3) = (-\infty,0,0)$. This choice was made so that one can consider a sphere $\Omega$ of radius $\zeta_f \rightarrow \infty$, corresponding to the whole space at some given Euclidean time $x^4$.

Because of this positioning of $x_R$\zc , an intermediary closed surface that scans $\Omega$ is actually formed by two parts. The first one is the surface of a thin cylinder, of radius $\epsilon$, that comes from $x_R$ at infinity to some intermediary point $(x^1,x^2,x^3) = (-\zeta,0,0)$, over the $x^1$-axis. This thin cylinder then connects, at this intermediary point, to the surface of a sphere of radius $\zeta$, which is centered at the origin. The whole surface, formed by these two parts, is given below.

\newpage

\vspace*{-0.65cm}

\renewcommand{\figurename}{Figure}

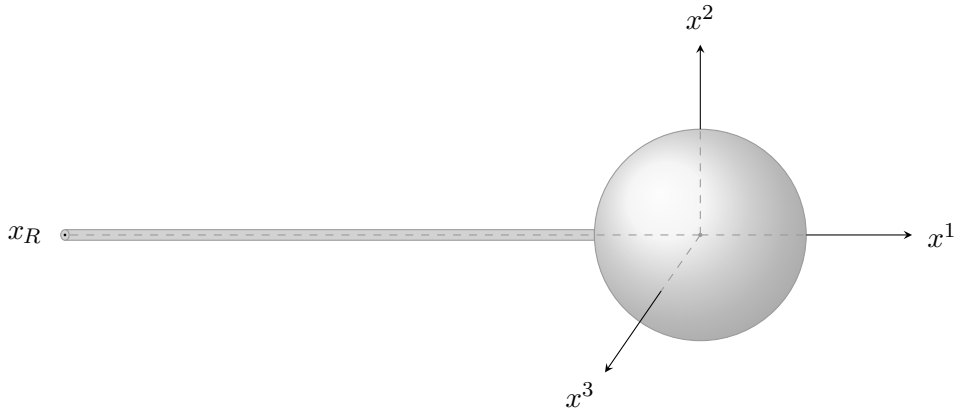
\begin{figure}[H]

\centering

\begin{tikzpicture}[scale=1.4,>=stealth,draw=gray!80]

\def\r{0.05}

\filldraw[fill=gray!35] (-6,-\r) rectangle (-1,\r);

\filldraw[fill=gray!35] (-6,0) ellipse (0.04 and \r);

\fill (-6,0) circle (0.0125);
\node[left] at (-6.125,0) {$x_{\hspace{-0.2mm}R}$};

\shade[ball color=gray!20, opacity=0.5] (0,0) circle (1);
\draw (0,0) circle (1);

\fill[black!40] (0,0) circle (0.02);

\draw[black!40,dashed] (0,0) -- (-6,0);
\draw[black!40,dashed] (0,0) -- (1,0);
\draw[black,->] (1,0) -- (2,0) node[right] at (2.05,0) {$x^1$};

\draw[black!40,dashed] (0,0) -- (0,1);
\draw[black,->] (0,1) -- (0,1.8) node[above] at (0,1.85) {$x^2$};

\draw[black!40,dashed] (0,0) -- (-0.37,-0.53);
\draw[black,->] (-0.37,-0.53) -- (-0.9,-1.3) node[below left] {$x^3$};

\end{tikzpicture}

\vspace{0.1cm}

\caption{Intermediary surface scanning $\Omega$, as constructed in \cite{directtest}. The point $x_R$ is at $(-\infty,0,0)$.}
\label{fig:parameterization}

\end{figure}

With this intermediary surface, one then simply inflate it to scan $\Omega$, until the sphere reaches $\zeta = \zeta_f$\zc . In fact, this expansion must start from the infinitesimal closed surface at the reference point $x_R$\zc . In other words, before inflating the surface of the sphere, one must first inflate the thin cylinder until it reaches the origin $(x^1,x^2,x^3) = (0,0,0)$. Nevertheless, this first step is unimportant since the thin cylinder, as its radius $\epsilon$ goes to zero, does not sweep any volume or area. The most important step is indeed the expansion of the sphere, starting from $\zeta=0$.

\phantom{Paragraph}

For the surface of Figure \ref{fig:parameterization}, one must then scan it with loops based at the point $x_R$\zc . This scanning is performed in three steps, with loops which we shall refer to as types I, II and III. Also, each loop may be divided into different segments, which we now explain below.

\begin{itemize}

\item \textit{Loops of type I:} these loops are responsible for scanning the thin cylinder, before reaching the surface of the sphere. They are parameterized by $\tau$ going from $-\infty$ to $-\pi/2$, with $\tau=-\infty$ corresponding to the infinitesimal loop at $x_R$ and $\tau = -\pi/2$ corresponding to the loop just before reaching the surface of the sphere. For an arbitrary $\tau$ in this interval, the corresponding loop is formed by the three segments below.

\begin{itemize}[label=\scalebox{0.75}{$\blacktriangleright$},topsep=0pt] 
\item \textit{Segment I.1:} this is a straight line on the surface of the thin  cylinder, going from $x_R$ at infinity to some arbitrary coordinate $x^1$ before the surface of the sphere.
\item \textit{Segment I.2:} this segment corresponds to a circle on the surface of the thin cylinder, contained on the plane $x^2x^3$ at the arbitrary coordinate $x^1$ reached by the previous segment. This circle goes around the cylinder, being in fact the segment that scans its surface, as the parameter $\tau$ varies from $-\infty$ to $-\pi/2$. Note however that such scanning will leave an infinitesimal `gap' on the surface of the cylinder, which must be scanned to close the surface. This will be perfomed by the loops of type III.
\item \textit{Segment I.3:} this last segment is another straight line on the surface of the cylinder, which returns to $x_R$ at infinity in order to close the loop.
\end{itemize}

\vspace{-1mm}

\item \textit{Loops of type II:} these loops are responsible for scanning the surface of the sphere, and thus are the most important ones. They are given by the $\tau$-parameter varying from $-\pi/2$ to $\pi/2$, and are also composed of three segments, as we now explain.

\begin{itemize}[label=\scalebox{0.75}{$\blacktriangleright$},topsep=0pt] 
\item \textit{Segment II.1:} this is a straight line on the surface of the thin cylinder, going from $x_R$ to the coordinate $x^1 = -\zeta$, where the cylinder connects with the spherical surface.
\item \textit{Segment II.2:} this is a circular segment that goes around the surface of the sphere, performing its scanning as $\tau$ varies from $-\pi/2$ to $\pi/2$. It starts and ends at the point that connects with the cylinder, and it is contained on the plane making an angle $\tau$ with the plane $x^1x^2$ (which is, also, perpendicular to the plane $x^1x^3$\zu\zu).
\item \textit{Segment II.3:} this final segment is again a straight line on the surface of the cylinder, returning to $x_R$ at infinity in order to close the loop.
\end{itemize}

\vspace{-1mm}

\item \textit{Loops of type III:} these last loops are responsible for closing the surface, as mentioned before. The scanning of the surface of the cylinder by the loops of type I leaves an open `gap', which must be scanned by an arc of circle traversing it, in the same way that the circular segments I.2 scan the rest of the surface. Nevertheless, in the limit where the segments I.1 and I.3 become infinitesimally close, this arc of circle will be infinitesimally small. So, in practice, we are left with the loops of type III consisting only of two segments, as indicated below. The parameter $\tau$ varies from $\pi/2$ to $\infty$, with $\tau = \pi/2$ being the loop just after the surface of the sphere, and $\tau=\infty$ being the final loop of the scanning, at \nlb $x_R$\zc\nlb{}.

\begin{itemize}[label=\scalebox{0.75}{$\blacktriangleright$},topsep=0pt] 
\item \textit{Segment III.1:} this is a straight line on the surface of the thin cylinder, going from $x_R$ at infinity to some arbitrary coordinate $x^1$ before the surface of the sphere.
\item \textit{Segment III.2:} this is another straight line on the surface of the cylinder, returning from this arbitrary coordinate $x^1$ to $x_R$ at infinity, in order to close the loop.
\end{itemize}

\end{itemize}

\vspace{1mm}

So, these are all the segments of all the loops necessary to scan the intermediate surface of Figure \ref{fig:parameterization}. For the scanning of $\Omega$, one just inflates the spherical surface, from $\zeta=0$ to $\zeta=\zeta_f$\zc .

For a visual representation of all the loops and segments above, one can consult the figures presented in \cite{directtest}, where this parameterization was first constructed. The parametric equations for each one of these segments, taken from \cite{directtest}, are given here in Appendix \ref{apx:parameterization equations}.

\subsection{Obtaining the Wilson lines}
\label{subsec:wilson lines}

Now that we have the parametric equations for all the segments of all the loops necessary to scan $\Omega$, we make use of the BPST instanton solution for the gauge fields $A_\mu$\zc , and substitute it on the Wilson line definition, given in (\ref{eq: wilson line definition}), to calculate $W$. The solution for $A_\mu$\zc , obtained in 1975 \cite{bpstinstanton} by Belavin, Polyakov, Schwartz and Tyupkin, can be written as:

\begin{equation}\label{eq: instanton solution}
A_i =  -\frac{2}{e} \hspace{0.3mm} \frac{1}{x^2 + 1} \hspace{0.3mm} \big( \hspace{0.3mm} \varepsilon_{ijk} \hspace{0.6mm} x^j \hspace{0.3mm} T_k - \kappa \hspace{0.5mm} x^4 \hspace{0.3mm} T_i \hspace{0.3mm} \big) \hspace{0.6cm} , \hspace{0.5cm} A_4 = - \frac{2\kappa}{e} \frac{x^k}{x^2 + 1} \hspace{0.3mm} T_k
\end{equation}

\vspace{0.2cm}

\noindent{}where $x^2 \equiv x_\mu \zt x^\mu$ with an Euclidean metric, $T_i$ are the generators of the $\mathfrak{su}(2)$ algebra, satisfying $\big[ \zt T_i \zc\zu , \zu T_j \zu\zu \big] \hspace{-0.2mm} = i \zc \varepsilon_{ij\zu\zu k} \zc T_k$\zc , and where the constant $\kappa$ is the same $\kappa = \pm 1$ appearing in the self-duality equation (\ref{eq: self-duality equation}). For $\kappa=1$ we have the instanton (or 1-instanton) solution, and for $\kappa=-1$ the anti-instanton. We will keep this constant $\kappa$ in our expressions, and therefore our following results will be valid for both the instanton and anti-instanton solutions.

It must be mentioned that the most general form of (\ref{eq: instanton solution}) would also involve five more constants, which are not present here. Four of them correspond to a constant vector $a^\mu$, which determines the position of the instanton (or anti-instanton). In expression (\ref{eq: instanton solution}) above we have set $a^\mu = 0$ for $\mu = 1,2,3,4$\zc , which means that we are considering our solutions centered at the origin of Euclidean spacetime. The other constant is generally called $\lambda$, and it has the interpretation of the `size' of the solutions. Nevertheless, since the pure Yang-Mills theory that is being considering is conformally invariant, we have made the scale transformation $x^\mu \rightarrow \lambda \zt x^\mu$ to remove this constant and obtain (\ref{eq: instanton solution}). This scale transformation that was performed establishes indeed the interpretation of $\lambda$ as the size of the instanton (or anti-instanton).

\phantom{Paragraph}

The integration volume $\Omega$ that we are considering for this paper is spatial in Euclidean spacetime. Therefore, we are only going to need the $A_i$ components of (\ref{eq: instanton solution}), and the quantity $A_\mu \hspace{-0.4mm} \cdot \hspace{-0.2mm} dx^\mu/d\sigma$ of (\ref{eq: wilson line definition}) reduces to $A_i \hspace{-0.2mm} \cdot \hspace{-0.0mm} dx^i/d\sigma$. So, let us start by considering the straight line segments of the previous subsection. These correspond to six of the eight segments present in the parameterization, and since they are all straight lines, we can calculate them all at once.

Consulting Appendix \ref{apx:parameterization equations}, we see that for all these segments $x^2 = 0$ and $x^3 = -\epsilon$. Also, we see that $dx^1/d\sigma = \pm \zt 1$, where the upper sign refers to the segments that are coming from $x_R$\zc{\nlb}, and the lower sign to the segments that are returning to it. So, we are left with $A_i \hspace{-0.2mm} \cdot \hspace{-0.0mm} dx^i/d\sigma = \pm \zt A_1$\zc{\nlb}, where we then substitute (\ref{eq: instanton solution}) and take the limit as $\epsilon \rightarrow 0$ for the thin cylinder, to obtain:

\begin{equation}\label{eq: A x dsigma straight lines}
A_\mu \hspace{0.1mm} \frac{dx^\mu}{d\sigma} \bigg|_{\hspace{0.3mm}\text{straight lines}} = \pm \frac{2\kappa}{e} \hspace{0.2mm} \frac{x^4}{(x^4)^2 + (x^1)^2 + 1} \hspace{0.5mm} T_1
\end{equation}

\vspace{0.25cm}

\noindent{}which is valid for the segments I.1, I.3, II.1, II.3, III.1 and III.2, and where the coordinate $x^1$ must be substituted by the correct expression for each segment, consulting Appendix \ref{apx:parameterization equations}.

Substituting this expression on the Wilson line definition (\ref{eq: wilson line definition}) we obtain:

\begin{equation}\label{eq: wilson line equation straight lines}
\frac{dW_{\zc\text{s\zt l}}^{\zt (\pm)}}{d\sigma} + i \zt\zt T_{\zc\text{s\zt l}}^{\zt (\pm)} \zt\zt W_{\zc\text{s\zt l}}^{\zt (\pm)}  = 0 \hspace{1cm} \text{where} \hspace{1cm} T_{\zc\text{s\zt l}}^{\zt (\pm)} \equiv \pm \zc \frac{2 \zt \kappa \zt x^4}{(x^4)^2 + (x^1)^2 + 1} \zc T_1
\end{equation}

\vspace{0.2cm}

In this expression, the subscript `sl' refers to the calculation on the straight-line segments, and the superscript $(\pm)$ refers to the plus and minus signs of the quantity $T_{\zc\text{s\zt l}}^{\zt (\pm)}$, whose meaning was just explained above. Now, note that $T_{\zc\text{s\zt l}}^{\zt (\pm)}$ is in the $T_1${\zt}-{\zt}direction only, which means that it commutes with itself. Because of that, the path-ordering of (\ref{eq: wilson line definition}) is unimportant, and we have the Wilson line $W_{\zc\text{s\zt l}}^{\zt (\pm)}$ on the straight lines being given by the following exponential of $T_1$\zt :

\vspace{-0.35cm}

\begin{equation}\label{eq: wilson line straight lines}
W_{\zc\text{s\zt l}}^{\zt (\pm)}  = \exp \bigg( \hspace{-1.3mm} - \hspace{-0.3mm} i \zc 2 \zt \kappa \zt x^4 \zt\zu w_{\zc\text{s\zt l}}^{\zt (\pm)} \zt\zu T_1\zu \bigg) \hspace{.9cm} \text{where} \hspace{.9cm} w_{\zc\text{s\zt l}}^{\zt (\pm)} \equiv \pm \hspace{-0.5mm} \int \hspace{-0.3mm} d\sigma \zt\zt \frac{1}{(x^4)^2 + (x^1)^2 + 1} 
\end{equation}

\vspace{0.2cm}

\noindent{}and where the coordinate $x^1$, as well as the integration limits, depend on which segment is being considered. For each segment, the $w_{\zc\text{s\zt l}}^{\zt (\pm)}$ integral above can be easily solved by using:

\begin{equation}\label{eq: integral to solve}
\int \hspace{-0.3mm} dx \zc\zt \frac{1}{a + (b \pm x)^2} = \pm \zt \frac{1}{\sqrt{a}} \zu\zu \arctan \hspace{-0.5mm} \bigg( \frac{b \pm x}{\sqrt{a}} \bigg) + c
\end{equation}

\vspace{0.25cm}

\noindent{}where $a$ and $b$ are real constants, with $a > 0$, and where $c$ is the integration constant.

\phantom{Paragraph}

So, by consulting Appendix \ref{apx:parameterization equations} and substituting the parametric equations for the straight-line segments on the expressions above, one obtains for the segments II.1 and II.3, for example:

\begin{equation}\label{eq: wilson ii.1}
W(\text{II.1}) = \exp \Bigg\{ \hspace{-1mm} - \hspace{-0.3mm} i \zc\zu\zu \frac{2\zu\kappa \zt x^4}{\sqrt{(x^4)^2 + 1}} \bigg[ \arctan \hspace{-0.2mm} \bigg( \frac{-\zeta}{\sqrt{(x^4)^2 + 1}} \zu\zu \bigg) + \frac{\pi}{2} \zc \bigg] \zc T_1 \zu\zu \Bigg\}
\end{equation}

\vspace{0.05cm}

\begin{equation}\label{eq: wilson ii.3}
W(\text{II.3}) = W(\text{II.1})^{-1}
\end{equation}

\vspace{0.25cm}

These will be the most important straight-line segments, since they conjugate segment II.2 which is on the surface of the sphere. Note that $W(\text{II.3}) = W(\text{II.1})^{-1}$, which is expected since the segment II.3 is the same as II.1, but going backwards. With respect to (\ref{eq: wilson ii.1}), we define:

\begin{equation}\label{eq: omega and beta definitions}
\omega \equiv \frac{\pi \kappa \zt x^4}{\sqrt{(x^4)^2 + 1}} \hspace{0.8cm} , \hspace{0.75cm} \eta \equiv - \zc \frac{2\zu\kappa \zt x^4}{\sqrt{(x^4)^2 + 1}} \zt \arctan \hspace{-0.3mm} \bigg( \frac{\zeta}{\sqrt{(x^4)^2 + 1}} \zu\zu \bigg)
\end{equation}

\vspace{0.225cm}

\noindent{}with which we may write $W(\text{II.1})$ and $W(\text{II.3})$ in the more compact forms:

\begin{equation}\label{eq: wilson lines ii.1 and ii.3}
W(\text{II.1}) = e^{-i \zt \eta \zt T_1} \zt e^{-i \zt \omega \zt T_1} \hspace{0.8cm} , \hspace{0.75cm} W(\text{II.3}) = e^{i \zt \omega \zt T_1} \zt e^{i \zt \eta \zt T_1}
\end{equation}

\vspace{0.25cm}

For the other straight lines we obtain expressions very similar to (\ref{eq: wilson ii.1}). Nevertheless, we are not going to need them. What will be important are the relations analogous to (\ref{eq: wilson ii.3}):

\begin{equation}\label{eq: wilson other straight lines}
W(\text{I.3}) = W(\text{I.1})^{-1} \hspace{0.85cm} , \hspace{0.8cm} W(\text{III.2}) = W(\text{III.1})^{-1}
\end{equation}

\vspace{0.2cm}

The next segment that we will consider is segment I.2, which is the circular path going around the surface of the thin cylinder, as explained in the previous subsection. Consulting Appendix \ref{apx:parameterization equations}, we see that $dx^1/d\sigma = 0$, $dx^2/d\sigma = \epsilon \cos\sigma$ and $dx^3/d\sigma = \epsilon \sin\sigma$. Therefore, we have that $A_i \hspace{-0.2mm} \cdot \hspace{-0.0mm} dx^i/d\sigma \propto \epsilon$, which goes to zero as $\epsilon \rightarrow 0$.\footnote{Actually, this limit must involve a more careful analysis since we will have $\epsilon$ being present both on the numerator and on the denominator of $A_i \hspace{-0.2mm} \cdot \hspace{-0.0mm} dx^i/d\sigma$. Also, we will have the presence of the coordinate $x^1$, which can assume infinite values for this segment. Nevertheless, such careful analysis shows indeed that $A_i \hspace{-0.2mm} \cdot \hspace{-0.0mm} dx^i/d\sigma \rightarrow 0$.} Thus, we obtain for $W(\text{I.2})$:

\begin{equation}\label{eq: wilson i.2}
A_\mu \hspace{0.1mm} \frac{dx^\mu}{d\sigma}\bigg|_{\hspace{0.3mm}\text{I.2}} = 0 \hspace{.8cm} \Rightarrow \hspace{.75cm} W(\text{I.2}) = \mathbb{1}
\end{equation}

\vspace{0.25cm}

The only segment left to consider is segment II.2. As already mentioned, this is the most important segment of all the parameterization, since it is the one effectivelly scanning the surface of the spheres. Also, it is the segment whose Wilson line demands the most steps to calculate.

The first step is to substitute $A_i$ and $dx^i/d\sigma$ into $A_i \hspace{-0.2mm} \cdot \hspace{-0.0mm} dx^i/d\sigma$. Doing that we obtain:

\vspace{-0.25cm}

\begin{equation}\label{eq: wilson sphere 1}
\begin{aligned}
A_\mu \hspace{0.1mm} \frac{dx^\mu}{d\sigma}\bigg|_{\hspace{0.3mm}\text{II.2}} &= \frac{2}{e} \hspace{0.2mm} \frac{\zeta\cos\tau}{(x^4)^2 + \zeta^2 + 1} \bigg\{ \zeta\cos\tau \sin\tau \hspace{0.5mm} (1 - \cos\sigma) \hspace{0.5mm} T_1 + \zeta\sin\tau \sin\sigma \hspace{0.5mm} T_2 \hspace{0.5mm} + \\[2.7mm]
&\hspace{-1.1cm}+ \zeta \big[ \sin^2\hspace{-0.3mm}\tau \hspace{0.5mm} (1 - \cos\sigma) - 1 \hspace{0.3mm} \big] \hspace{0.3mm} T_3 + \kappa \hspace{0.5mm} x^4 \big( \hspace{-0.5mm} \cos\tau \sin\sigma \hspace{0.5mm} T_1 + \cos\sigma \hspace{0.5mm} T_2 + \sin\tau \sin\sigma \hspace{0.5mm} T_3 \hspace{0.2mm} \big) \hspace{-0.2mm} \bigg\}
\end{aligned}
\end{equation}

\vspace{0.2cm}

\noindent{}which can be greatly simplified by making use of the following result for the $\mathfrak{su}(2)$ algebra:

\begin{equation}\label{eq: adjoint action su(2)}
e^{i \zu \theta \zu T_i} \zt\zt T_j \zt\zt e^{-i \zu \theta \zu T_i} = \cos\theta \hspace{0.3mm} \hspace{0.3mm} T_j - \sin\theta \hspace{0.3mm} \hspace{0.3mm} \varepsilon_{ijk} \hspace{0.3mm} T_k \hspace{.8cm} , \hspace{.75cm} i \neq j
\end{equation}

\vspace{0.25cm}

With that, one may write the following quantities as conjugations of generators of $\mathfrak{su}(2)$:

\vspace{0.1cm}

\begin{equation}\label{eq: simplification conjugation 1}
\begin{gathered}
\cos\tau \sin\tau \hspace{0.5mm} (1 - \cos\sigma) \hspace{0.5mm} T_1 + \sin\tau \sin\sigma \hspace{0.5mm} T_2 \hspace{0.5mm} + \big[ \sin^2\hspace{-0.3mm}\tau \hspace{0.5mm} (1 - \cos\sigma) - 1 \hspace{0.3mm} \big] \hspace{0.3mm} T_3 = \\[5mm]
= - e^{i \zt \tau \zt T_2}\zt e^{i\zt\sigma\zt T_3}\zt e^{-i\zt\tau\zt T_2}\zt\zt T_3 \zt\zt e^{i\zt\tau\zt T_2}\zt e^{-i\zt\sigma\zt T_3}\zt e^{-i\zt\tau\zt T_2}
\end{gathered}
\end{equation}


\begin{equation}\label{eq: simplification conjugation 2}
\begin{gathered}
\cos\tau \sin\sigma \hspace{0.5mm} T_1 + \cos\sigma \hspace{0.5mm} T_2 + \sin\tau \sin\sigma \hspace{0.5mm} T_3 = e^{i \zt \tau \zt T_2}\zt e^{i\zt\sigma\zt T_3}\zt\zt T_2 \zt\zt e^{-i\zt\sigma\zt T_3}\zt e^{-i\zt\tau\zt T_2}
\end{gathered}
\end{equation}

\vspace{0.3cm}

\noindent{}which then can be used to write expression (\ref{eq: wilson sphere 1}) in the much interesting form:

\vspace{-0.2cm}

\begin{equation}\label{eq: wilson sphere 2}
A_\mu \hspace{0.1mm} \frac{dx^\mu}{d\sigma}\bigg|_{\hspace{0.3mm}\text{II.2}} = \frac{2}{e} \hspace{0.2mm} \frac{\zeta\cos\tau}{(x^4)^2 + \zeta^2 + 1} \zt\zt e^{i \zt \tau \zt T_2}\zt e^{i\zt\sigma\zt T_3}\zt e^{-i\zt\tau\zt T_2}\zt \big( \hspace{0.2mm} \kappa \zt x^4 \zt T_2 - \zeta \zt T_3 \hspace{0.3mm} \big) \zt e^{i\zt\tau\zt T_2}\zt e^{-i\zt\sigma\zt T_3}\zt e^{-i\zt\tau\zt T_2}
\end{equation}

\vspace{0.2cm}

The interest in writing $A_\mu \hspace{-0.2mm} \cdot \hspace{-0.0mm} dx^\mu/d\sigma$ in this format is that now the $\sigma$-dependence is totally separated in the conjugation by $e^{i\zt\sigma\zt T_3}$, whereas in expression (\ref{eq: wilson sphere 1}) it was present in almost all of the terms. In fact, with the expression above, one can actually perform a gauge transformation to completely \textit{remove} the $\sigma$-depende. Considering the gauge transformation given by:

\vspace{-0.25cm}

\begin{equation}\label{eq: gauge transformation 1}
A_\mu \hspace{0.1mm} \frac{dx^\mu}{d\sigma} \rightarrow \bar{A}_\mu \hspace{0.1mm} \frac{dx^\mu}{d\sigma} = g A_\mu \zc g^{-1} \zu \frac{dx^\mu}{d\sigma} + \frac{i}{e} \zt( \partial_\mu \zt g) \zt g^{-1} \zu \frac{dx^\mu}{d\sigma} =g \bigg( \hspace{-0.7mm} A_\mu \frac{dx^\mu}{d\sigma} \zu\zu \bigg) \zu g^{-1} + \frac{i}{e} \bigg( \frac{\partial g}{\partial\sigma} \bigg) \zu g^{-1}
\end{equation}

\vspace{0.25cm}

\noindent{}and making the following choice for the group element $g$, with its inverse $g^{-1}$:

\begin{equation}\label{eq: gauge transformation 2}
g = e^{i \zt \tau \zt T_2}\zt e^{-i\zt\sigma\zt T_3}\zt e^{-i\zt\tau\zt T_2} \hspace{0.75cm} \Rightarrow \hspace{0.75cm} g^{-1} = e^{i \zt \tau \zt T_2}\zt e^{i\zt\sigma\zt T_3}\zt e^{-i\zt\tau\zt T_2}
\end{equation}

\vspace{0.25cm}

\noindent{}it is simple to show, using (\ref{eq: gauge transformation 1}), that expression (\ref{eq: wilson sphere 2}) transforms itself into:

\vspace{-0.3cm}

\begin{equation}\label{eq: wilson sphere 3}
\bar{A}_\mu \hspace{0.1mm} \frac{dx^\mu}{d\sigma} \bigg|_{\hspace{0.3mm}\text{II.2}} = \frac{1}{e} \Bigg[ \frac{2 \zc\zt \zeta \hspace{-0.3mm} \cos\tau}{(x^4)^2 + \zeta^2 + 1} \zu\zu  \big( \hspace{0.2mm} \kappa \zt x^4 \zt T_2 - \zeta \zt T_3 \hspace{0.3mm} \big) + \cos\tau \zt\zt T_3 - \sin\tau \zt\zt T_ 1 \zu \Bigg] \equiv \frac{1}{e} \zt\zu T_{\zt\text{s}}
\end{equation}

\vspace{0.225cm}

Note that the quantity $T_{\zt\text{s}}$ defined above does not in fact depend on $\sigma$. Therefore, the path-ordering to calculate the Wilson line is, again, unimportant, and from (\ref{eq: wilson line definition}) we obtain:

\begin{equation}\label{eq: wilson sphere 4}
\overbar{W}(\text{II.2}) = e^{-i \zt 2\pi \zu\zu T_{\zt\text{s}}}
\end{equation}

\vspace{0.25cm}

\noindent{}where the factor of $2\pi$ comes from the $\sigma$-interval in the parametric equations (\ref{eq_apx1: segment ii.2}). Note that we have obtained the transformed Wilson line $\overbar{W}(\text{II.2})$, since we have used the transformed expression (\ref{eq: wilson sphere 3}). It is of interest here to return to the previous gauge, which can be easily done by remebering that $W(x)$ transforms as $W(x) \rightarrow\overbar{W}(x) = g(x) W(x) g_{\zt i}^{-1}$. In the calculation above, $\overbar{W}(\text{II.2})$ was determined at the end of segment II.2, which corresponds to the parameter $\sigma_{\hspace{-0.4mm}f}=2\pi$. For the element $g_{\zt i}$ of the transformation of $W(x)$ indicated above, we have that it corresponds to the beginning of the segment, which for segment II.2 is given by the parameter $\sigma_i = 0$. Therefore, by identifying $g\zt(\sigma = 2\pi) \equiv g_f$ and $g \zt (\sigma = 0) \equiv g_{\zt i}$\zc , we obtain from (\ref{eq: gauge transformation 2}):

\begin{equation}\label{eq: gauge transformation 3}
g_f = e^{i \zt \tau \zt T_2}\zt e^{-i\zt 2\pi\zt T_3}\zt e^{-i\zt\tau\zt T_2} = \pm \mathbb{1} \hspace{0.85cm} , \hspace{0.75cm} g^{-1}_{\zt i}  = e^{i \zt \tau \zt T_2}\zt e^{i\zt 0 \zt T_3}\zt e^{-i\zt\tau\zt T_2} = \mathbb{1}
\end{equation}

\vspace{0.25cm}

\noindent{}where we have used $e^{-i \zt 2\pi \zt T_3} = \pm \mathbb{1}$, with the top sign corresponding to the case of an integer-spin representation of $\mathfrak{su}(2)$, and the lower sign to a half-integer-spin representation. Employing then the transformation $W(x) \rightarrow\overbar{W}(x) = g(x) W(x) g_{\zt i}^{-1}$\zc , one easily obtains the Wilson line:

\begin{equation}\label{eq: wilson sphere 5}
W(\text{II.2}) = \pm \zu\zu e^{-i \zt 2\pi \zu\zu T_{\zt\text{s}}}
\end{equation}

\vspace{0.25cm}

\noindent{}with the same interpretation as indicated above for the upper or lower sign.

\phantom{Paragraph}

The result (\ref{eq: wilson sphere 5}) above can be further simplified. By rearranging $T_{\zt\text{s}}$ in (\ref{eq: wilson sphere 3}), one obtains:

\begin{equation}\label{eq: simplification 1}
T_{\zt\text{s}} = \cos\tau \zt \Bigg[ \zt \frac{(x^4)^2 + 1 - \zeta^2}{(x^4)^2 + 1 + \zeta^2} \zt\zt T_3 + \frac{2 \zt \zeta \zu\zu \kappa \zc x^4}{(x^4)^2 + 1 + \zeta^2} \zt\zt T_2  \Bigg] - \sin\tau \zc T_1
\end{equation}

\vspace{0.25cm}

Next, with the terms inside of the square brackets, one defines the following quantities:

\begin{equation}\label{eq: definition alpha and K}
\cos\upsilon \equiv \frac{1}{K} \zc \frac{(x^4)^2 + 1 - \zeta^2}{(x^4)^2 + 1 + \zeta^2} \hspace{0.8cm} , \hspace{0.75cm} \sin\upsilon \equiv \frac{1}{K} \zc \frac{2 \zt \zeta \zu\zu \kappa \zc x^4}{(x^4)^2 + 1 + \zeta^2}
\end{equation}

\vspace{0.25cm}

\noindent{}where the expression for the function $K$ introduced above may be obtained by squaring both terms and using $\sin^2\hspace{-0.5mm}\upsilon + \cos^2\hspace{-0.5mm}\upsilon = 1$. Doing so, and simplifying the result, one obtains:

\begin{equation}\label{eq: expression K}
K = \zt \sqrt{ \zc\zt 1 - \frac{4\zeta^2}{[ \zt (x^4)^2 + 1 + \zeta^2 \zt ]^{\zt 2}} \zc}
\end{equation}

\vspace{0.225cm}

Having defined (\ref{eq: definition alpha and K}), one then uses property (\ref{eq: adjoint action su(2)}) so that $T_{\zt\text{s}}$ can be written as:

\begin{equation}\label{eq: simplification 2}
T_{\zt\text{s}} = e^{i\zt\upsilon\zt T_1} \zt \big( K\hspace{-0.3mm}\cos\tau \zc T_3 - \sin\tau \zc T_1 \zu\zu \big) \zt e^{-i\zt\upsilon\zt T_1}
\end{equation}

\vspace{0.225cm}

Repeating the same procedure with the terms inside the brackets, we now define:

\vspace{-0.35cm}

\begin{equation}\label{eq: definition gamma and F}
\cos\gamma \equiv \frac{1}{F} \zt K \hspace{-0.3mm} \cos\tau \hspace{0.7cm} , \hspace{0.65cm} \sin\gamma \equiv \frac{1}{F} \zt \sin\tau \hspace{0.7cm} , \hspace{0.65cm} F = \sqrt{K^{\zu\zu2} \hspace{-0.1mm} \cos^{\zt\zu2}\hspace{-0.5mm}\tau + \sin^{\zu\zu2}\hspace{-0.6mm}\tau}
\end{equation}

\vspace{0.25cm}

\noindent{}and use property (\ref{eq: adjoint action su(2)}) one more time, so that the quantity $T_{\zt\text{s}}$ above may be written as:

\begin{equation}\label{eq: simplification 3}
T_{\zt\text{s}} = e^{i\zt\upsilon\zt T_1} \zt e^{i\zt\gamma\zt T_2} \zt F \zt T_3 \zt\zt\zt e^{-i\zt\gamma\zt T_2} \zt e^{-i\zt\upsilon\zt T_1}
\end{equation}

\vspace{0.25cm}

Lastly, recording the expression in (\ref{eq: wilson sphere 5}), the final step is to use the general property:

\begin{equation}\label{eq: exponential property}
e^{\zu e^{\zu L} \zt T \zt e^{-L}} = e^L \zt e^T \zt e^{-L}
\end{equation}

\vspace{0.225cm}

\noindent{}and thus write the Wilson line $W(\text{II.2})$ on the surface of the sphere as the conjugation:

\begin{equation}\label{eq: wilson line ii.2}
W(\text{II.2}) = \pm \zc e^{i\zt\upsilon\zt T_1} \zt e^{i\zt\gamma\zt T_2} \zt e^{-i \zt 2\pi \zu F \zt\zu T_3} \zt e^{-i\zt\gamma\zt T_2} \zt e^{-i\zt\upsilon\zt T_1}
\end{equation}

\vspace{0.25cm}

This is an important result of the present paper. With it, we complete the determination of the Wilson lines on all the segments of the parameterization of the spatial spheres.

\phantom{Paragraph}

What we must do now, before going to the next section, is combine the Wilson lines obtained above to have them calculated on the whole loops. Starting with the loops of type I, we combine its segments following $W_c\zc (\zu\text{I}\zu) = W(\text{I.3}) \zt W(\text{I.2}) \zt W(\text{I.1})$, where we are making use of the index $c$ to denote a closed path, as introduced in the text above (\ref{eq: alpha equation wilson line}). In (\ref{eq: wilson i.2}), we have obtained $W(\text{I.2}) = \mathbb{1}$, and from (\ref{eq: wilson other straight lines}) we have $W(\text{I.3}) = W(\text{I.1})^{-1}$. Thus, we obtain $W_c\zc (\zu\text{I}\zu)$ given by:

\begin{equation}\label{eq: wilson loop i}
W_c\zc (\zu\text{I}\zu) = \mathbb{1}
\end{equation}

\vspace{0.25cm}

Next, for the loops of type II, we combine the results (\ref{eq: wilson lines ii.1 and ii.3}) and (\ref{eq: wilson line ii.2}), following the same ordering as indicated above. For the outermost conjugations, we define the following function:

\begin{equation}\label{eq: rho conjugation definition}
\xi \equiv \omega + \eta + \upsilon
\end{equation}

\vspace{0.25cm}

\noindent{}with $\omega$ and $\eta$ defined in (\ref{eq: omega and beta definitions}), and $\upsilon$ defined in (\ref{eq: definition alpha and K}). Doing so, we obtain $W_c\zc (\text{II})$ given by:

\begin{equation}\label{eq: wilson loop ii}
W_c\zc (\text{II}) = \pm \zc e^{i\zt\xi\zt T_1} \zt e^{i\zt\gamma\zt T_2} \zt e^{-i \zt 2\pi \zu F \zt\zu T_3} \zt e^{-i\zt\gamma\zt T_2} \zt e^{-i\zt\xi\zt T_1}
\end{equation}

\vspace{0.25cm}

\noindent{}where the upper sign must be used for an integer-spin representation of $\mathfrak{su}(2)$ and the lower sign for a half-integer-spin representation. As we will see below, this sign will not be important.

Finally, with the result in (\ref{eq: wilson other straight lines}) for the segments III.1 and III.2, we obtain directly:

\begin{equation}\label{eq: wilson loop iii}
W_c\zc (\text{III}) = \mathbb{1}
\end{equation}

\vspace{0.25cm}

Consulting equation (\ref{eq: alpha equation wilson line}), in particular its left-hand side (on which we are more interested), we see that the next quantity we shall calculate is the term $W_c^{-1}\hspace{-0.5mm} \cdot \hspace{-0.1mm} dW_c/d\tau$, for each type of loop. Thus, we immediately see that we will have no contribution from the loops of type I and III, given in (\ref{eq: wilson loop i}) and (\ref{eq: wilson loop iii}) above, which as we saw are responsible for scanning the surface of thin cylinder. The only loops that will contribute are the loops of type II, given in (\ref{eq: wilson loop ii}), which are the ones effectively scanning the surface of the sphere. For these loops, as indicated in the previous subsection, we have the $\tau$-parameter varying from $-\pi/2$ to $\pi/2$. Therefore, this is the interval of $\tau$ that is going to be relevant for the integral on the left-hand side of (\ref{eq: alpha equation wilson line}).

\section{Gauge-invariant results}
\label{sec:results}
\setcounter{equation}{0}

\subsection{Magnetic fluxes and enclosed magnetic charges}
\label{subsec:charges and fluxes}

In this subsection, the quantity that we will be primarily interested is the gauge-invariant magnetic flux (\ref{eq: alpha fluxes definition}), which is obtained from the left-hand side of the integral equation of order $\alpha$ (\ref{eq: alpha equation}), corresponding to the non-Abelian version of Gauss law for the magnetic fields.

For the case of the instanton (and anti-instanton), given in (\ref{eq: instanton solution}), this integral equation corresponds also to the non-Abelian Gauss law for the electric fields, because of self-duality. Therefore, the magnetic flux that we shall plot for these solutions, which from Gauss law correspond to the enclosed magnetic charges on the integration volumes considered, will also represent the \textit{electric} flux for these solutions, being equal to its enclosed \textit{electric} charges. Nevertheless, since we are considering the $\alpha$-equation, we shall sometimes refer to such results simply as magnetic.

\phantom{Paragraph}

For the left-hand side of the $\alpha$-equation (\ref{eq: alpha equation}), let us make use of the notation $L_\alpha$ introduced in the text above equation (\ref{eq: variation left alpha}), and of the results (\ref{eq: variation wilson final}) and (\ref{eq: alpha equation wilson line}), to write:

\begin{equation}\label{eq: left side alpha wilson}
L_\alpha \equiv \int_{\tau_i}^{\tau_f} \hspace{-0.5mm} d\tau \hspace{-0.3mm} \int_{\sigma_i}^{\sigma_f} \hspace{-0.3mm} d\sigma \hspace{0.2mm} F_{\mu\nu}^{\hspace{0.3mm}W} \hspace{0.2mm} \frac{dx^\mu}{d\sigma} \frac{dx^\nu}{d\tau} \bigg|_{\zeta = \zeta_f} \hspace{-0.75mm} = \frac{1}{ie} \int_{\tau_i}^{\tau_f}\hspace{-0.5mm} d\tau \hspace{0.3mm} W_c^{-1} \hspace{0.3mm} \frac{dW_c}{d\tau} \bigg|_{\hspace{0.2mm}\zeta = \zeta_f}
\end{equation}

\vspace{0.2cm}

From (\ref{eq: alpha fluxes definition}), we see that we must calculate $L_\alpha$ in order to evaluate the fluxes $\Phi_B^{(\alpha \zu , \zu N \zu)}$. In particular, from the last equality above, we see that this can be done by simply substituting the quantity $W_c\zc (\text{II})$ regarding the loops of type II, as given in (\ref{eq: wilson loop ii}). The loops of types I and III will not be necessary because the quantities (\ref{eq: wilson loop i}) and (\ref{eq: wilson loop iii}) will not contribute to (\ref{eq: left side alpha wilson}).

With that being said, by substituting (\ref{eq: wilson loop ii}) and its inverse into $W_c^{-1}\hspace{-0.5mm} \cdot \hspace{-0.1mm} dW_c/d\tau$, making use of the property (\ref{eq: adjoint action su(2)}) for the $\mathfrak{su}(2)$ algebra and simplifying the expression, one obtains:

\vspace{-0.1cm}

\begin{equation}\label{eq: wilson result 1}
\begin{aligned}
W_c^{-1} \zt \frac{dW_c}{d\tau} \zt \bigg|_{\zt \text{II}} &= i \zt e^{i\zt\xi\zt T_1} \zc \Bigg\{ \bigg[ \zt \frac{d\gamma}{d\tau} \zu \sin \hspace{-0.5mm} \big(2\pi \hspace{-0.2mm} F\big) \hspace{-0.2mm} \cos\gamma + 2\pi \zt \frac{dF}{d\tau} \zu \sin\gamma \zt \bigg] \zt T_1 \zc - \\[2.25mm] 
&\hspace{-0.5cm}- \frac{d\gamma}{d\tau} \zt \big[ \zu\zu 1 - \cos\hspace{-0.1mm}\big(2\pi \hspace{-0.2mm} F\big)\zu\zu\big] \zc  T_2 + \hspace{-0.2mm} \bigg[ \zt \frac{d\gamma}{d\tau} \zu \sin \hspace{-0.5mm} \big(2\pi \hspace{-0.2mm} F\big) \hspace{-0.2mm} \sin\gamma - 2\pi \zt \frac{dF}{d\tau} \hspace{-0.2mm} \cos\gamma \zt \bigg] \zt T_3 \zt  \Bigg\} \zt e^{-i\zt\xi\zt T_1}
\end{aligned}
\end{equation}

\vspace{0.225cm}

Note that the $\pm$ signs of (\ref{eq: wilson loop ii}) canceled out in the expression, so that the result above is valid for any representation of $\mathfrak{su}(2)$. Also, note that we have factored out the conjugation by $e^{i\zt\xi\zt T_1}$, which comes from (\ref{eq: wilson loop ii}). The reason for it is that this conjugation does not depend on the parameter $\tau$. Indeed, by consulting the definition of $\xi$ in equation (\ref{eq: rho conjugation definition}), and checking out the functions $\omega$, $\eta$ and $\upsilon$ that are present in this quantity, one sees that $\xi = \xi \zc (\zeta \zt ; x^4)$. Therefore, when substituting (\ref{eq: wilson result 1}) into (\ref{eq: left side alpha wilson}), this conjugation by $e^{i\zt\xi\zt T_1}$ will drop out the $\tau$-integral.

More than that, being the gauge-invariant flux (\ref{eq: alpha fluxes definition}) given by a trace, we see that this conjugation will actually cancel out by the cyclic property, and thus will not contribute to the quantity $\Phi_B^{(\alpha \zu , \zu N \zu)}$. This is a rather important result if we remember that the quantities $\omega$ and $\eta$, which are present inside $\xi$, actualy come from the straight-line segments II.1 and II.3, given in equation (\ref{eq: wilson lines ii.1 and ii.3}). Therefore, the result above means that there will be no contribution at all from the straight-line segments of the parameterization for the gauge-invariant results to be obtained from (\ref{eq: alpha fluxes definition}). The segments connecting $x_R$ are important for the loop-based scheme on which the integral formulation is built, but here we see that they will not contribute to the gauge-invariant results to be extracted from the integral equations. The only segment that will contribute for these results is segment II.2, which is the one effectively scanning the surface of the spheres.

\phantom{Paragraph}

Having said that, we denote the quantity (\ref{eq: wilson result 1}) as being given by the following definition:

\begin{equation}\label{eq: wilson result 2}
W_c^{-1} \zt \frac{dW_c}{d\tau} \zt \bigg|_{\zt \text{II}} \equiv e^{i\zt\xi\zt T_1} \zu \big( \zu\zu \widetilde{w}^{\zc 1} \hspace{0.2mm} T_1 + \widetilde{w}^{\zc 2} \hspace{0.2mm} T_2 + \widetilde{w}^{\zc 3} \hspace{0.2mm} T_3 \zt \big) \zt e^{-i\zt\xi\zt T_1}
\end{equation}

\vspace{0.2cm}

\noindent{}where the coefficients $\widetilde{w}^{\zc i}$ must be determined from expression (\ref{eq: wilson result 1}), and where the tilde notation, to be employed in other definitions, indicates that we have factored out the conjugation by $e^{i\zt\xi\zt T_1}$. In fact, considering equation (\ref{eq: left side alpha wilson}), which relates $L_\alpha$ to the quantity (\ref{eq: wilson result 1}), we may also define:

\begin{equation}\label{eq: left side alpha tilde expansion}
L_\alpha \equiv e^{i\zt\xi\zt T_1} \zt \widetilde{L}_\alpha \zc\zt e^{-i\zt\xi\zt T_1} = e^{i\zt\xi\zt T_1} \zu \Big( \zt \widetilde{L}^{\zt (1)}_\alpha \hspace{0.4mm} T_1 + \widetilde{L}^{\zt (2)}_\alpha \hspace{0.4mm} T_2 + \widetilde{L}^{\zt (3)}_\alpha \hspace{0.4mm} T_3 \zt \Big) \zt\zu e^{-i\zt\xi\zt T_1}
\end{equation}

\vspace{0.225cm}

\noindent{}where the coefficients $\widetilde{L}^{\zt (i)}_\alpha$ above will thus be given in terms of $\widetilde{w}^{\zc i}$. With this definition, the discussion made on the previous paragraph translates itself into the following equality for (\ref{eq: alpha fluxes definition}):

\begin{equation}\label{eq: alpha fluxes tilde}
\Phi_B^{(\alpha \zu , \zu N \zu)} = \frac{1}{N} \Tr \big( \zu\zu L_\alpha^N \zt \big) = \frac{1}{N} \Tr \hspace{-0.3mm} \Big( \zu\zu \widetilde{L}_\alpha^N \zu\zu \Big)
\end{equation}

\vspace{0.25cm}

So, by making use of (\ref{eq: definition gamma and F}), one obtains for the derivatives present in expression (\ref{eq: wilson result 1}):

\begin{equation}\label{eq: derivatives gamma and F}
\frac{d\gamma}{d\tau} = \frac{K}{F^2} \hspace{1cm} , \hspace{0.9cm} \frac{dF}{d\tau} = \frac{\big(1 - K^2 \zu\zu \big)}{F} \zu\zu \sin\hspace{-0.4mm}\tau \zu \cos\hspace{-0.2mm}\tau
\end{equation}

\vspace{0.25cm}

\noindent{}which can then be substituted into this expression, giving, after some simplifications:

\vspace{-0.25cm}

\begin{equation}\label{eq: coefficients w tilde}
\begin{gathered}
\widetilde{w}^{\zt 1} = i \zt\zt\zu 2\pi \hspace{-0.2mm} \cos\hspace{-0.2mm}\tau \zc \Bigg\{ \zt 1 + \frac{K^2}{F^2} \zt \bigg[ \zt \frac{\sin \hspace{-0.5mm} \big( 2\pi \hspace{-0.2mm} F \big)}{2\pi \hspace{-0.2mm} F} - 1 \zt \bigg] \Bigg\} \hspace{0.75cm} , \hspace{0.7cm} \widetilde{w}^{\zc 2} = -\zt i \zt \frac{K}{F^2} \big[ \zu\zu 1 - \cos\hspace{-0.3mm}\big( 2 \pi \hspace{-0.2mm} F \big) \zu\zu \big] \\[4.8mm]
\widetilde{w}^{\zc 3} = i \zt\zt\zu 2\pi \hspace{-0.1mm} K \hspace{-0.3mm} \sin\hspace{-0.1mm}\tau \zc \Bigg\{ \zt 1 + \frac{1}{F^2} \zt \bigg[ \zt \frac{\sin \hspace{-0.5mm} \big( 2\pi \hspace{-0.2mm} F \big)}{2\pi \hspace{-0.2mm} F} - 1 \zt \bigg] \Bigg\}
\end{gathered}
\end{equation}

\vspace{0.225cm}

With these coefficients, one can then compare (\ref{eq: wilson result 2}) with (\ref{eq: left side alpha wilson}) and (\ref{eq: left side alpha tilde expansion}) to obtain, finally: 

\vspace{-0.05cm}

\begin{align}
\widetilde{L}^{\zt (1)}_\alpha &= \frac{2\pi}{e} \hspace{-0.4mm} \int_{-\pi/2}^{\pi/2} \zc d\tau \hspace{0.3mm} \cos\hspace{-0.2mm}\tau \zc \Bigg\{ \zt 1 + \frac{K^2}{F^2} \zt \bigg[ \zt \frac{\sin \hspace{-0.5mm} \big( 2\pi \hspace{-0.2mm} F \big)}{2\pi \hspace{-0.2mm} F} - 1 \zt \bigg] \Bigg\} \zu\zu \Bigg|_{\hspace{0.2mm}\zeta = \zeta_f} \label{eq: L1 tilde} \\[5.7mm]
\widetilde{L}^{\zt (2)}_\alpha &= - \zt \frac{K}{e} \hspace{-0.3mm} \int_{-\pi/2}^{\pi/2} \zc d\tau \hspace{0.3mm}  \frac{1}{F^2} \big[ \zu\zu 1 - \cos\hspace{-0.3mm}\big( 2 \pi \hspace{-0.2mm} F \big) \zu\zu \big] \zu\zu \bigg|_{\hspace{0.2mm}\zeta = \zeta_f} \label{eq: L2 tilde} \\[5.7mm]
\widetilde{L}^{\zt (3)}_\alpha &= \frac{2\pi \hspace{-0.3mm} K}{e} \hspace{-0.4mm} \int_{-\pi/2}^{\pi/2} \zc d\tau \hspace{0.3mm}  \sin\hspace{-0.1mm}\tau \zc \Bigg\{ \zt 1 + \frac{1}{F^2} \zt \bigg[ \zt \frac{\sin \hspace{-0.5mm} \big( 2\pi \hspace{-0.2mm} F \big)}{2\pi \hspace{-0.2mm} F} - 1 \zt \bigg] \Bigg\} \zu\zu \Bigg|_{\hspace{0.2mm}\zeta = \zeta_f} \hspace{-0.75mm} = \hspace{0.25mm} 0 \label{eq: L3 tilde}
\end{align}

\vspace{0.45cm}

On the expressions above, the $\tau$-intervals goes from $-\pi/2$ to $\pi/2$ because we are considering the loops of type II. Also, we have the vanishing of $\widetilde{L}^{\zt (3)}_\alpha$ because its integrand is of odd parity.

\phantom{Paragraph}

With expressions (\ref{eq: L1 tilde}) to (\ref{eq: L3 tilde}) above, we can finally evaluate the gauge-invariant magnetic fluxes (\ref{eq: alpha fluxes tilde}) for the instanton (and anti-instanton) solution. Until now, all of our calculations were performed for any representation of the algebra $\mathfrak{su}(2)$, and the results (\ref{eq: L1 tilde}) to (\ref{eq: L3 tilde}) obtained above are, in particular, independent of any representation. Nevertheless, a representation must eventually be choosed, to evaluate the traces on expression (\ref{eq: alpha fluxes tilde}) and to plot the results.

Here, for simplicity, we shall consider the bidimensional representation of $\mathfrak{su}(2)$, where its generators are given by $1/2$ of the Pauli matrices: $T_i = \sigma_i / 2$\zt . Therefore, the $\widetilde{L}_\alpha$ matrix obtained with the coefficients above will be $2\times 2$. For such matrices, it can be shown that its eigenvalues can be expressed entirely in terms of $\Tr \hspace{-0.3mm} \big( \zu\zu \widetilde{L}_\alpha \zu\zu \big)$ and $\Tr \hspace{-0.3mm} \big( \zu\zu \widetilde{L}_\alpha^{\zt 2} \zu\zu \big)$. However, $\Tr T_i = 0$, for $i=1,2,3$, and thus we see that the only important quantity for these eingenvalues is $\Tr \hspace{-0.3mm} \big( \zu\zu \widetilde{L}_\alpha^{\zt 2} \zu\zu \big)$. Here, we have not defined the fluxes $\Phi_B^{(\alpha \zu , \zu N \zu)}$ in terms of eigenvalues, but instead in terms of traces. Nevertheless, from the discussion above, we shall consider here only the quantity $\Phi_B^{(\alpha \zu , \zu 2 \zu)}$, corresponding to the trace $\Tr \hspace{-0.3mm} \big( \zu\zu \widetilde{L}_\alpha^{\zt 2} \zu\zu \big)$. We will not plot the fluxes $\Phi_B^{(\alpha \zu , \zu N \zu)}$ for the traces of higher orders in $N$. 

\phantom{Paragraph}

With that being said, and to simplify the notation for the following results, let us denote:

\begin{equation}\label{eq: alpha flux}
\Phi_B \equiv \Phi_B^{(\alpha \zu , \zu 2 \zu)}
\end{equation}

\vspace{0.25cm}

From now on, this is what we call the gauge-invariant magnetic flux of the instanton (and anti-instanton) solution, for a spherical surface $\partial\Omega$ of radius $\zeta_f$\zc . From self-duality, this corresponds also to the \textit{electric} flux through $\partial\Omega$. In addition, since we are dealing with a Gauss law, this $\Phi_B$ can also be identified with the magnetic (or electric) charge enclosed in $\Omega$.

Considering (\ref{eq: alpha flux}) above, and expanding the quantity $\Phi_B^{(\alpha \zu , \zu 2 \zu)}$ using (\ref{eq: alpha fluxes tilde}), one obtains:

\begin{equation}\label{eq: alpha flux expanded}
\Phi_B = \frac{1}{2} \Tr \hspace{-0.3mm} \Big( \zt \widetilde{L}^{\zt (i)}_\alpha \hspace{0.4mm} T_i \zc\zt \widetilde{L}^{\zt (j)}_\alpha \hspace{0.4mm} T_j \Big) = \frac{\scalebox{0.9}{$\mathcal{N}$}}{2} \zt \bigg[ \zu\zu \Big( \widetilde{L}^{\zt (1)}_\alpha \Big)^{\hspace{-0.5mm}2} \hspace{-0.4mm} + \Big( \widetilde{L}^{\zt (2)}_\alpha \Big)^{\hspace{-0.5mm}2} \hspace{-0.4mm} + \Big( \widetilde{L}^{\zt (3)}_\alpha \Big)^{\hspace{-0.5mm}2} \zc\zu \bigg]
\end{equation}

\vspace{0.25cm}

\noindent{}where $\scalebox{0.9}{$\mathcal{N}$} = \Tr \hspace{-0.4mm} \big( \zu\zu T_1 \zt T_1 \big) = \Tr \hspace{-0.4mm} \big( \zu\zu T_2 \zc T_2 \big) = \Tr \hspace{-0.4mm} \big( \zu\zu T_3 \zc T_3 \big)$. Actually, this is the quantity that effectively depends on the choice of representation. The factor $1/2$ above comes from the definition of $\Phi_B^{(\alpha \zu , \zu 2 \zu)}$, and the coefficients $\widetilde{L}^{\zt (i)}_\alpha$ are themselves independent of representation, as already mentioned. For the case of the bidimensional representation of $\mathfrak{su}(2)$, where $T_i = \sigma_i / 2$, we have $\scalebox{0.9}{$\mathcal{N}$} = 1/2$. Thus, using from (\ref{eq: L3 tilde}) that $\widetilde{L}^{\zt (3)}_\alpha = 0$, we obtain for the instanton (and anti-instanton) solution:

\begin{equation}\label{eq: alpha flux final}
\Phi_B = \frac{1}{4} \zt \bigg[ \zu\zu \Big( \widetilde{L}^{\zt (1)}_\alpha \Big)^{\hspace{-0.5mm}2} \hspace{-0.4mm} + \Big( \widetilde{L}^{\zt (2)}_\alpha \Big)^{\hspace{-0.5mm}2} \zc\zu \bigg]
\end{equation}

\vspace{0.225cm}

In fact, one can easily check that this quantity does not dependend on the constant $\kappa$, so that we obtain the same expression above for both the instanton and anti-instanton.

\phantom{Paragraph}

The gauge-invariant magnetic flux (\ref{eq: alpha flux final}) depends on the radius $\zeta_f$ of $\partial\Omega$, which we shall denote by $r$, and on the Euclidean time $x^4$. Considering first $x^4 = 0$, and making $e=1$ for the gauge coupling constant, we obtain for the instanton and anti-instanton solutions:

\newpage

\vspace*{-0.5cm}

\begin{figure}[H]

\centering

\includegraphics[scale=0.6]{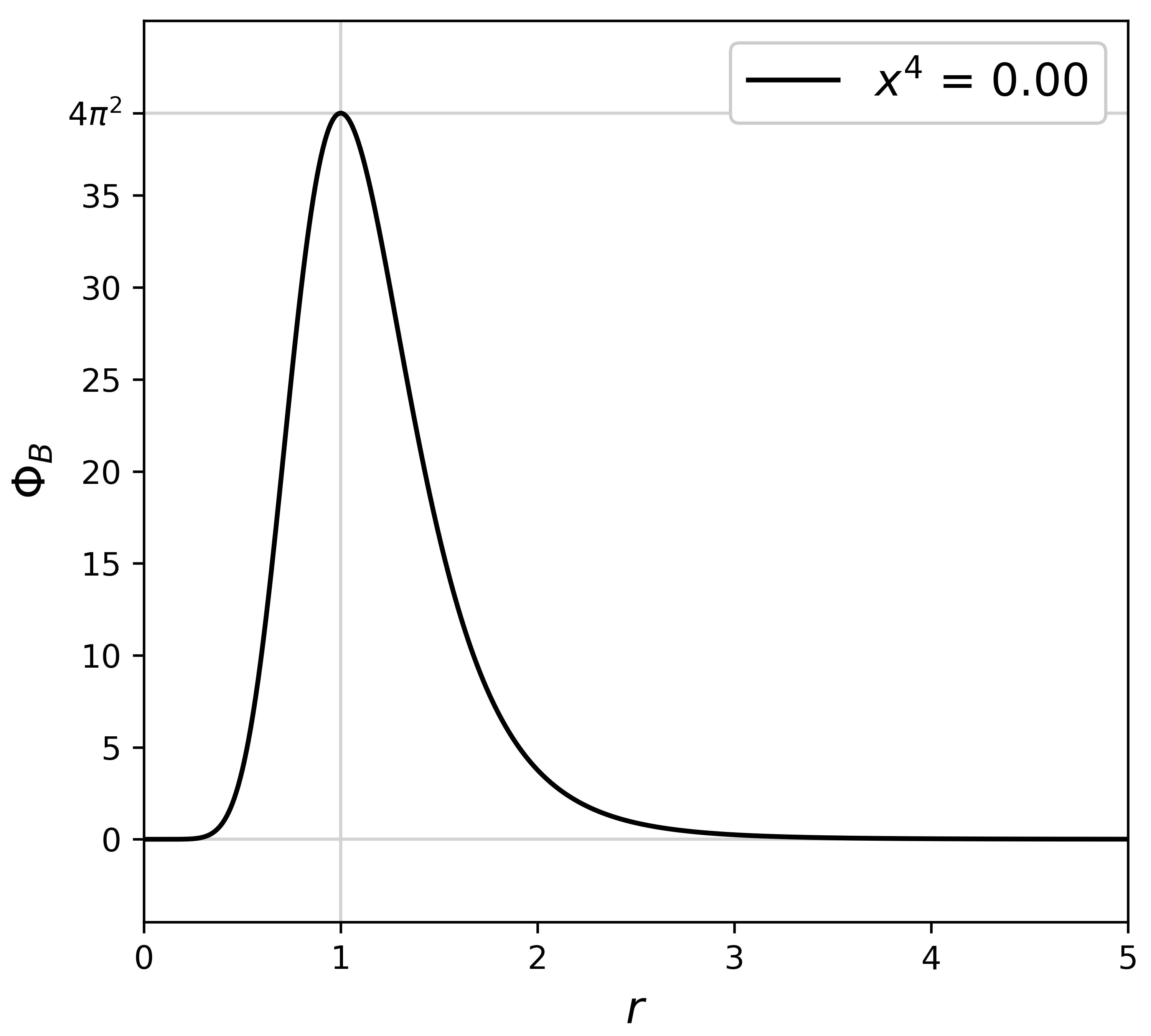}

\caption{Gauge-invariant magnetic flux (enclosed magnetic charge) for the instanton at $x^4 = 0$.}
\label{fig:alpha flux t eq 0}

\end{figure}

From this graph we observe that the flux goes to zero as $r$ increases, indicating that $\Phi_B = 0$ as $r \rightarrow \infty$. This result can be shown analytically, for any Euclidean time $x^4$, and it means that the instanton and anti-instanton solutions, when considering all space, have no net magnetic and electric charges coming from the non-Abelian Gauss law (\ref{eq: alpha equation}). This result, regarding the total charge of these solutions, have already been obtained in \cite{ym1}. Also, as mentioned before, when considering all space we have that these net charges are conserved in time

Despite that, for $r \sim 1$, we observe in the graph above a non-zero flux for the solutions, which corresponds to an enclosed magnetic (and electric) charge for the spherical surfaces of radii $r$ considered above. In particular, for $r=1$ and $x^4 = 0$ we obtain a maximum flux for the solutions, which using (\ref{eq: alpha flux final}) can be shown analytically to give $\Phi_B \zu\zu (\zu\zu r=1 , x^4 = 0) = (\frac{4\pi}{e})^2 \hspace{-0.3mm}/4$, if we return with the gauge coupling constant $e \neq 1$. Indeed, for this radius and Euclidean time, the quantity $K$ defined in (\ref{eq: expression K}) is equal to zero, an therefore we obtain $\widetilde{L}^{\zt (2)}_\alpha = 0$ for the coefficient in (\ref{eq: L2 tilde}) and $\widetilde{L}^{\zt (1)}_\alpha = 4\pi/e$ for the coefficient in (\ref{eq: L1 tilde}). Thus, the quantity $\widetilde{L}_\alpha$ defined in (\ref{eq: left side alpha tilde expansion}) is pointing in the $T_1$-direction only, and is given by $\widetilde{L}_\alpha = \frac{4\pi}{e} \zt T_1$\zt .

As we will mention soon, this particular result for $r=1$ and $x^4 = 0$ satisfies the reparameteri\-zation-invariance condition, and thus this magnetic (or electric) flux can be considered to be observable, since it is also gauge-invariant. Therefore, although the instanton and anti-instanton solutions have no net charge when considering all space, this observable flux for $r=1$ and $x^4 = 0$ indicates that these solutions have an internal charge configuration, coming from the non-Abelian Gauss law (\ref{eq: alpha equation}). Also, note that this radius $r=1$ is precisely at the size $\lambda$ of the solutions, since we have performed the scale trans\-formation $x^\mu \rightarrow \lambda \zt x^\mu$\zc .

\phantom{Paragraph}

We can also plot the flux (\ref{eq: alpha flux final}) for other Euclidean times. First, by consulting the quantity $K$ in (\ref{eq: expression K}), one sees that the we have $(x^4)^2$. Therefore, the coefficients in (\ref{eq: L1 tilde}) and (\ref{eq: L2 tilde}) are the same for $x^4$ and $-x^4$, indicating that the flux (\ref{eq: alpha flux final}) is symmetric in Euclidean time:

\vspace{0.2cm}

\begin{figure}[H]

\centering

\includegraphics[width=0.32\linewidth]{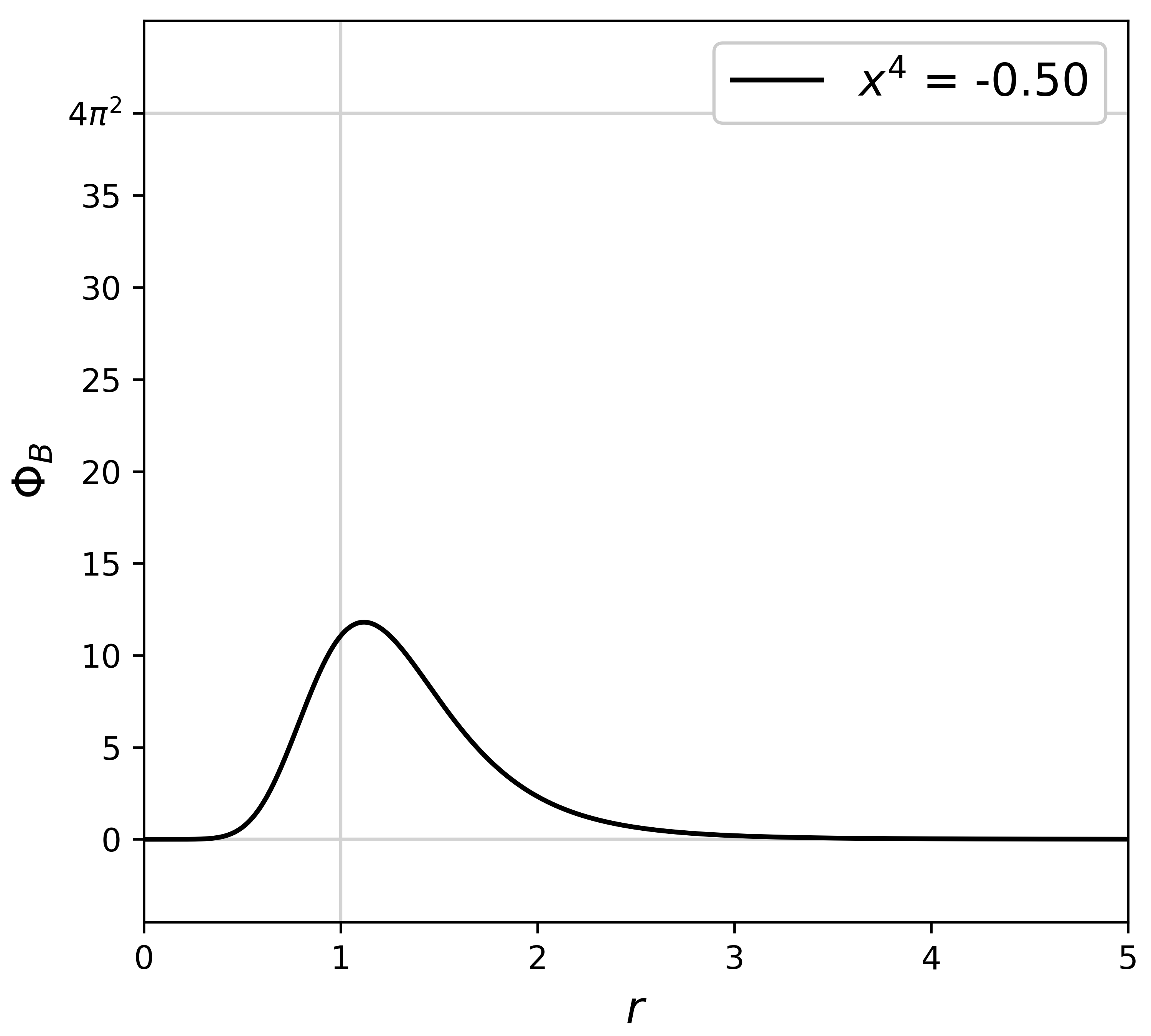}
\includegraphics[width=0.32\linewidth]{alpha_flux_t_eq_0.png}
\includegraphics[width=0.32\linewidth]{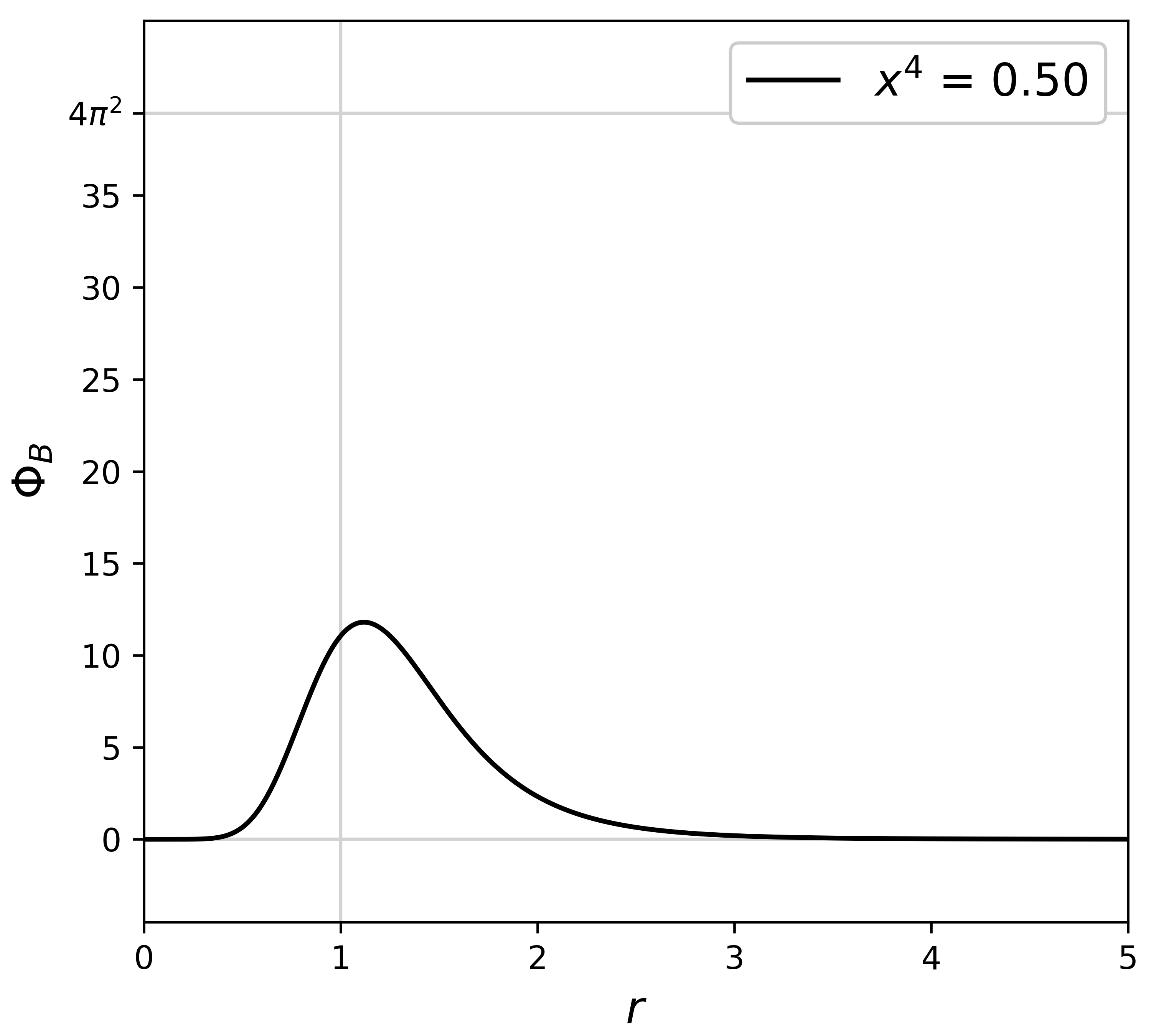}

\caption{Euclidean time symmetry of the gauge-invariant magnetic flux for the instanton.}
\label{fig:time symmetry}

\end{figure}

Now, for the Euclidean time evolution of the flux (\ref{eq: alpha flux final}) for the instanton and anti-instan\-ton solutions, we can plot the following graphs, going from $x^4 = 0$ to $x^4 = 1$:

\vspace{0.2cm}

\begin{figure}[H]

\centering

\includegraphics[width=0.32\linewidth]{alpha_flux_t_eq_0.png}
\includegraphics[width=0.32\linewidth]{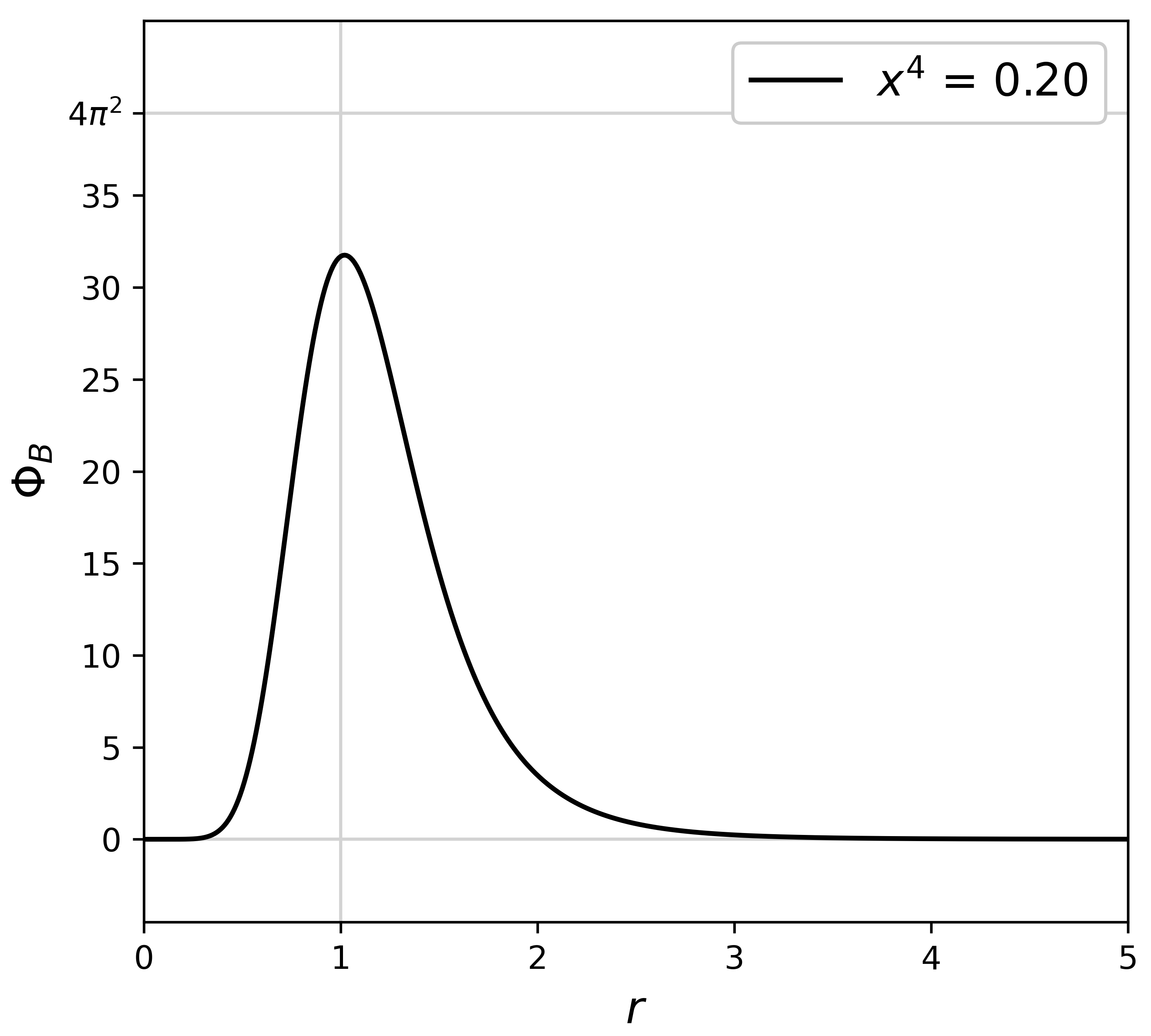}
\includegraphics[width=0.32\linewidth]{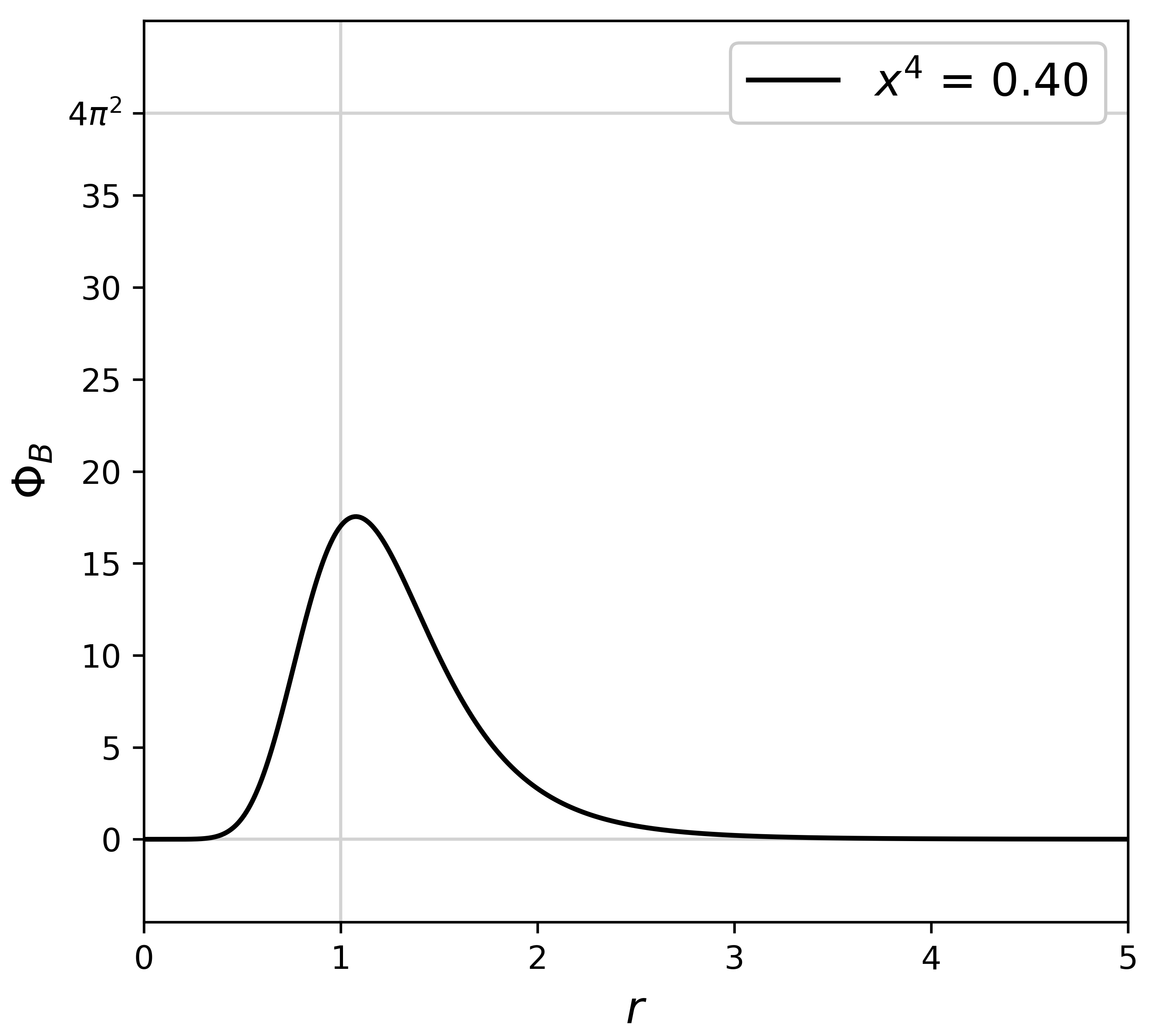}

\vspace{0.125cm}

\includegraphics[width=0.32\linewidth]{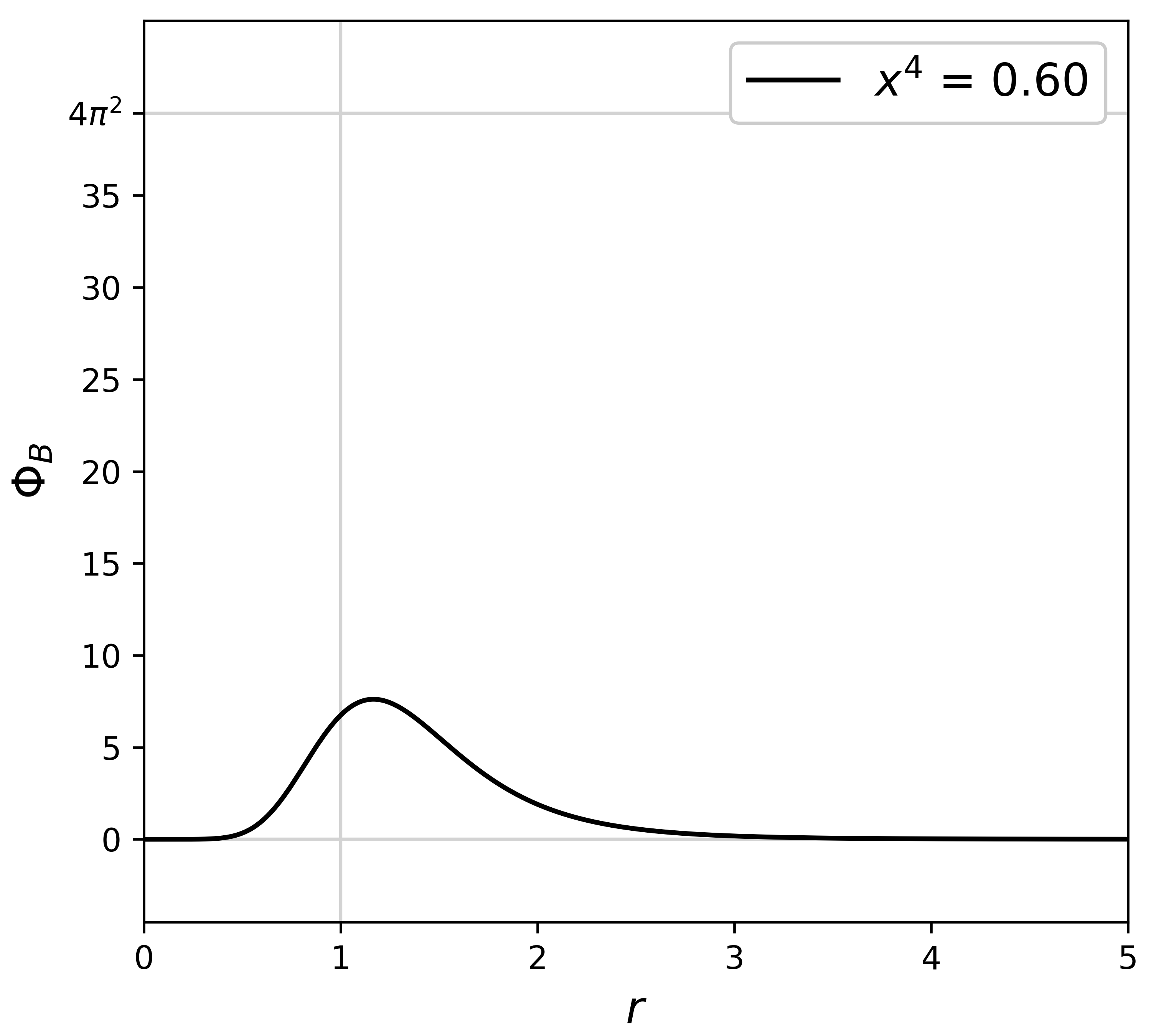}
\includegraphics[width=0.32\linewidth]{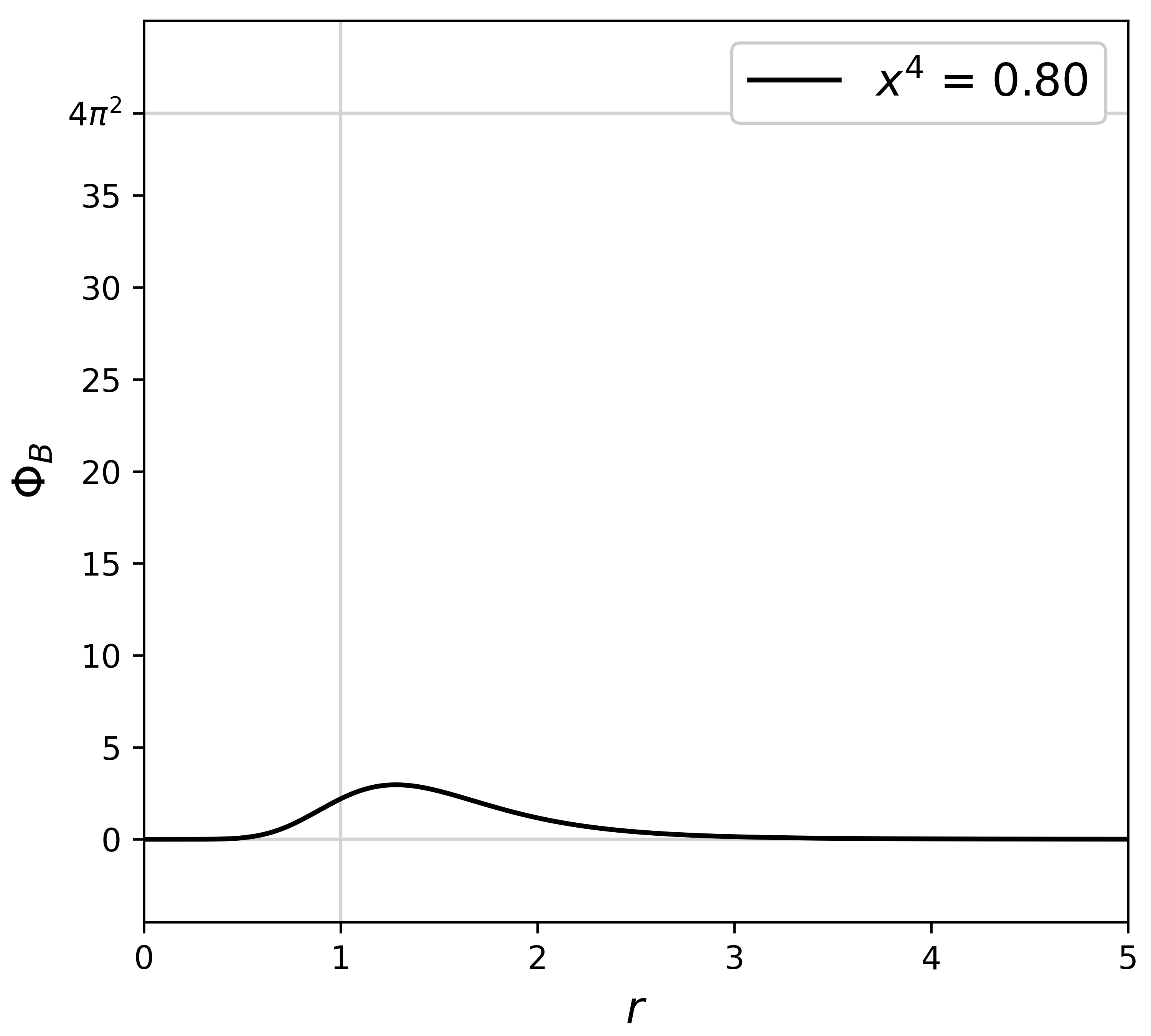}
\includegraphics[width=0.32\linewidth]{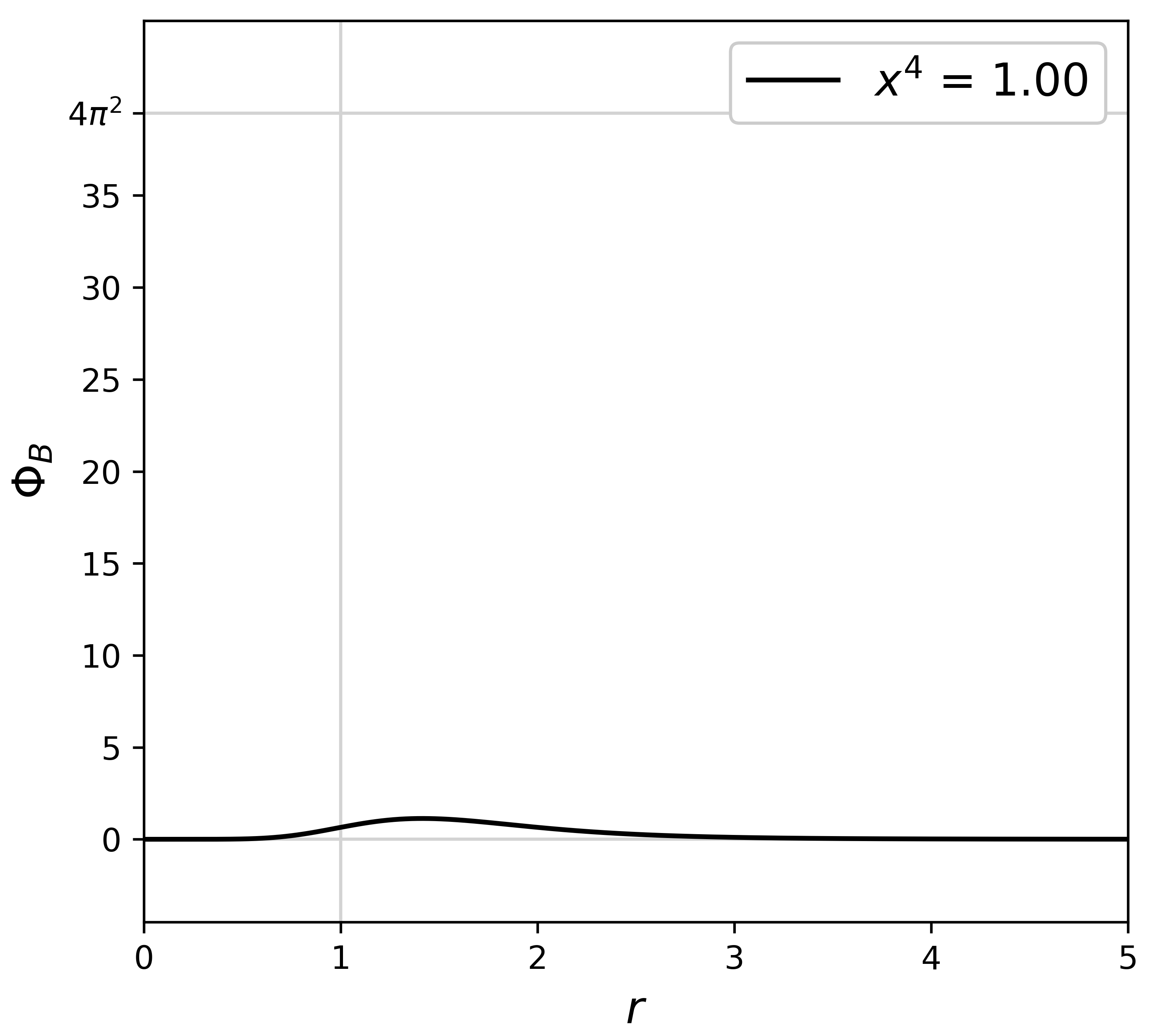}

\caption{Euclidean time evolution of the gauge-invariant magnetic flux for the instanton.}
\label{fig:time evolution}

\end{figure}

Here we see that the magnetic (or electric) flux goes to zero as the Euclidean time increases. In fact, from the symmetry indicated in Figure \ref{fig:time symmetry}, we see that there will be no enclosed magnetic (or electric) charge, for any radius $r$, for the instanton and anti-instanton solutions at $x^4 \rightarrow - \infty$ and $x^4 \rightarrow \infty$.\footnote{This can be proven analytically through simple calculations with the coefficients (\ref{eq: L1 tilde}) and (\ref{eq: L2 tilde}).} As the Euclidean time approaches $x^4 = 0$ the solutions develop an internal char\-ge configuration, which acquires its maximum at $x^4=0$, $r=1$. This evolution is in accordance with the temporal behavior of the solutions, since they are in fact localized at $x^4 = 0$.

\phantom{Paragraph}

For the reparameterization-invariance of the results above, one needs to evaluate the expression for $\delta L_\alpha$ in (\ref{eq: variation left alpha final}). Note in this equation that the first term inside the commutator is in fact the quantity $W_c^{-1}\hspace{-0.5mm} \cdot \hspace{-0.1mm} dW_c/d\tau$ that we previously calculated, which is obtained by employing (\ref{eq: variation wilson final}) in the $\tau$-direction. Also, note that the second term inside the commutator is almost this same quantity, but without integrating in $\sigma^\prime$ since we do not know the arbitrary parameter $b \zt (\sigma^\prime,\tau)$.

Therefore, the evaluation of (\ref{eq: variation left alpha final}) for the instanton (and anti-instanton) solution is very similar to the calculations performed in the last pages. The only thing different is that now we also need the field strenght $F_{\mu\nu}$\zc , in order to evaluate the $\sigma^\prime$-term mentioned above. For the instanton and anti-instanton solutions given in (\ref{eq: instanton solution}), the spatial components of $F_{\mu\nu}$ are:

\begin{equation}\label{eq: field strenght instanton}
F_{ij} = \frac{4}{e} \zu \frac{1}{(x^2 + 1)^{\zu\zu 2}} \zc\zu\zu \varepsilon_{ijk} \zt T_k
\end{equation}

\vspace{0.225cm}

\noindent{}where again $x^2 \equiv x_\mu \zt x^\mu$ with an Euclidean metric. Thus, by employing the field strenght $F_{ij}$ above and the Wilson lines determined in Subsection \ref{subsec:wilson lines}, evaluating the commutator of (\ref{eq: variation left alpha final}) and simplifying the obtained expression, one arrives at the following equation for $\delta L_\alpha$\zt :

\begin{equation}\label{eq: variation alpha simplified}
\delta L_\alpha = - \zu \frac{2\pi}{e} \zu\zu K \big( \zu\zu 1 - K^{ \zu\zu 2} \zu\zu \big) \zc e^{i\zt\xi\zt T_1} \zu \big( \zu\zu C^{\zu 1} \zu\zu T_1 + C^{\zu\zu 2} \zu\zu T_2 + C^{\zu\zu3} \zu\zu T_3 \zt \big) \zt\zu e^{-i\zt\xi\zt T_1}
\end{equation}

\vspace{0.225cm}

\noindent{}where the coefficients $C^{\zt i}$ are given by integrals in $\tau$, containing integrals in $\sigma^\prime$ which cannot be solved because of the arbitrary parameter $b \zt (\sigma^\prime,\tau)$. We again have a conjugation by $e^{i\zt\xi\zt T_1}$, and the factor $-2\pi/e$ is actually of no importance, since it can be absorved into the coefficients $C^{\zt i}$.

What is particularly important in the expression above is the term $K \big( \zu 1 - K^{ \zu\zu 2} \zu \big)$. By consulting (\ref{eq: expression K}), on sees that this quantity is equal to zero in only four cases: (i) $r=0$ with $x^4$ arbitrary, (ii) $r=1$ with $x^4 = 0$, (iii) $r \rightarrow \infty$ with $x^4$ arbitrary and (iv) $x^4 \rightarrow \pm \infty$ with $r$ arbitrary. For all these cases $\delta L_\alpha = 0$, which means that the corresponding fluxes are reparameterization-invariant, and thus observable. Nevertheless, as one can see, the truly interesting result is only case (ii), which has already been discussed below Figure \ref{fig:alpha flux t eq 0}. Case (i) corresponds to a surface of null radius, which has no flux and no enclosed charge, case (iii) corresponds to a surface of infinite radius enclosing all space\footnote{The reparameterization-invariance for $r \rightarrow \infty$ is actually a very important result because it means that the total charge (which is conserved), obtained from the Yang-Mills integral equations, is in fact observable. Never\-theless, for the instanton and anti-instanton solutions, this total charge is zero, so the result is not too \nlb interesting.}, for which the flux is always zero and case (iv) corresponds to the Euclidean time limits $x^4 \rightarrow \pm \infty$, for which there is no flux, for any surface of radius $r$.

\phantom{Paragraph}

The graphs presented above correspond to the flux $\Phi_B \equiv \Phi_B^{(\alpha \zu , \zu 2 \zu)}$, and not exactly to $L_\alpha$\zc{\nlb}. For this flux, we have derived the reparameterization-invariance condition as being given by (\ref{eq: variation alpha fluxes}), and not by $\delta L_\alpha$\zc . Therefore, it could occur that the trace in $\delta \zu\zu \Phi_B^{(\alpha \zu , \zu 2 \zu)} = \Tr \zt ( L_\alpha \zc \delta L_\alpha)$ was equal to zero, even though the quantity $\delta L_\alpha$ did not vanish. For such cases, the condition $\delta \zu\zu \Phi_B = 0$ would be satisfied, and the corresponding fluxes would then also be considered observable. 

Nevertheless, considering the expressions for $L_\alpha$ and $\delta L_\alpha$\zc , and most importantly the directions that these quantities take in the algebra, this does not happen. The only cases for which $\delta \zu\zu \Phi_B = 0$ are indeed the ones indicated above, for which $\delta L_\alpha = 0$. For the remaining radii and Euclidean times, the flux $\Phi_B$ is not reparameterization-invariant, and the corresponding results are defined on the generalized loop space $\mathcal{L}^{(2)}$ only, as explained in Subsection \ref{subsec:fluxes and charge density}.

\phantom{Paragraph}

In the next subsection, we will present the gauge-invariant magnetic charge density, defined in equation (\ref{eq: charge density definition}), for the instanton and anti-instanton solutions. We shall end our pre\-sentation with the discussion of the reparameterization-invariance condition for this quantity.

\subsection{The magnetic charge density}
\label{subsec:charge density}

In Subsection \ref{subsec:fluxes and charge density}, we defined our magnetic charge density $\rho^{\zu\zu (\hspace{-0.1mm} N \zu)}$ in terms of the gauge-invariant enclosed charge $Q^{\zu (\hspace{-0.1mm} N \zu)}_{enc} \equiv \Phi_B^{(\alpha \zu , \zu N \zu)}$. In the last subsection, we have focused on the flux $\Phi_B^{(\alpha \zu , \zu 2 \zu)}$, since for $N=1$ we obtain $\Phi_B^{(\alpha \zu , \zu 1 \zu)} = 0$. Also, by considering the bidimensional representation of $\mathfrak{su}(2)$, and arguing that the eigenvalues of a $2 \times 2$ matrix $A$ can be determined solely by the traces $\Tr \hspace{-0,3mm} A$ and $\Tr \hspace{-0.3mm} \big(A^2 \zu\zu \big)$, we have decided to consider only the flux $\Phi_B^{(\alpha \zu , \zu 2 \zu)}$, for $N=2$.

Therefore, we are dealing only with the enclosed magnetic charge $Q^{\zu (\hspace{-0.1mm} 2 \zu)}_{enc} \equiv \Phi_B^{(\alpha \zu , \zu 2 \zu)}$, and thus, from (\ref{eq: charge density definition}) and (\ref{eq: alpha flux}), we shall define our gauge-invariant magnetic charge density as the quantity:

\begin{equation}\label{eq: alpha charge density}
\rho_{\hspace{-0.2mm}\scalebox{0.55}{$B$}} \equiv \frac{1}{4\pi r^2}\frac{d \zt \Phi_{\hspace{-0.2mm}B}}{dr}
\end{equation}

\vspace{0.225cm}

\noindent{}where we have substituted $\rho^{\zu\zu (\hspace{-0.1mm} 2 \zu)}$ by $\rho_{\hspace{-0.2mm}\scalebox{0.55}{$B$}}$\zt , for convenience. Note that this quantity will correspond also to the \textit{electric} charge density for the instanton and anti-instanton solutions, because of self-duality. Despite this, we shall sometimes refer to it only as magnetic, for simplicity.

\phantom{Paragraph}

Having plotted all the graphs of the previous subsection for the flux $\Phi_{\hspace{-0.2mm}B}$\zc ,  one sees that the gauge-invariant charge density $\rho_{\hspace{-0.2mm}\scalebox{0.55}{$B$}}$ can be obtained very easily by employing (\ref{eq: alpha charge density}). Con\-si\-dering first $x^4 = 0$, and making again $e=1$, we obtain for the instanton and anti-instanton solutions:

\newpage

\vspace*{-0.6cm}

\begin{figure}[H]

\centering

\includegraphics[scale=0.6]{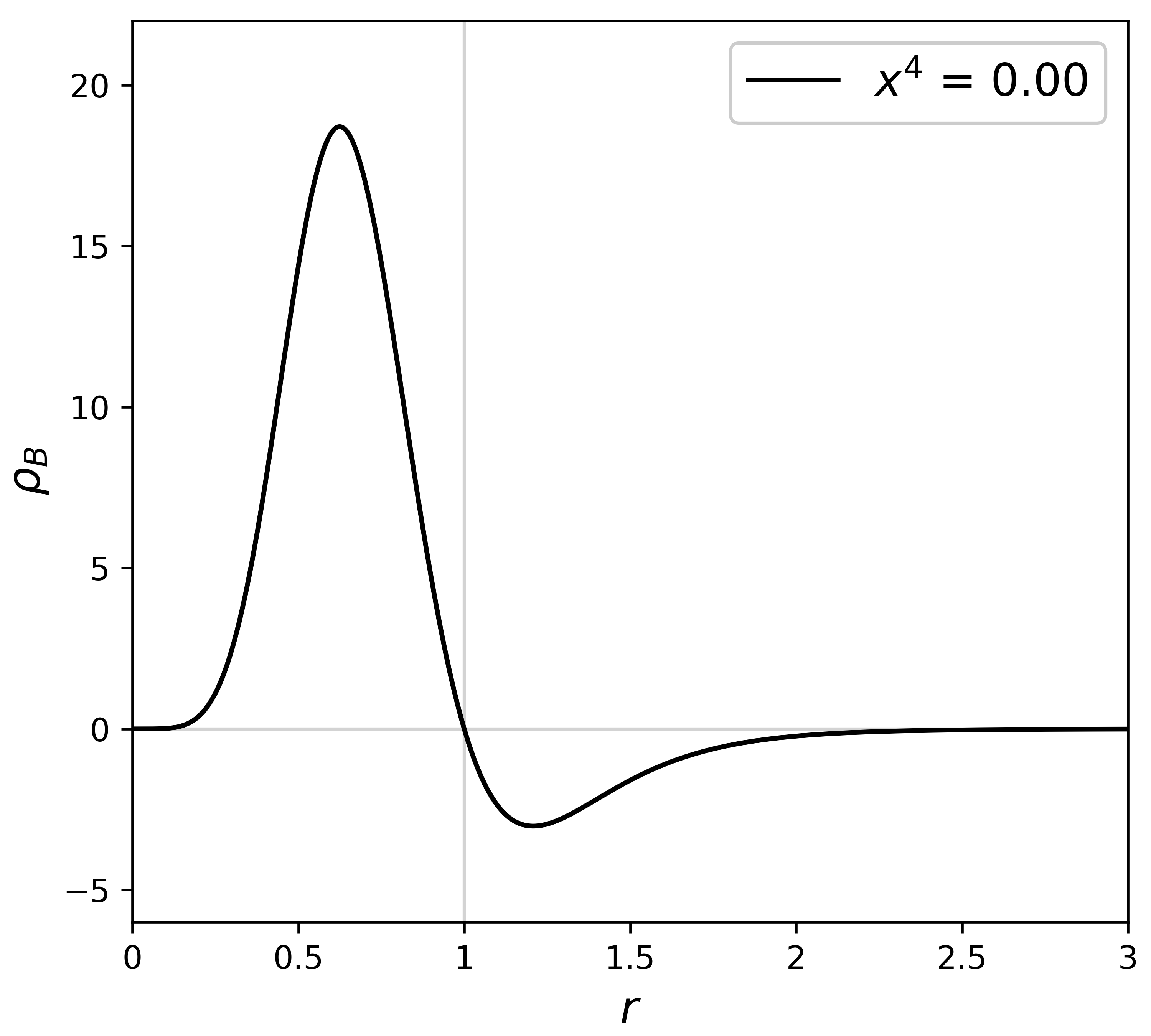}

\caption{Gauge-invariant magnetic charge density for the instanton at $x^4 = 0$.}
\label{fig:alpha density t eq 0}

\end{figure}

This graph must be compared with the one at Figure \ref{fig:alpha flux t eq 0}. On that graph, which is also at the Euclidean time $x^4=0$, one sees that the instanton enclosed magnetic charge starts to rise, from $0$ to a maximum value of $4\pi^2$, as the radii of the spherical surfaces considered increases from $r=0$ to $r=1$. Since this maximum value of $\Phi_{\hspace{-0.2mm}B}$ is observable, this means that the instanton must have some positive charge density on this region. As we can see, a positive charge density profile, for $r$ between 0 and 1, is indeed being depicted on the graph above, despite the fact that the curve obtained is not reparameterization-invariant, as we are going to see. Considering again the graph at Figure \ref{fig:alpha flux t eq 0}, one sees that, after $r=1$, the instanton enclosed magnetic charge starts to decrease, and it eventually vanishes as $r \rightarrow \infty$. This null total charge, corresponding to $\Phi_{\hspace{-0.2mm}B} = 0$ at $r \rightarrow \infty$, is also an observable result, which thus mean that the instanton must also have some negative charge density, on the region after $r=1$. As we can see, such a negative charge density profile is also being depicted on the graph above, despite the fact that the curve obtained is not reparameterization-invariant either. Therefore, although this condition is not satisfied, meaning that the gauge-invariant charge density $\rho_{\hspace{-0.2mm}\scalebox{0.55}{$B$}}$ is not observable, from the discussion made above this quantity may still be of interest, as it indicates the overall internal charge configuration of the instanton and anti-instanton solutions. The reparameterization-invariance condition for the magnetic (or electric) charge density (\ref{eq: alpha charge density}) will be properly presented in a moment.

\phantom{Paragraph}

In Figure \ref{fig:time symmetry}, we observed the Euclidean time symmetry of the gauge-invariant flux $\Phi_{\hspace{-0.2mm}B}$ for the instanton (and anti-instanton) solution. This temporal behavior clearly carries over to the charge density $\rho_{\hspace{-0.2mm}\scalebox{0.55}{$B$}}$\zc , and there is no necessity to plot it here. What is more interesting to show is the Euclidean time evolution of the charge density. Using the same Euclidean time interval of Figure \ref{fig:time evolution}, we obtain that the evolution of the gauge-invariant magnetic (or electric) charge density $\rho_{\hspace{-0.2mm}\scalebox{0.55}{$B$}}$ for the instanton and anti-instanton solutions is given by:

\newpage

\vspace*{-0.525cm}

\begin{figure}[H]

\centering

\includegraphics[width=0.32\linewidth]{alpha_density_t_eq_0.png}
\includegraphics[width=0.32\linewidth]{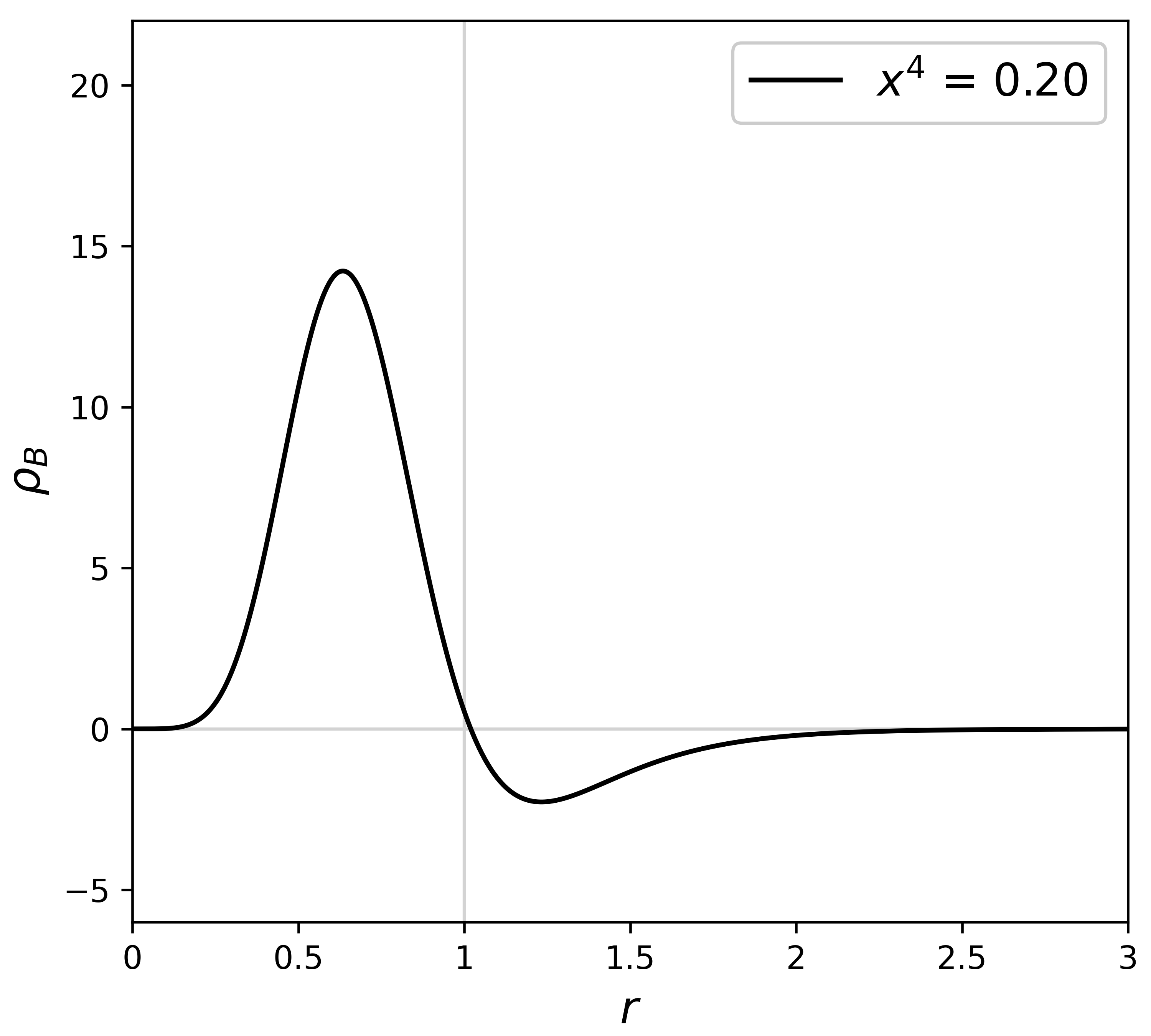}
\includegraphics[width=0.32\linewidth]{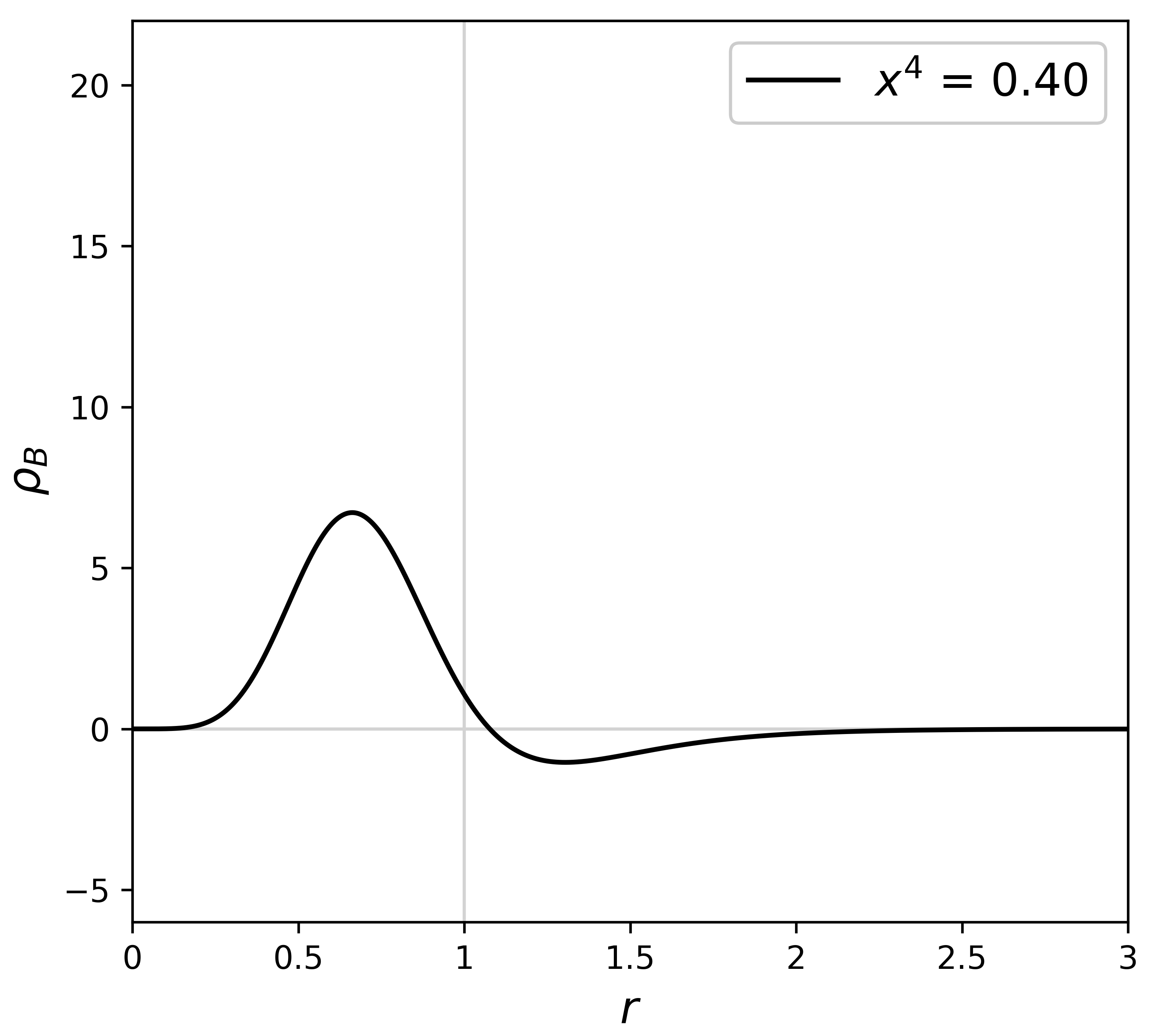}

\vspace{0.125cm}

\includegraphics[width=0.32\linewidth]{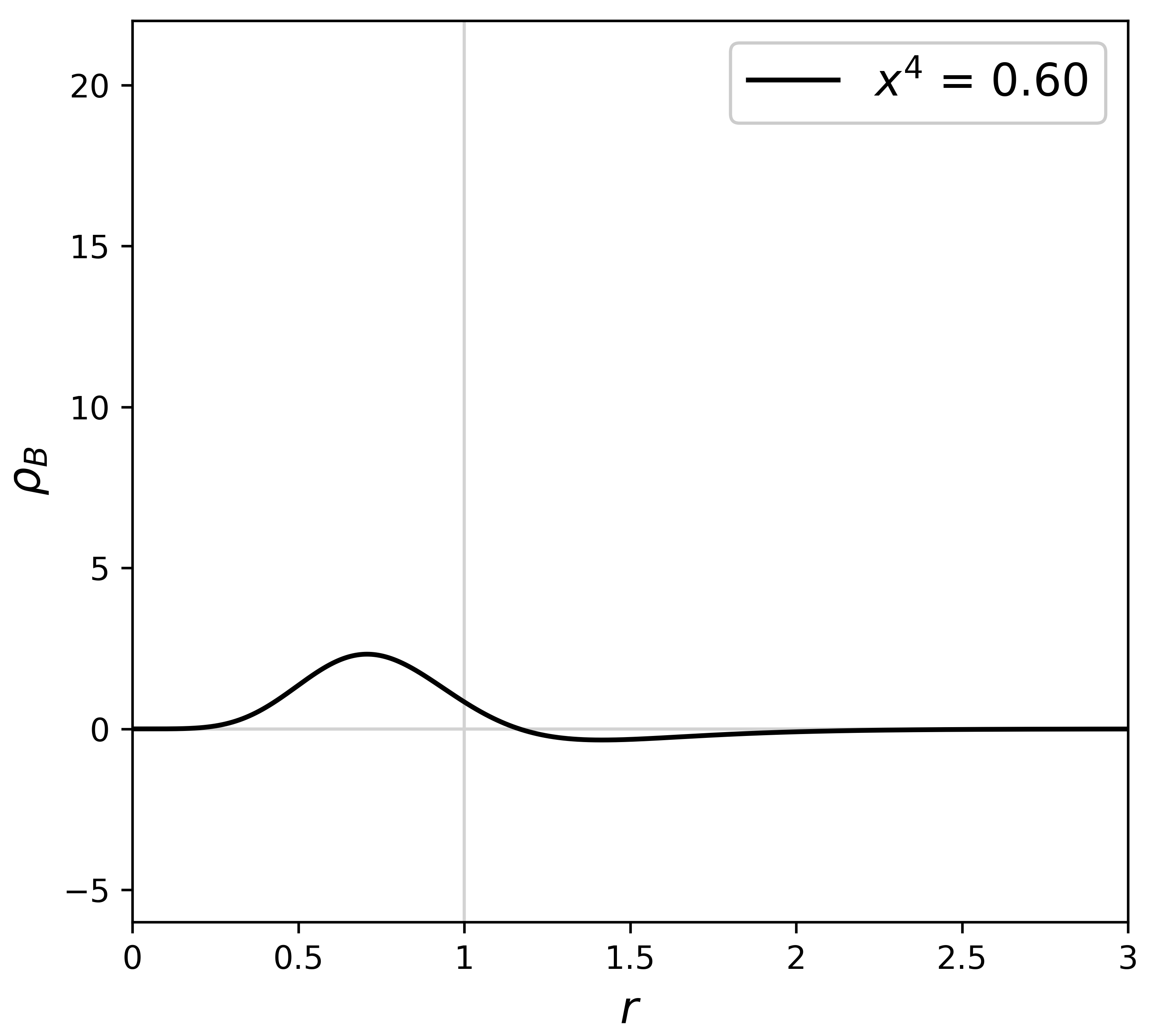}
\includegraphics[width=0.32\linewidth]{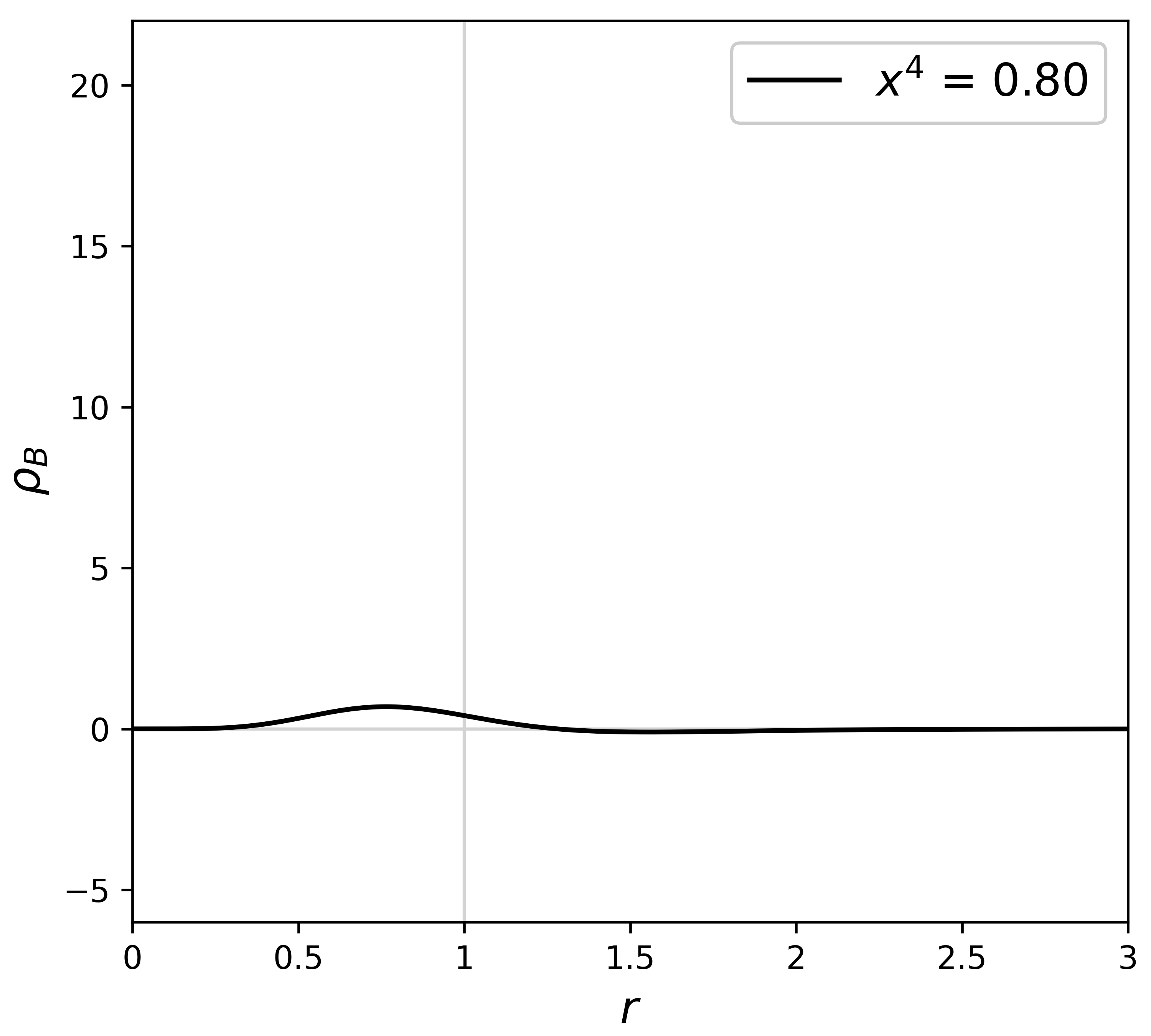}
\includegraphics[width=0.32\linewidth]{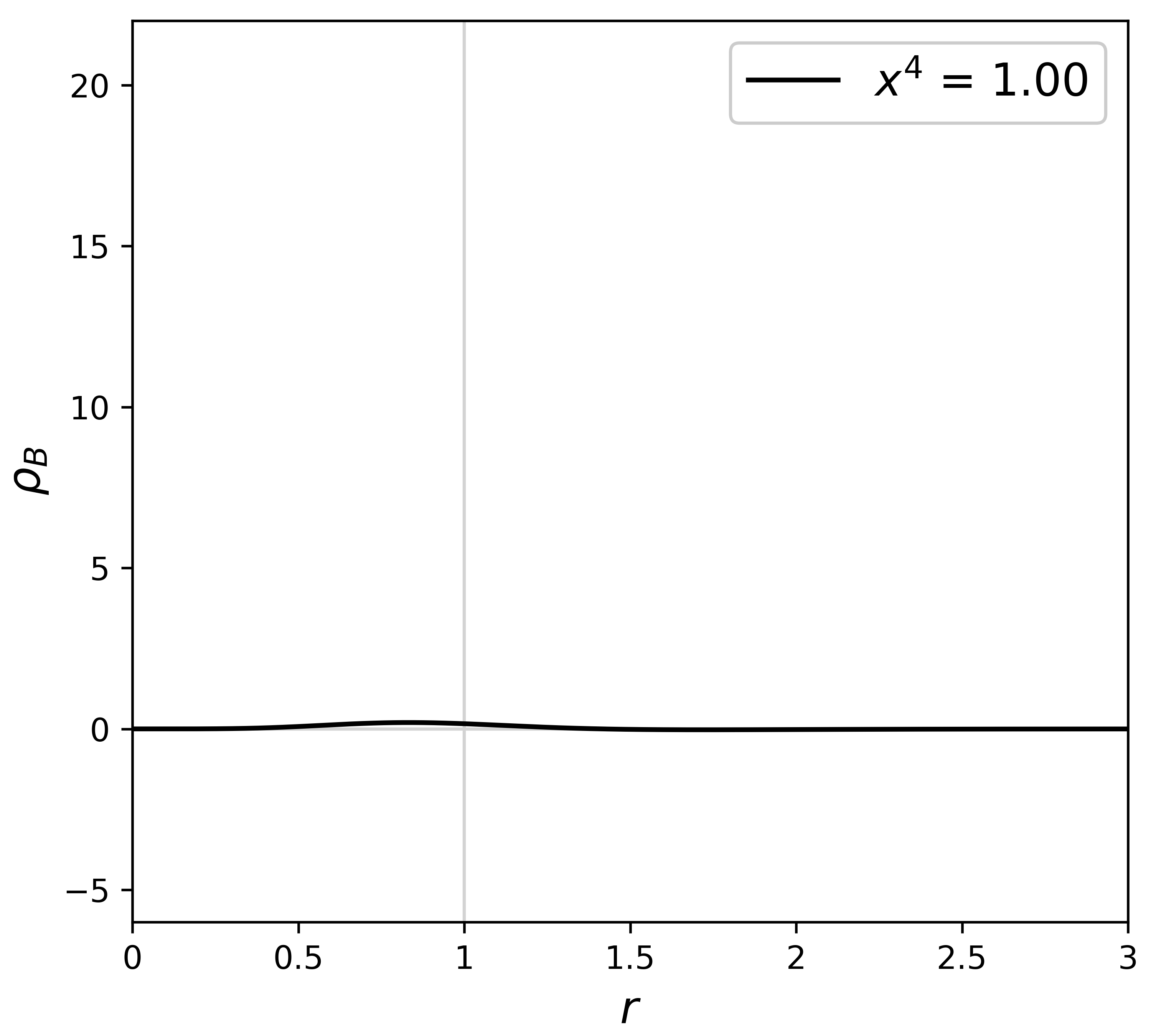}

\caption{Euclidean time evolution of the gauge-invariant magnetic charge density for the ins\-tanton.}
\label{fig:density time evolution}

\end{figure}

As we see, the charge density will eventually vanish for $x^4 \rightarrow \infty$ (and also for $x^4 \rightarrow -\infty$, from the Euclidean time symmetry of the solutions), which is in accordance with the evolution of the flux $\Phi_{\hspace{-0.2mm}B}$ plotted in Figure \ref{fig:time evolution}. Also, note that the points of $\rho_{\hspace{-0.2mm}\scalebox{0.55}{$B$}} = 0$, occurring at the \nlb tran\-sitions between the positive and negative regions, correspond to the points of maxima of the graphs of Figure \ref{fig:time evolution}. This is to be expected because, in accordance with equation (\ref{eq: alpha charge density}), the graphs for $\rho_{\hspace{-0.2mm}\scalebox{0.55}{$B$}}$ above are obtained by differentiating the graphs of Figure \ref{fig:time evolution} (with a factor of $1/4\pi r^2$).

\phantom{Paragraph}

To end our discussion, let us now present what is obtained for the reparameterization-invariance condition of the quantity $\rho_{\hspace{-0.2mm}\scalebox{0.55}{$B$}}$ for the instanton and anti-instanton solutions. As \nlb ex\-plained before, the gauge-invariant magnetic (or electric) charge density that we are considering in this paper is the quantity $\rho_{\hspace{-0.2mm}\scalebox{0.55}{$B$}} \equiv \rho^{\zu\zu (\hspace{-0.1mm} 2 \zu)}$. Therefore, from equation (\ref{eq: charge density variation}) for $\delta \zu \rho^{\zu\zu (\hspace{-0.1mm} N \zu)}$, we see that the expression we must verify for the reparameterization-invariance condition of $\rho_{\hspace{-0.2mm}\scalebox{0.55}{$B$}}$ is given by:

\begin{equation}\label{eq: charge density variation n eq 2}
\delta \zu \rho_{\hspace{-0.1mm}\scalebox{0.55}{$B$}} = \frac{1}{4\pi r^2} \Tr \hspace{-0.3mm} \Bigg( \frac{d \zu L_\alpha}{dr} \zt \delta L_\alpha + L_\alpha \zu \frac{d \zt \delta \hspace{-0.2mm} L_\alpha}{dr} \hspace{-0.4mm}  \Bigg)
\end{equation}

\vspace{0.225cm}

Apart from the special case of $r=0$, where a more careful analysis must be employed, the factor of $1/4\pi r^2$ above is of no importance for the verification of $\delta \zu \rho_{\hspace{-0.2mm}\scalebox{0.55}{$B$}} = 0$. If this condition is to be satisfied, the most important quantity will indeed be the trace term above.


As previously presented, we have already obtained the expressions for the quantities $L_\alpha$ and $\delta L_\alpha$ present in (\ref{eq: charge density variation n eq 2}), in the equations (\ref{eq: left side alpha tilde expansion}) and (\ref{eq: variation alpha simplified}), respectively. What we then need to do, is calculate the derivative of these two quantities, multiply them together and take the trace of (\ref{eq: charge density variation n eq 2}). This is a very long calculation to perform, considering the numerous quantities present in $L_\alpha$ and, particularly, in $\delta L_\alpha$\zc . Nevertheless, the overall structure of the expression \nlb to be obtained can be established by a simple analyis, involving the terms that make up $L_\alpha$ and \nlb $\delta L_\alpha$\zc .

Starting with $L_\alpha$\zc , let us first mention that the conjugation by $e^{i\zt\xi\zt T_1}$, that appear on this quantity, will also be present on the other three terms of (\ref{eq: charge density variation n eq 2}), since it appears in $\delta L_\alpha$\zc , by equation (\ref{eq: variation alpha simplified}). Therefore, despite the presence of the derivatives in (\ref{eq: charge density variation n eq 2}), it can be easily shown that these conjugations by $e^{i\zt\xi\zt T_1}$ will cancel out, by the cyclic property of the trace. Having said that, returning then to the quantity $L_\alpha$ or, more appropriately, to the quantity $\widetilde{L}_\alpha$ of (\ref{eq: left side alpha tilde expansion}), one sees by its coefficients (\ref{eq: L1 tilde}) and (\ref{eq: L2 tilde}) that this quantity is expressed in terms of the functions $K$ and $F$, given in (\ref{eq: expression K}) and (\ref{eq: definition gamma and F}), respectively. In fact, on equation (\ref{eq: definition gamma and F}), one sees that the function $F$ is itself also determined in terms of $K$. Therefore, since we have:

\begin{equation}\label{eq: F derivative zeta}
\frac{dF}{d \zu \zeta} = \frac{K}{F} \frac{dK}{d\zu\zeta} \cos^{\zt\zu2}\hspace{-0.5mm}\tau
\end{equation}

\vspace{0.225cm}

\noindent{}and since we have a general dependence of $\widetilde{L}_\alpha$ with $K$, we conclude that the derivative of this quantity, which appears in the first term of (\ref{eq: charge density variation n eq 2}), will acquire a global factor of $dK/d\zu\zeta$. Here, remember that the radius $r$ that is being written in (\ref{eq: charge density variation n eq 2}) is, in reality, the parameter $\zeta$ above.

Now, for the quantity $\delta L_\alpha$\zc , or $\delta \widetilde{L}_\alpha$\zc , if we factor out the conjugation by $e^{i\zt\xi\zt T_1}$, we have that the derivative of this quantity will give two types of terms. The first one will also have a global factor of $dK/d\zu\zeta$\zt , since $\delta \widetilde{L}_\alpha$ is itself also predominantly expressed in terms of $K$. However, for the quantity $\delta \widetilde{L}_\alpha$\zc , we will have another quantity whose derivative with respect to $r$ (or $\zeta$) must be evaluated: the arbitrary parameter $b$ of (\ref{eq: arbitrary variation}), regarding the reparameterization. On this equation, we have defined this parameter as depending on $\sigma$ and $\tau$ only, since the arbitrary deformations of the loops are restricted to a surface, and therefore to a constant $\zeta$. Nevertheless, being the parameter $b$ completely arbitrary, there is no reason, in principle, to not consider it depending also on the radius $\zeta$. This does not mean that the deformations will no longer be restricted to the surfaces, but instead, that the arbitrary reparameterizations may be differently performed if we are considering surfaces of different radii. So, we are thus led to consider:

\begin{equation}\label{eq: b dependence}
b = b \zt ( \zu \sigma,\tau,\zeta \zu\zu)
\end{equation}

\vspace{0.25cm}

\noindent{}and therefore we see that the derivative of $\delta \widetilde{L}_\alpha$ will acquire a second type of term, depending on the quantity $d\zu b/d\zu\zeta$. Consulting equation (\ref{eq: variation alpha simplified}) for $\delta L_\alpha$\zc , we see that such terms will depend on $db/d\zu\zeta$ through some new coefficients $D^{\zu\zu i}$, and that there will be a global factor of $K \big( \zu\zu 1 - K^{ \zu\zu 2} \zu\zu \big)$.

\phantom{Paragraph}

Considering then all the terms mentioned on the previous paragraphs, one can indeed show that the quantity $\delta \zu \rho_{\hspace{-0.1mm}\scalebox{0.55}{$B$}}$\zc , calculated with (\ref{eq: charge density variation n eq 2}), will have an overall structure given by:

\begin{equation}\label{eq: reparam density}
\delta \zu \rho_{\hspace{-0.1mm}\scalebox{0.55}{$B$}} = K \zu \zu \frac{dK}{d\zu\zeta} \zu\zu \mathcal{F} \zu\zu (b) + K \big( \zu\zu 1 - K^{ \zu\zu 2} \zu\zu \big) \zc \mathcal{G} \bigg( \frac{d\zu b}{d\zu\zeta} \bigg)
\end{equation}

\vspace{0.225cm}

\noindent{}where $\mathcal{F}$ and $\mathcal{G}$ are complicated functions, depending on the arbitrary quantities $b$ and $d\zu b/d\zu\zeta$, respectively. For the quantity $K \cdot dK/d\zu\zeta$ above one obtains, by employing (\ref{eq: expression K}):

\begin{equation}\label{eq: K derivative}
K \zu \zu \frac{dK}{d\zu\zeta} = -4\zt\zeta \zu \frac{1 - \zeta^2 + (x^4)^2}{[ \zu\zu\zu 1 + \zeta^2 + (x^4)^2 \zt]^{\zu\zu 3}}
\end{equation}

\vspace{0.225cm}

With this result, and given the presence of $K \big( \zu\zu 1 - K^{ \zu\zu 2} \zu\zu \big)$ in (\ref{eq: reparam density}), one sees that the reparamete\-rization-invariance condition $\delta \zu \rho_{\hspace{-0.1mm}\scalebox{0.55}{$B$}} = 0$ will be satisfied (only) for the same four cases indicated below equation (\ref{eq: variation alpha simplified}) for the reparameterization-invariance of $\Phi_B$\zc : (i) $r=0$ with $x^4$ arbitrary\footnote{As we have mentioned, the case $r=0$ requires a more careful analysis because of the factor $1/4\pi r^2$ in (\ref{eq: charge density variation n eq 2}). Nevertheless, we shall not pursue such discussion, since the result obtained for $r=0$ is of little physical interest.}, (ii) $r=1$ with $x^4 = 0$, (iii) $r \rightarrow \infty$ with $x^4$ arbitrary and (iv) $x^4 \rightarrow \pm \infty$ with $r$ arbitrary. For the case of the gauge-invariant flux, case (ii) was the most interesting observable result. Here, the corresponding gauge-invariant charge density is also observable, but of much less physical interest since it corresponds to a null value, as indicated, for instance, in Figure \ref{fig:alpha density t eq 0}.

If the arbitrary parameter $b$ could be considered as depending only on $\sigma$ and $\tau$, and not on $\zeta$, the second term of (\ref{eq: reparam density}) would not be present, and by (\ref{eq: K derivative}) we would have another class of observable results, corresponding to the radii and Euclidean times satisfying $1 - \zeta^2 + (x^4)^2 = 0$. In fact, it can be shown that this relation correspond to the null points at the transitions between the positive and negative regions of charge density, which can be seen in Figure \ref{fig:density time evolution}. Nevertheless, there is no reason for not considering the general dependence (\ref{eq: b dependence}), and thus this aforementioned class of results is not reparameterization-invariant, and thus not observable.

\phantom{Paragraph}

For the radii and Euclidean times not covered by the four cases indicated above, we have $\delta \zu \rho_{\hspace{-0.1mm}\scalebox{0.55}{$B$}} \neq 0$, for the instanton and anti-instanton solutions. Thus, the corresponding gauge-invariant magnetic (or electric) charge densities are not reparameterization-invariant, and are defined on the generalized loop space $\mathcal{L}^{(2)}$ only. Despite that, following the discusssion made after Figure \ref{fig:alpha density t eq 0}, such results may still be of interest as they indicate the overall internal charge configuration of the instanton and anti-instanton. Such a magnetic (or electric) charge configuration must indeed be present for these solutions, since we obtained the observable flux $\Phi_B$ for $r=1$, $x^4=0$.

\section{Conclusions}
\label{sec:conclusions}
\setcounter{equation}{0}

In this paper, we have obtained some gauge-invariant results by making use of the integral equations of Yang-Mills theories, obtained for the first time by our group, in 2012 \cite{ym1,ym2}. Working with the integral equation of order $\alpha$, given in (\ref{eq: alpha equation}), we have considered a purely  spatial  volume $\Omega$ on Euclidean space-time, for which equation (\ref{eq: alpha equation}) corresponds to the non-Abelian version of Gauss law for the magnetic fields. With such equation, we then considered the self-dual instanton and anti-instanton solutions given in (\ref{eq: instanton solution}). Because of self-duality, the $\alpha$-equation (\ref{eq: alpha equation}) corresponds also to the non-Abelian version of Gauss law for the electric fields. Now, regarding the aforementioned integration volume, we have considered $\Omega$ as corresponding to spheres, centered at the origin and of radius $r$, whose parameterization was constructed in \cite{directtest}. Then, by making use of (\ref{eq: alpha equation}), we have plotted the gauge-invariant magnetic (or electric) fluxes $\Phi_{\hspace{-0.2mm}B}$ calculated on the spherical surfaces $\partial\Omega$, which by the Gauss law can be identified with the gauge-invariant enclosed magnetic (or electric) charges on $\Omega$. Such results can be observed, for instance, in Figure \ref{fig:time evolution}. Also, we have demonstrated that the magnetic (or electric) flux $\Phi_{\hspace{-0.2mm}B}$\zc , calculated for a spherical surface of radius $r=1$, at the Euclidean time $x^4=0$, is observable. It was pointed out that this specific radius corresponds in fact to the size $\lambda$ of the instanton (and anti-instanton) solution. This observable flux thus reveals an internal charge configuration for the instanton and anti-instanton solutions, coming from the non-Abelian Gauss law (\ref{eq: alpha equation}).

Making use of the gauge-invariant flux $\Phi_{\hspace{-0.2mm}B}$ that comes from the integral equation of order $\alpha$, we have defined a gauge-invariant magnetic charge density, by equation (\ref{eq: charge density definition}). Because of self-duality, this also corresponds to an electric charge density for the instanton and anti-instanton solutions. We have plotted the magnetic (or electric) charge density $\rho_{\hspace{-0.2mm}\scalebox{0.55}{$B$}}$ for different Euclidean times, and the results can be seen, for instance, in Figure \ref{fig:density time evolution}. Also, we have demonstrated that these results are not reparameterization-invariant, with the exception of some specific critical points, in particular in one where the charge is maximum and the density is zero. Consequently, such results are not observable, being defined on the generalized loop space $\mathcal{L}^{(2)}$ only. Despite that, it was argued that the gauge-invariant charge density $\rho_{\hspace{-0.2mm}\scalebox{0.55}{$B$}}$ may still be of interest, since it indicates the overall internal charge configuration that was shown to exist for the instanton and anti-instanton solutions.

\section*{Acknowledgements}

The authors are grateful to H. Malavazzi for valuable discussions regarding the results of this paper, and to P. A. Faria da Veiga and D. M. Schmidtt for pertinent comments. CAdS acknowledges the financial support of FAPESP (Fundação de Amparo à Pesquisa do Estado de São Paulo), grants 2025/24377-3 and 2023/13750-0. LAF acknowledges the financial support of FAPESP, grant 2025/09036-5, and of CNPq (Conselho Nacional de Desenvolvimento Científico e Tecnológico), grant 307833/2022-4.

This study was financed, in part, by the São Paulo Research Foundation (FAPESP), Brasil. Process Numbers: 2025/24377-3, 2023/13750-0 and 2025/09036-5.

\appendix

\section{The parametric equations for the spatial spheres}
\label{apx:parameterization equations}
\setcounter{equation}{0}

In this appendix we list the parametric equations for all the segments of the loops of types I, II and III, presented in Subsection \ref{subsec:parameterization} for the parameterization of the spatial spheres. These expressions were taken from \cite{directtest}, where this parameterization was first constructed.

\phantom{Paragraph}

For the \textit{loops of type I}, where $\tau$ varies from $-\infty$ to $-\pi/2$, we have:

\begin{itemize}[label=\scalebox{0.75}{$\blacktriangleright$}] 

\item \textit{Segment I.1.} Straight line going from $x_R$ to some arbitrary point on the thin cylinder:

\begin{equation}\label{eq_apx1: segment i.1}
\left\{
\begin{aligned}
x^1 &= \tau + \sigma - \zeta + \frac{\pi}{2} \\
x^2 &= 0 \hspace{3.2cm} -\infty \leq \sigma \leq 0 \\
x^3 &= -\epsilon
\end{aligned}
\right.
\end{equation}

\vspace{0.2cm}

\item \textit{Segment I.2.} Circular path going around the surface of the thin cylinder:

\begin{equation}\label{eq_apx1: segment i.2}
\left\{
\begin{aligned}
x^1 &= \tau - \zeta + \frac{\pi}{2} \\
x^2 &= \epsilon\sin\sigma \hspace{1.7cm} 0 \leq \sigma \leq 2\pi \\
x^3 &= -\epsilon \cos\sigma
\end{aligned}
\right.
\end{equation}

\vspace{0.2cm}

\item \textit{Segment I.3.} Straight line going back to $x_R$\zc , in order to close the loop:

\begin{equation}\label{eq_apx1: segment i.3}
\left\{
\begin{aligned}
x^1 &= \tau + 2\pi - \sigma - \zeta + \frac{\pi}{2} \\
x^2 &= 0 \hspace{4.3cm} 2\pi \leq \sigma \leq \infty \\
x^3 &= -\epsilon
\end{aligned}
\right.
\end{equation}

\end{itemize}

\vspace{2mm}

Now, for the \textit{loops of type II}, where $\tau$ varies from $-\pi/2$ to $\pi/2$, we have:

\begin{itemize}[label=\scalebox{0.75}{$\blacktriangleright$}] 

\item \textit{Segment II.1.} Straight line going from $x_R$ to the point attached to the sphere:

\begin{equation}\label{eq_apx1: segment ii.1}
\left\{
\begin{aligned}
x^1 &= \sigma - \zeta \\
x^2 &= 0 \hspace{1.6cm} -\infty \leq \sigma \leq 0 \\
x^3 &= -\epsilon
\end{aligned}
\right.
\end{equation}

\vspace{0.2cm}

\item \textit{Segment II.2.} Circular path on the surface of the sphere:

\begin{equation}\label{eq_apx1: segment ii.2}
\left\{
\begin{aligned}
x^1 &= \zeta \big[ \cos^{\hspace{0.2mm}2}\hspace{-0.5mm}\tau \hspace{0.5mm} (1 - \cos\sigma) - 1 \big] \\
x^2 &= \zeta\cos\tau\sin\sigma \hspace{3.2cm} 0 \leq \sigma \leq 2\pi \\
x^3 &= \zeta\cos\tau\sin\tau \hspace{0.5mm} (1 - \cos\sigma)
\end{aligned}
\right.
\end{equation}

\vspace{0.2cm}

\item \textit{Segment II.3.} Straight line going back to $x_R$\zc , in order to close the loop:

\begin{equation}\label{eq_apx1: segment ii.3}
\left\{
\begin{aligned}
x^1 &= - \sigma + 2\pi - \zeta \\
x^2 &= 0 \hspace{3.0cm} 2\pi \leq \sigma \leq \infty \\
x^3 &= -\epsilon
\end{aligned}
\right.
\end{equation}

\end{itemize}

\vspace{2mm}

Finally, for the \textit{loops of type III}, where $\tau$ varies from $\pi/2$ to $\infty$, we have:

\begin{itemize}[label=\scalebox{0.75}{$\blacktriangleright$}] 

\item \textit{Segment III.1.} Straight line going from $x_R$ to some arbitrary point on the thin cylinder:

\begin{equation}\label{eq_apx1: segment iii.1}
\left\{
\begin{aligned}
x^1 &= \frac{\pi}{2} - \tau - \zeta + \sigma \\
x^2 &= 0 \hspace{3.25cm} -\infty \leq \sigma \leq 0 \\
x^3 &= -\epsilon
\end{aligned}
\right.
\end{equation}

\vspace{0.2cm}

\item \textit{Segment III.2.} Straight line going back to $x_R$\zc , in order to close the loop:

\begin{equation}\label{eq_apx1: segment iii.2}
\left\{
\begin{aligned}
x^1 &= \frac{\pi}{2} - \tau - \zeta - \sigma \\
x^2 &= 0 \hspace{3.35cm} 0 \leq \sigma \leq \infty \\
x^3 &= -\epsilon
\end{aligned}
\right.
\end{equation}

\end{itemize}

\vspace{2mm}

It must also be mentioned that the parameter $\zeta$ varies from $0$ to $\zeta_f$ during the scanning of $\Omega$, with $\zeta_f$ being any radius $0 \leq \zeta_f < \infty$. For $\zeta_f \rightarrow \infty$, $\Omega$ corresponds to the whole space.

\label{lastpage}
\end{document}